\documentclass{emulateapj}
\usepackage{graphicx}

\begin{document}

\title{Three-dimensional simulations of the non-thermal broad-band emission from young supernova remnants including efficient particle acceleration}

\shorttitle{Non-thermal emission from supernova remnants undergoing particle acceleration}
\shortauthors{Ferrand et al.}

\author{Gilles Ferrand}
\affil{Department of Physics and Astronomy, University of Manitoba, Winnipeg, MB R3T 2N2, Canada}
\email{gferrand@physics.umanitoba.ca}

\author{Anne Decourchelle}
\affil{Laboratoire AIM (CEA/Irfu, CNRS/INSU, Universit{\'e} Paris VII), CEA Saclay, b{\^a}t. 709, F-91191  Gif sur Yvette, France}
\email{anne.decourchelle@cea.fr}

\and

\author{Samar Safi-Harb\altaffilmark{1}}
\affil{Department of Physics and Astronomy, University of Manitoba, Winnipeg, MB R3T 2N2, Canada}
\email{samar@physics.umanitoba.ca}
\altaffiltext{1}{Canada Research Chair}

\submitted{accepted for publication in ApJ, May 3, 2014}

\begin{abstract}

Supernova remnants are believed to be the major contributors to Galactic cosmic rays. In this paper, we explore how the non-thermal emission from young remnants can be used to probe the production of energetic particles at the shock (both protons and electrons). 
Our model couples hydrodynamic simulations of a supernova remnant with a kinetic treatment of particle acceleration. We~include two important back-reaction loops upstream of the shock: energetic particles can (i)~modify the flow structure and (ii)~amplify the magnetic field. As the latter process is not fully understood, we use different limit cases that encompass a wide range of possibilities. We~follow the history of the shock dynamics and of the particle transport downstream of the shock, which allows us to compute the non-thermal emission from the remnant at any given age. We~do this in 3D, in order to generate projected maps that can be compared with observations.
We~observe that completely different recipes for the magnetic field can lead to similar modifications of the shock structure, although to very different configurations of the field and particles. We show how this affects the emission patterns in different energy bands, from radio to X-rays and $\gamma$-rays. High magnetic fields ($>100 \mu$G) directly impact the synchrotron emission from electrons, by restricting their emission to thin rims, and indirectly impact the inverse Compton emission from electrons and also the pion decay emission from protons, mostly by shifting their cut-off energies to respectively lower and higher energies.

\end{abstract}



\section{Introduction
\label{Introduction}}

Supernovae are the final stage of stellar evolution for white dwarfs and massive stars. Supernova remnants (SNRs), produced by the interaction of the ejected stellar material with the ambient medium \citep{Chevalier1977a}, are a key link between the life of stars and the evolution of the interstellar medium (ISM). Moreover, they are believed to be the main production sites of Galactic cosmic-rays, up to the so-called ``knee" in their spectrum (at a few PeV). Indeed the supersonic ejecta generate a strong, magnetized shock wave, that can efficiently accelerate charged particles through the process of {\em diffusive shock acceleration} \citep{Drury1983a}. A variety of multi-wavelength observations supports this scenario, especially at high energies; see \cite{Vink2012a} and \cite{Helder2012a} for recent reviews, and \cite{Ferrand2012b} for an up-to-date catalogue of X-ray and $\gamma$-ray observations of all Galactic SNRs. However, if the presence of relativistic particles in SNRs is well established, their nature and their actual energy distribution is still unclear. In the last decade, several remnants were finally detected in $\gamma$-rays, but it proved difficult to disentangle the leptonic and hadronic contributions from inverse Compton scattering and neutral pion decay, respectively (e.g. \citealt{Gabici2008a}). In the GeV $\gamma$-ray band, the characteristic pion-decay spectral signature could be detected with certainty only in a couple of middle-aged remnants \citep{Ackermann2013b}, and does not require the underlying proton spectrum to extend above the TeV energies. In the TeV $\gamma$-ray band, the hadronic scenario is favoured in several remnants because of the presence of molecular clouds that make good targets for protons (e.g. \citealt{Aharonian2008a,Aharonian2008b}), but again, due to their advanced age, these objects are not believed to still be Pevatron accelerators, as required to power the Galactic cosmic rays up to ``knee" energies. Younger, potentially more active objects (with faster, stronger shocks), were finally detected as well, amongst the historical remnants. One of the most interesting targets is Tycho's SNR (G120.1+1.4), a remnant of a type~Ia supernova, that has a nearly spherical shape and evolves in a relatively uniform ISM. In this object, most models point to a hadronic origin of the $\gamma$-ray emission, and imply modest to very high efficiencies of acceleration at the shock \citep{Tang2011b,Morlino2012a,Berezhko2013a,Zhang2013a,Slane2014a}. 

When modeling the emission from a SNR, it is important to have a joint description of the remnant and of particle acceleration (energetic protons and electrons). If particle acceleration is as efficient as is believed, especially regarding ions, then this must impact the dynamics of the flow and of the magnetic field -- and in turn the way acceleration works. The back-reaction of the energetic particles on the shock is two-fold: (i) the pressure of particles can set the fluid in motion \citep{Malkov2001c}, and (ii) their current can amplify the magnetic field \citep{Schure2012a}. Both effects happen immediately ahead of the shock, in the so-called precursor, where energetic particles are free to wander in the yet un-shocked medium. To compute the emission from the SNR, it is mandatory to solve jointly the (magneto-)hydrodynamic equations describing the evolution of the fluid and the kinetic equations describing the acceleration of the particles. Various numerical approaches are possible to tackle this difficult problem. Direct computations of the coupled system are expensive, and require special treatments to be useful for the large scale modeling of SNRs; in particular, \cite{Kang2006a} performed simulations in a comoving grid expanding with the (forward) shock, while \cite{Zirakashvili2012a} introduced a change of variables that allows for the simultaneous handling of the forward and reverse shocks. An alternative to explicitly solving the differential equation describing the transport of particles is to resort to Monte-Carlo approaches, where the particles are treated in a stochastic way while the fluid is treated with standard numerical methods; for instance \cite{Schure2010a} used stochastic differential equations coupled with an hydrodynamic code. It is possible to simplify the problem by using a two-fluid approximation, that is by relying on a hydrodynamic description for both the thermal and relativistic components of the system, with the drawback of losing the energy dependence of the diffusion coefficient; this allowed \cite{Kosenko2011a} to make parametric studies. We advocate here a hybrid approach, where a hydrodynamic description of the SNR evolution is coupled to a kinetic model of acceleration. \cite{Decourchelle2000a} coupled a simple acceleration model, in which the spectrum of particles is linear by parts, to a code computing self-similar hydrodynamic solutions. \cite{Ellison2004a} coupled the same acceleration model to a code solving the Euler equations. \cite{Ellison2007a} switched to a more realistic acceleration model, and recently \cite{Lee2012a} improved its treatment of the magnetic field. In this approach, the fluid simulated by the hydrodynamic code (that can be called a ``pseudo-fluid") accounts for both the thermal and non-thermal populations, yet the kinetic model ensures that the mix of the two follows the expectations from non-linear DSA at any given time and position. This allows for a realistic description of some important aspects of the acceleration mechanism, while having a relatively small overhead compared to a purely hydrodynamic simulation. 

Our overall aim in our ongoing modeling effort is to constrain the acceleration properties in a SNR from its broadband thermal and non-thermal emission. In our preliminary work \citep[hereafter paper~I]{Ferrand2010a}, we presented the morphological evolution of the remnant, taking into account both the development of the instabilities and the shrinking of the shocked region caused by the presence of energetic particles, in a time-dependent manner. In our subsequent paper \citep[hereafter paper~II]{Ferrand2012g}, we computed the thermal X-ray emission from the shocked material, showing how it is affected by the presence of energetic particles. In the present work, we compute the {\em non-thermal} broadband emission from the accelerated particles (from radio to $\gamma$-rays), again discussing non-linear effects in the interpretation of observations -- and now including different hypotheses for the evolution of the magnetic field. An important aspect in our modeling is to take into account the actual morphology of the remnant, by running three-dimensional simulations, whereas the modeling efforts previously cited that couple the time evolution of particle acceleration with the hydrodynamics are one-dimensional, based on the simplifying assumption of spherical symmetry. Even under the assumptions of uniform ejecta and ISM, the assumption of symmetry does not hold for young remnants: the interface between the shocked ISM and the shocked ejecta (called the contact discontinuity) is subject to Rayleigh-Taylor instabilities, which produce a distinctive pattern of fingers and holes \citep{Chevalier1992a,Chevalier1995a}. It is mandatory to take this effect into account to properly quantify the size of the shrinked shocked ISM and to reproduce spatially-resolved images of the thermal emission, as observed with X-ray satellites (see e.g. \cite{Warren2005a} and \cite{Cassam-Chenai2007a} for Tycho's SNR). In this work we consider that particles are accelerated at the forward shock, and so we concede that the non-thermal emission, that mostly follows the distribution of the shocked ISM, has a simpler geometry than the thermal emission, that mostly follows the distribution of the ejecta. The impact of the instabilities occurring inside the shell may be more subtle, although it should be investigated as well. In any case, we want to be able to compute all kinds of emission within a single model, which offers the best possible description of the SNR evolution. 
\cite{Kang2012a} showed the importance of the geometry of the downstream region in the computation of the emission, and \cite{Kang2013d} further showed the importance of properly taking into account the dynamical evolution of the SNR shock to obtain the particle population at a given age. In this work, we compute the continuous distribution of particles in both space and time, as consistently as we can in the framework of our pseudo-fluid approach. 
\cite{Atoyan2012a} demonstrated the importance of multi-zone versus single-zone modeling when interpreting the $\gamma$-ray emission from Tycho's SNR and trying to assess the presence or absence of energetic protons in its shell. Whereas they used an ad-hoc parametrization of the zones, our model naturally produces the fine-grained radial structure of the shocked region. Moreover, we would like to emphasize that, to be able to reproduce the observations, one needs to compute the emissivity as seen in projection in the plane of the sky. Then each pixel of the map is obtained by integrating all the variations that occur along the line of sight, that are due both to geometry effects (non-uniform plasma conditions in the interior) and age effects (non-uniform conditions for energetic particles and the magnetic field at the shock).

The outline of the paper is as follows. Our method is presented in Section~\ref{sec:model}. We briefly recall how we couple a hydrodynamic code with a kinetic model of non-linear acceleration, and we detail the treatment of the magnetic field and of the accelerated particles in the shocked ISM. Our results are presented in Section~\ref{sec:results}. We first show slices of some useful hydrodynamic quantities, to demonstrate the effect of the particles' back-reaction on the shock structure and on the downstream magnetic field, and how this impacts the energy losses of particles. We then show projected maps of the various non-thermal emission mechanisms in different energy bands, relevant to currently available instruments. Finally we present broad-band spectra, integrated over the whole remnant, to compare the signatures of the different emissions of the two kinds of particles (electrons vs. protons). Our findings are discussed in Section~\ref{sec:discussion}. We summarize the main observational diagnostics of efficient acceleration of particles (electrons and protons) inside young SNRs, and we review the main physical assumptions underlying our model. Our conclusions are presented in Section~\ref{sec:conclusion}.


\section{Model}
\label{sec:model}

In this section we present our numerical model. We recall how we couple a hydrodynamic code and a kinetic model to run simulations of a SNR undergoing efficient particle acceleration (Section~\ref{sec:model-SNR}). We discuss the fate of the magnetic field upstream of the shock, where it is affected by particle-driven instabilities (Section~\ref{sec:model-MFA}), as well as downstream of the shock, where it is advected with the flow, together with the energetic particles (Section~\ref{sec:model-down}). Finally we present the code used to compute the non-thermal emission of the particles at any point inside the shell (Section~\ref{sec:model-em}).

\subsection{Remnant evolution with particle acceleration}
\label{sec:model-SNR}

Particles are assumed to be accelerated at the blast wave of SNRs through the process of diffusive shock acceleration, in which charged particles gain energy by repeatedly crossing the velocity jump. Here we consider acceleration at the forward shock only, as in our previous work (see paper~II for a discussion of the possible role of the reverse shock). Throughout this paper, we are working with the isotropic distribution function of particles in phase space $f(t,\mathbf{x},p)$, a function of time~$t$, position~$\mathbf{x}$, and momentum~$p$. In the test-particle regime, the final spectrum is a power-law $f(p) \propto p^{-s}$ of index $s=3R/(R-1)$ where $R$ is the compression ratio of the shock (for a strong shock $R=4$ and thus $s=4$). This index in turns determines the spectrum of the emitted photons. But an important feature of efficient acceleration is its non-linearity, which leads to a variety of spectra, hence the need for a consistent treatment of the thermal and non-thermal populations. Here we briefly outline our approach, which was introduced in paper~I and detailed in paper~II.

\subsubsection{Hydrodynamics}
\label{sec:model-hydro}

The evolution of the remnant is described using the code RAMSES \citep{Teyssier2002a}. It~is used to solve the Euler equations on a 3D Cartesian grid, which is comoving with the blast wave \citep{Fraschetti2010a}. The tree-based grid is adaptively refined around the forward shock that propagates in the ISM, the reverse shock that propagates backwards in the ejecta, and the contact discontinuity that separates the ejecta from the ISM. These two media are a single fluid within the hydrodynamic solver, with their respective position known thanks to a passive scalar~$f_{\rm ej}$ tracing the ejecta. 

Although our numerical setup is very efficient in terms of computational resources (thanks to the adaptive mesh refinement and comoving grid),  the shock fronts are affected by the so-called carbuncle instability \citep{Fraschetti2010a} because they are quasi-stationary. In paper~II focused on the thermal emission, the effect was noticeable but could be neglected, given that most of the radiation came from the ejecta. In this paper focused on the non-thermal emission, it appeared mandatory to improve the numerical stability of the forward shock front, given that most of the radiation comes from its immediate vicinity. To this end, we implemented a smoothing procedure within the code: in the medium located just upstream of the shock, the density and pressure are replaced by their azimuthally-averaged values. This leads to a much cleaner shock front, while still preserving the overall remnant's evolution, and without disturbing the flow in the downstream region. However, this restricts our ability to investigate the propagation of the forward shock in a medium that would vary strongly with azimuth.

\subsubsection{Particle acceleration}
\label{sec:model-accel}

Diffusive shock acceleration is computed at each time step using the semi-analytical kinetic model of Blasi (\citeyear{Blasi2002a}; \citeyear{Blasi2004a}), as explained in papers~I and~II. Given the shock properties (velocity and upstream conditions), this model jointly solves the particle spectrum~$f(p)$ and the fluid velocity profile~$u(x)$. Note that it can only find stationary solutions, so, assuming that the acceleration process can be described as a succession of quasi steady-states, we re-compute the results after each hydrodynamic time-step, given the current flow conditions and the maximum energy particles have reached at that time. 

We recall that the injection of protons is parametrized with the thermal leakage recipe of \cite{Blasi2005a}, controlled by the free parameter $\xi$ defined as 
\begin{equation}
\label{eq:p_inj}
\xi = \frac{p_{\rm inj}}{p_{\rm th}}\:,
\end{equation}
where $p_{\rm inj}$ is the injection momentum of particles and $p_{\rm th}$ is the mean thermal momentum immediately downstream of the shock. By definition, particles at $p_{\rm inj}$ are somewhat more energetic than the average, so that their Larmor radius is a few times larger than the width of the collisionless shock. As a result, $\xi$ is expected to be in the range 2--4, which corresponds to a fraction of particles injected at the shock front, $\eta$, ranging from roughly $10^{-5}$ to $10^{-1}$. From a theoretical perspective, the injection mechanism for electrons is less clear than for protons, but, apart from the vicinity of the minimum and maximum momenta, particles of same rigidity $pc/|q|$ will have the same spectral shape ($p$~is the particle momentum and $q$ its charge). So, as commonly done, we normalize their spectrum with respect to the protons' spectrum:
\begin{equation}
\label{eq:Kep}
f_e(m_p c) = K_{ep} \times f_p(m_p c)
\end{equation}
with the normalization $K_{ep} \ll 1$ being a free parameter.

The maximum energy of protons is limited by both the finite age and finite size of the remnant, according to the diffusion time- and length-scales \citep{Ptuskin2003a}, under the common assumption of Bohm diffusion. For electrons, the maximum energy is limited by radiative losses (more in Section~\ref{sec:model-down-part}), and is computed according to the procedure outlined in \cite{Morlino2009a}. 

\subsubsection{Hydro-kinetic coupling}
\label{sec:model-coupling}

When ions are efficiently accelerated, their pressure modifies the shock's structure: a precursor is formed ahead of the shock front, where the ambient medium is pre-accelerated and pre-heated by the streaming energetic particles \citep{Malkov2001c}. Then it is useful to make the distinction between the unperturbed ISM far upstream of the shock front (denoted hereafter by subscript~0), the perturbed ISM immediately upstream of the shock front (subscript~1), and the perturbed and compressed ISM immediately downstream of the shock front (subscript~2). We can now define two compression ratios: $R_\mathrm{sub}$ over the actual discontinuity (from medium~1 to medium~2), and $R_\mathrm{tot}$ over the whole structure (from medium~0 to medium~2). The compression ratio $R_\mathrm{sub}$ (resp. $R_\mathrm{tot}$) is expected to be lower (resp. higher) than in the standard adiabatic case $R=4$. 

In our model, we implement this back-reaction of particles on the shock with a convenient pseudo-fluid approach: the fluid evolved by the hydrodynamic code in the shocked region accounts for both the thermal and non-thermal populations. This is made possible through the use of an effective adiabatic index~$\gamma_{\mathrm{eff}}$, which is a function of both time and space. This variable index is set according to the acceleration model of Section~\ref{sec:model-accel}, and embodies the effect of DSA on the flow outlined in the previous paragraph. In effect, it makes the fluid more compressible, so that the shocked region gets denser and narrower, as required (see \citealt{Decourchelle2000a} and paper~I). This hydro-kinetic coupling is applied to the initial SNR profiles, computed from the self-similar solutions of \cite{Chevalier1983a}, and then after each hydrodynamic update in RAMSES.

Of course, ideally one would like to directly integrate the transport equation of particles over time (as done e.g. in \citealt{Kang2013d}), however such simulations are computationally expensive, and still unpractical in 3D for realistic applications. In our work we choose to focus on the description of the SNR evolution. We note that both \cite{Morlino2012a} and \cite{Slane2014a} used the same model of acceleration in their modeling of Tycho. The former coupled it to prescribed, analytical, un-modified solutions of the shock evolution, while the later coupled it self-consistently to a hydro code as we do. Both \cite{Kang2013d} and \cite{Slane2014a} assumed spherical symmetry, which is appropriate after the remnant has entered the Sedov-Taylor phase of expansion. Our setup can accurately describe the dynamics of the shell and its interior from a young age, and as it evolves through different stages.

\subsection{Magnetic field amplification}
\label{sec:model-MFA}

Both theoretical and observational evidence imply that energetic particles can substantially amplify the magnetic field upstream of the shock, by exciting magnetic waves (see \citealt{Schure2012a} and \citealt{Bykov2012a} for recent reviews and references therein). Because the magnetic turbulence controls the diffusion of the particles, this constitutes a {\em second} back-reaction loop. This effect needs to be included when computing the non-thermal emission, because the fate of the magnetic field is critical for the emission of energetic electrons. As we shall see, it also impacts the intensity of the shock modification, in a way that is similar to changing the injection level. Because the latter is poorly constrained, we had deliberately not included a discussion of the impact of the magnetic field in paper~II focused on the thermal emission. In this paper, we explicitly describe its evolution in the precursor under different scenarios.

The shock precursor (the region where streaming energetic particles can affect the incoming flow) is not modeled in the hydrodynamic part of the simulation. It is modeled inside the kinetic acceleration model, under the assumption that the diffusion coefficient $D(p)$ is a fast and strictly rising function of momentum~$p$. If the shock travels at speed $u_S$, particles of momentum~$p$ can travel a certain distance
\begin{equation}
x_p \simeq \frac{D(p)}{u_S}
\end{equation}
ahead of the shock front, where the fluid has a density~$\rho_p$ and a velocity~$u_p$. The shock modification then follows from the condition of conservation of the total momentum between the far unperturbed upstream medium and any point~$x_p$ in the precursor: 
\begin{equation}
P_{\mathrm{dy},p} + P_{\mathrm{th},p} + P_{\mathrm{cr},p} + P_{\mathrm{w},p} = P_{\mathrm{dy},0} + P_{\mathrm{th},0} + P_{\mathrm{cr},0} + P_{\mathrm{w},0} \:, 
\end{equation}
where $P_{\mathrm{dy}} = \rho u^2$ is the dynamical pressure, $P_{\mathrm{th}}$ is the thermal pressure, $P_{\mathrm{cr}}$ is the particle pressure computed from their distribution~$f(p)$:
\begin{equation}
P_\mathrm{cr} = \int_{p} \frac{pv}{3}\:f(p)\:4\pi p^{2}\mathrm{d}p
\end{equation}
with $p$ and $v$ being the momentum and velocity of the particles, and $P_{\mathrm{w}} = {\delta\mathbf{B}^{2}}/{8\pi}$ is the pressure of the magnetic waves. It is the purpose of the subsequent sections to provide estimates of $P_{\mathrm{w},p}$ and $P_{\mathrm{th},p}$.

\subsubsection{Field amplification}
\label{sec:model-MFA-amplification}

Various instabilities have been invoked to explain magnetic filed amplification (MFA) in the vicinity of SNR shocks, both resonant and non-resonant, operating at different time- and length-scales. Here we use a simple model for the growth of the waves' pressure, proposed by \cite{Caprioli2009a} for the most commonly considered resonant instability:
\begin{equation}
\label{eq:Pw,p-res}
\frac{P_{\mathrm{w},p}}{\rho_{0}u_{0}^{2}}=\frac{1-\zeta}{4M_{A,0}}\, \left(\frac{u_p}{u_0}\right)^{-3/2}\left(1-\left(\frac{u_p}{u_0}\right)^{2}\right)\
\end{equation}
where $0 \le \zeta \le 1$ is a free parameter (that is also related to field damping and plasma heating, see Section~\ref{sec:model-MFA-damping}). This formula was derived under the assumption of large sonic and Alfv{\'e}nic Mach numbers in the unperturbed upstream medium. We recall that the sonic Mach number is defined as
\begin{equation}
\label{eq:MS}
M_{S}=\frac{u_S}{c_S}
\end{equation}
where $c_S$ is the sound velocity in the medium considered, defined as
\begin{equation}
\label{eq:cS}
c_{S}=\sqrt{ \frac{\gamma_{\rm th} P_{\rm th}}{\rho}}\:,
\end{equation}
and that the Alfv{\'e}nic Mach number is defined as
\begin{equation}
\label{eq:MA}
M_{A}=\frac{u_S}{v_{A}}
\end{equation}
where $v_{A}$ is the Alfv{\'e}n velocity in the medium considered, defined as (in Gaussian units)
\begin{equation}
\label{eq:vA}
v_{A}=\frac{B}{\sqrt{4\pi\rho}}\:.
\end{equation}
The total magnetic field at point~$x_p$ is then given by
\begin{equation}
\label{eq:Bp}
B_{p}^2 = B_0^2 + 8\pi P_{\mathrm{w},p}
\end{equation}
where $B_0$ is the ordered component of the ambient field.

We note that formula~(\ref{eq:Pw,p-res}) was also used by \cite{Lee2012a}, and that a different version was proposed in \cite{Caprioli2012a} and used by \cite{Kang2012b,Kang2013b}. The later was obtained by expressing the Alfv{\'e}n speed using the amplified $\delta B(x)$ instead of the ambient~$B_0$. This may seem more consistent, however when the field gets turbulent the very definition of $v_A$ becomes questionable as we shall see in the next section. We note that \cite{Caprioli2012a} finally opted to set the field everywhere in the precursor equal to the value predicted by the formula immediately upstream of the shock, which may seem a bit artificial. We would like to recall that, anyway, all such formulas are only heuristic formulas, derived to encapsulate the overall behaviour of the magnetic turbulence. Addressing MFA directly would require a different numerical treatment, and is well beyond the scope of this paper -- our point here is just to show how the inclusion of MFA can affect the computation of the emission from a realistic SNR. To complement our work, we note that recently \cite{Kang2013d} performed a comparison of different results obtained for MFA in the precursor: the two aforementioned recipes from \cite{Caprioli2012a}, and two more models for $B(x)$ that are parametrizations from simulations made by other groups. They find that the last three models do not match observations of real SNRs.

\subsubsection{Alfv{\'e}nic drift}
\label{sec:model-MFA-drift}

When the magnetic field is substantially amplified, the common assumption of largely super-Alfv{\'e}nic flows may no longer hold true. It is important to recall that the charged energetic particles are tied to the flow because they are being scattered off the magnetic waves, so that the velocity that matters to them is not the fluid velocity~$u$ but rather the waves velocity $u_w$. In the MFA picture, Alfv{\'e}n waves are generated by particles that are counter-streaming the flow, so that 
\begin{equation}
\label{eq:uw}
u_{w,p} = u_p - v_{A,p}
\end{equation}
at point $x_p$.\footnote{We are here only considering Alfv{\'e}nic drift in the upstream medium. Downstream, we assume that the magnetic turbulence is isotropic so that on average $u_{w,2} = u_2$. It is known that taking $u_{w,2} = u_2 + v_{A,2}$ strongly affects the spectra, but the physical justification for such a choice is far less clear than in the precursor.} This value is used in our acceleration model for both the transport of particles and the growth of waves. The difference between $u$ and the actual $u_w$ is commonly referred to as the Alfv{\'e}nic drift. It had long been neglected on the assumption that $v_A \ll u_S$, but it was recently realized that it should not be neglected when MFA is efficient and the magnetic field gets much higher than in the ISM. Because of the Alfv{\'e}nic drift, particles experience smaller velocity jumps that lead to steeper spectra and less modified shocks. This turns out to be an important ingredient in the modeling of non-linear acceleration at SNR shocks \citep{Ptuskin2010d,Caprioli2011b,Caprioli2012a,Morlino2012a,Kang2012b,Kang2013b}. However, one difficulty is that, when the field gets highly turbulent through MFA, it is not clear whether the velocity of the perturbations should still be taken as the Alfv{\'e}n velocity defined by Equation~(\ref{eq:vA}), and if so whether it should be computed in the amplified magnetic field. Following others \citep{Zirakashvili2008a,Lee2012a,Kang2012b}, we take  the Alfv{\'e}nic drift to be the Alfv{\'e}n velocity $v_A(B_A)$ in an effective magnetic field $B_A$ varying between the ambient field $B_0$ and the total amplified field $B_p$: 
\begin{equation}
\label{eq:f_A}
B_A = B_0 + f_A \times (B_p - B_0)
\end{equation}
with free parameter $0 \le f_A \le 1$.

\subsubsection{Field damping and plasma heating}
\label{sec:model-MFA-damping}

MFA can also affect the plasma that is supporting the waves. If we were to assume adiabatic compression in the precursor, the thermal pressure at point $x_p$ would simply be given by
\begin{equation}
\label{eq:Pth,p-adiab}
\frac{P_{{\rm th},p}}{P_{{\rm th},0}} = \left(\frac{\rho_p}{\rho_0}\right)^{\gamma_{\rm th}} = \left(\frac{u_p}{u_0}\right)^{-\gamma_{\rm th}}
\end{equation}
where $\gamma_{\rm th} = 5/3$ is the adiabatic index of the thermal fluid. But several mechanisms can potentially alter the energy distribution in the precursor. In particular, if the magnetic waves discussed in Section~\ref{sec:model-MFA-amplification} are damped in the plasma, then the magnetic energy will be dissipated as heat in the plasma. \cite{Berezhko1999a} then proposed the following recipe (obtained for large $M_{A,0}$):
\begin{equation}
\label{eq:Pth,p-alfven}
\frac{P_{\mathrm{th},p}}{P_{\mathrm{th},0}} = \left(\frac{u_p}{u_0}\right)^{-\gamma_{\mathrm{th}}}\left[1+\zeta\left(\gamma_{\mathrm{th}}-1\right)\frac{M_{S,0}^{2}}{M_{A,0}}\left(1-\left(\frac{u_p}{u_0}\right)^{\gamma_{\mathrm{th}}}\right)\right]
\end{equation}
where the heating is parametrized by the parameter $\zeta$, that balances the factor ($1-\zeta$) in Equation~(\ref{eq:Pw,p-res}). This recipe, with $\zeta=1$, has often been used in kinetic models of DSA, long before MFA was established, in order to temper the effect of back-reaction and obtain reasonable shock modifications (see the discussion in \citealt{Amato2006a}). Later on, it was realized that amplification and damping are competing processes, and that either can back-react on the flow \citep{Caprioli2008a,Caprioli2009a} as explained in the next section.

\subsubsection{Field advection and magnetized shock}
\label{sec:model-MFA-advection}

If the pressure of magnetic waves reaching the sub-shock is high (case $\zeta \simeq 0$), then it must be included in the energy budget of the shock, and the standard Rankine-Hugoniot jump condition for the thermal pressure
\begin{equation}
\label{eq:Pth,2/Pth,1-RH}
\frac{P_{\mathrm{th},2}}{P_{\mathrm{th},1}}=\frac{\left(\gamma_{\mathrm{th}}+1\right)R_{\mathrm{sub}}}{\left(\gamma_{\mathrm{th}}+1\right)-\left(\gamma_{\mathrm{th}}-1\right)R_{\mathrm{sub}}}
\end{equation}
has to be modified according to
\begin{equation}
\label{eq:Pth,2/Pth,1-Pw}
\frac{P_{\mathrm{th},2}}{P_{\mathrm{th},1}}=\frac{\left(\gamma_{\mathrm{th}}+1\right)R_{\mathrm{sub}}-\left(\gamma_{\mathrm{th}}-1\right)\left(1-\left(R_{\mathrm{sub}}-1\right)^{3}\frac{P_{\mathrm{w},1}}{P_{\mathrm{th},1}}\right)}{\left(\gamma_{\mathrm{th}}+1\right)-\left(\gamma_{\mathrm{th}}-1\right)R_{\mathrm{sub}}}\:.
\end{equation}
This simplified formula was derived by \cite{Caprioli2008a,Caprioli2009a}, assuming the ordered magnetic field $B_0$ is initially parallel to the shock. The key point is that, upstream of the shock discontinuity, the pressure of magnetic waves $P_{\rm w}$ can become comparable to the thermal pressure $P_{\rm th}$, even though it is negligible compared to the dynamical pressure $P_\mathrm{dy} = \rho u^2$ both downstream and upstream of the shock. When $P_{{\rm w},1} \sim P_{{\rm th},1}$, according to Equation~(\ref{eq:Pth,2/Pth,1-Pw}) the thermal pressure is reduced, so that the effect of particle back-reaction is offset to some point: energetic particles tend to make the medium more compressible, but the magnetic field that they amplify tend to make it less so.

\subsection{Downstream transport}
\label{sec:model-down}

Particles accelerated at the shock front are eventually advected downstream, where they can be traced by their non-thermal radiation. In our model, this step is post-processed: it is computed at the end of the hydrodynamic run in order to limit the numerical requirements. In paper~II, we showed how to obtain the thermodynamic properties of the thermal plasma in the shocked region for the computation of the thermal emission. Similarly, in this section, we show how we reconstruct the quantities of interest for the computation of the non-thermal emission: particle spectra and magnetic field. 

\subsubsection{Energetic particles}
\label{sec:model-down-part}

The acceleration model provides the spectrum of particles $f_{\rm S}(p_{\rm S})$ (for both protons and electrons) immediately downstream of the forward shock at each time~$t_S$. We assume that particles are coupled to the gas inside the shocked region, which amounts to neglecting the diffusion of the particles of the highest energies. This assumption, together with the assumption of a series of quasi-steady states at the shock front, seems reasonable for the bulk of particles (that contributes most of the pressure), however it restricts our ability to describe the fate of the highest energy particles (that will eventually escape the remnant anyway). The particle spectra are thus advected towards the interior of the remnant, while they suffer energy losses. The spectrum~$f(p)$ at the final time~$t$ in a cell at position~$\mathbf{x}$ inside the shell is obtained from the relations \citep{Reynolds1998a,Cassam-Chenai2007a}:
\begin{equation}
\label{eq:f/f0}
\frac{f(p)}{f_{\rm S}(p_{\rm S})} = \left(\frac{\alpha^{1/3}}{\alpha^{1/3}-\Theta\: p/(mc)} \right)^4
\end{equation}
and
\begin{equation}
\label{eq:p/p0}
\frac{p}{p_{\rm S}} = \frac{\alpha^{1/3}}{1+\Theta\: p_{\rm S}/(mc)}
\end{equation}
with
\begin{equation}
\label{eq:alpha}
\alpha\left(\mathbf{x},t\right) = \frac{\rho\left(\mathbf{x},t\right)}{\rho\left(\mathbf{x_{\rm S}}(t_{\rm S}),t_{\rm S}\right)}
\end{equation}
and
\begin{equation}
\label{eq:theta}
\Theta\left(\mathbf{x},t\right)=\int_{t_{\rm S}}^{t}\frac{\lambda B^{2}\left(\mathbf{x},t'\right)}{mc^{2}}\,\alpha^{1/3}\left(\mathbf{x},t'\right) \:{\rm d}t'
\end{equation}
where $t_{\rm S}$ is the time at which the material in the cell was shocked and $\mathbf{x_{\rm S}}(t_{\rm S})$ is the position of the shock at that time. $\lambda$ is a constant which in Gaussian cgs units reads:
\begin{equation}
\label{eq:lambda}
\lambda = \frac{4q^{4}}{9m^{2}c^{3}}
\end{equation}
where $q$ is the charge of the particle ($+e$ or $-e$) and $m$ its mass ($m_p$ or $m_e$). Formulae~(\ref{eq:f/f0}) and~(\ref{eq:p/p0}) are strictly valid for relativistic particles of $p > mc$. The variable~$\alpha$ represents the adiabatic losses that shift the spectrum in energy without altering its shape (independently of the particle mass or charge). The variable~$\Theta$ represents the radiative losses (in the magnetic field) that alter the shape of the spectrum; the effect is important only for electrons given the ratio $m_p/m_e$. The synchrotron loss time in a constant magnetic field~$B$, defined so that $p(t_{\mathrm{sync}}-t_{0})=p(t_{0})/2$, is given for relativistic particles by 
\begin{equation}
\label{eq:t_sync}
t_{\mathrm{sync}}(p) = \frac{mc^{2}}{\lambda B^{2}} \times \frac{mc}{p} \quad ; \quad p \gg mc
\end{equation}
that is for electrons
\begin{equation}
\label{eq:t_sync-2}
t_{\mathrm{sync}} \simeq 12.5~\mathrm{yr} \times \left(\frac{E}{100~\mathrm{TeV}}\right)^{-1} \times \left(\frac{B}{100~\mathrm{\mu G}}\right)^{-2}\:.
\end{equation}

In practice, in the code RAMSES, we advect the following two quantities inside the shocked region as passive scalars:
\begin{equation}
\label{eq:tau_S}
\tau_{\rm S}\left(\mathbf{x},t\right) = \int_{t_{\rm S}}^t {\rm d}t'
\end{equation}
is the cell age that gives knowledge of~$t_{\rm S}$; and
\begin{equation}
\label{eq:tau_L}
\tau_{\rm L}\left(\mathbf{x},t\right) = \int_{t_{\rm S}}^t \left(\frac{B(\mathbf{x},t')}{B_0}\right)^2 \alpha^{1/3}(\mathbf{x},t') \:{\rm d}t'
\end{equation}
can be thought as a loss age that permits the computation of~$\Theta$ (for both protons and electrons).

\subsubsection{Magnetic field}
\label{sec:model-down-MF}

To compute the radiative losses of electrons inside the shell, as well as to compute their radiation, it is also necessary to have a description of the evolution of the magnetic field after it crossed the shock. 

The value of~$B$ immediately downstream of the shock is provided by the acceleration model. According to the discussion of Section~\ref{sec:model-MFA}, the pre-shock field consists of both the ambient field and the amplified field (Equation~\ref{eq:Bp}). Assuming that the ambient field $B_0$ is isotropized in the vicinity of the shock,\footnote{The treatment of~$B_0$ may not seem fully consistent with the models for the behavior of the magnetic field in the precursor presented in Section~\ref{sec:model-MFA}: formulas~(\ref{eq:uw}) and~(\ref{eq:Pth,2/Pth,1-Pw}) were nominally obtained by assuming that the magnetic field $B_0$ is initially parallel to the shock. However a fully consistent treatment of the magnetic field would be difficult, especially at this level of modeling. We note that \cite{Kang2012b,Kang2013b} used a similar approach: they assumed that the quadratic sum of the ordered and turbulent components of the field is isotropic immediately upstream of the shock. The aim of the aforementioned formulas is just to catch some important effects, that are believed to extend when the field gets fully turbulent: the fact that the shock jump is reduced, as seen by particles, because of the waves speed $u_w$, and the fact that the shock jump is reduced, as seen by the fluid, because of $P_{w,1}$.} then two out of its three components are compressed, by a factor taken to be $R_\mathrm{tot}$. For the turbulent component, the calculation shows that it is compressed by a factor of $R_\mathrm{sub}$ at the sub-shock \citep{Caprioli2009a}. The total magnetic field immediately downstream of the shock is then given by
\begin{equation}
\label{eq:B2_shock}
B_2^2 = \frac{1+2 R_\mathrm{tot}^2}{3} \times B_0^2 + R_\mathrm{sub}^2 \times 8 \pi P_\mathrm{w,1} 
\end{equation}
We note that when MFA is efficient ($\zeta \simeq 0$), the downstream $B_2$ is totally controlled by the turbulent component $P_{w}$, so that the exact fate of $B_0$ does not matter much. When MFA is quenched ($\zeta \simeq 1$), then $P_{w}$ is negligible, and the chosen formula still allows for some reasonable compression of $B_0$.

Downstream of the shock, we assume that the magnetic flux is frozen in the plasma. Its evolution can then be re-constructed analytically, without the need to activate the MHD solver of RAMSES. The two components of the field, its radial component~$B_{\mid\mid}$ (parallel to the shock normal) and its tangential component~$B_{\perp}$ (perpendicular to the shock normal) evolve separately according to \citep{Reynolds1998a}:
\begin{eqnarray}
\label{eq:B2_down}
B_{\mid\mid}\left(r\right) & = & B_{\mid\mid\, 2}\,\left(\frac{r}{r_{S}}\right)^{-2}\label{eq:Br_down}\\
B_{\perp}\left(r\right) & = & B_{\perp\, 2}\,\frac{\rho}{\rho_{2}}\,\frac{r}{r_{S}}\label{eq:Bt_down}
\end{eqnarray}
where $r$ is the current radius and $r_S$ the shock radius when $B$ was compressed.

This computation applies to the shocked ISM only, between the forward shock and the contact discontinuity. The evolution of the magnetic field is not computed at smaller radii, since the field is thought to be extremely low in the ejecta due to adiabatic expansion \citep{Ellison2005a}.

\subsection{Particle emission}
\label{sec:model-em}

Accelerated particles produce non-thermal emission over a very wide energy band: energetic electrons generate radio to X-ray photons through interaction with the magnetic field (synchrotron radiation) and $\gamma$-ray photons through interaction with ambient photons (inverse Compton effect), energetic protons generate $\gamma$-ray photons through the collisional creation and subsequent decay of neutral pions. Energetic protons can also produce secondary leptons, but their own radiation is usually negligible inside SNRs, and is not considered here. We also neglect the bremsstrahlung emission from accelerated electrons, which is usually found to be a sub-dominant process in SNRs (except possibly in the center of the shell, see \citealt{Atoyan2012a}). We already noted in Section~\ref{sec:model-down-part} that synchrotron emission limits the maximum energy of electrons through losses, and thus needs to be computed consistently. Apart from this, the computation of the particles' emission can be decoupled from the computation of the SNR's evolution and of DSA at the shock. In our simulations it is computed in post-processing, at any chosen age. 

To compute the non-thermal emission of particles, we use the general-purpose code AURA (Edmon, private communication), which is a rewrite of the code COSMICP presented in \cite{Edmon2011a}. It makes use of the formalism of \cite{Schlickeiser2002a} for leptonic processes, and of the formalism of \cite{Kelner2006a} and \cite{Kelner2008a} for hadronic processes. From the distribution function $f(p)$ of particles and the relevant ambient parameters, the code computes the volume emissivity of photons per unit frequency, for each radiative process of interest. Pion decay emission depends on the ambient density which is known from the hydrodynamic simulation. Inverse Compton emission depends on the background radiation, here taken to be the uniform cosmic microwave background (CMB). Synchrotron emission depends on the intensity of the magnetic field
which is modeled according to the previous sections. 

In this work, we focus on the non-thermal emission coming from the shocked ambient medium, due to particle acceleration at the forward shock. The computation of the non-thermal emission is thus done, cell by cell, in the shocked ISM only (between the outermost position of the contact discontinuity and the azimuthally averaged position of the forward shock). It is important to understand that, even though our acceleration model at any time is approximate, we keep track of the particle acceleration history over the whole simulation, in order to compute their emission at any point in the shocked ISM at the time considered. The photon spectra are then projected to obtain 2D maps of the emission that can be directly compared with observations. So even if the maps presented here are a snapshot of the SNR at a given age, they do integrate the particle spectra in both space (along the the line of sight) and time (along the shock history). 

\subsection{Model parameters}
\label{sec:parameters}

Our purpose in this paper is not to make a detailed modeling of a specific object, but rather to point out the important physical aspects to consider when interpreting observations. Therefore, we do not explore the full parameter space, but rather use a single set of parameters appropriate for a type Ia supernova such as Tycho's SNR (as in our study of the morphology in paper~I and our study of the thermal X-ray emission in paper~II). 

\subsubsection{Fiducial SNR}
\label{sec:parameters-SNR}

We are considering the remnant formed by low-mass ejecta in a homogeneous medium. We set the explosion energy to $E=10^{51}~{\rm erg}$, and the ejecta mass to $M_{\rm ej}=1.4$~solar~masses. We assume the ejecta to be initially spherically symmetric, and adopt a power-law profile with index $n=7$. We assume a uniform ambient medium (that is, a power-law profile with index $s=0$) of density $n_{H,0}=0.1\,{\rm cm^{-3}}$. We use these profiles to initialize the hydrodynamic code from self-similar solutions at a small age of $t=10$~years. We then let the code evolve the profiles until $t=500$~years. 

Particles are accelerated continuously in time at the forward shock. For the injection of protons, we take $\xi=3.5$, which produces reasonable injection fractions at the shock front of a few~$10^{-4}$. Similar values were used by other groups in their modeling of Tycho's SNR: \cite{Volk2008a} found a best fit of $\eta = 3 \times 10^{-4}$, \cite{Morlino2012a} used $\xi=3.7$ and \cite{Slane2014a} preferred~$\xi=3.6$. For the injection of electrons, we set $K_{ep}=10^{-2}$, as observed in the local cosmic-ray spectrum. We note that the previously mentioned models of Tycho's SNR need smaller values: \cite{Volk2008a} obtained $K_{ep}=(4-15)\times10^{-4}$, \cite{Morlino2012a} $K_{ep}=16\times10^{-4}$, and \cite{Slane2014a} $K_{ep}=(3-10)\times10^{-4}$. However \cite{Yuan2012a} proposed a model in which $K_{ep}=10^{-2}$ is a universal value that can fit all Galactic SNRs. In any case, as electrons play no dynamical role in our model, their spectra and thus their emissions could be re-scaled as needed. We assume an ambient magnetic field $B_0 = 5~\mu{\rm G}$ for the diffusion of particles, representative of the average Galactic field. 

\subsubsection{Limit cases for MFA}
\label{sec:parameters-MFA}

A~key aspect discussed in this paper is the amplification of the magnetic field by particles, which in our model depends on the free parameter~$\zeta$ (see Equations~(\ref{eq:Pw,p-res}) and~(\ref{eq:Pth,p-alfven})). In all our simulations, we considered the following two limit cases: 
\begin{itemize}
\item $\zeta=1$: The waves are all damped, so that the net magnetic field is limited to the ambient value, which is only enhanced at the shock front (by standard compression). The intensity of the back-reaction is limited by the pre-heating of the plasma in the precursor (see Section~\ref{sec:model-MFA-damping}).
\item $\zeta=0$: The waves grow freely until saturation of the amplification process, and are advected to the sub-shock. The intensity of the back-reaction is limited by the magnetic tension at the shock discontinuity (see Section~\ref{sec:model-MFA-advection}).
\end{itemize}

In addition, we consider two limit cases for the Alfv{\'e}nic drift, controlled by the parameter $f_A$ (see Equation~\ref{eq:f_A}): 
\begin{itemize}
\item $f_A=0$: The net upward velocity of the waves is computed in the ambient field, and is therefore very close to the local velocity of the flow. The inclusion of the drift is then only a minor correction in the computation of DSA, regardless of $\zeta$.
\item $f_A=1$: The net upward velocity of the waves is computed in the total magnetic field. In the case $\zeta=1$ this does not make any difference, since there is not net MFA. But in the case $\zeta=0$, the inclusion of the drift is an important correction in the computation of DSA, leading to steeper spectra and less modified shocks.
\end{itemize}

By comparing such extreme cases for MFA and Alfv{\'e}nic drift, we encompass a large range of possibilities, and hope to account for the uncertainties in the modeling of the behavior of the magnetic field that were mentioned thorough Section~\ref{sec:model-MFA}. We note that our approach is orthogonal to the one by \cite{Kang2013d}, who compared different MFA recipes, but arbitrarily set all free parameters to their median values ($\zeta=0.5$ and $f_A=0.5$).

\subsubsection{Numerical grid}
\label{sec:parameters-grid}

Finally, we have to specify the numerical resolution of the simulations.

Regarding acceleration, for the particles, the momentum grid is logarithmically sampled between $p_\mathrm{inj}$ and $p_\mathrm{max}$ with 20~points per decade. This is largely sufficient to get proper estimates of the shape of the spectrum and of the resulting pressure and back-reaction.

Regarding transport, for both the fluid and the particles, the spatial AMR grid has a formal resolution of $1024^3$, so that we produce emission maps of $1024^2$ pixels in each energy bin (as in papers~I and~II). Note that we only simulate one octant of the remnant, and therefore obtain projected maps of one quarter of the remnant. 
The code automatically refines the grid around the discontinuities in the profiles, and because of the Rayleigh-Taylor instabilities, most of the shocked region ends up being at the highest resolution of $1024^3$. The rest of the shocked region, and more, is at the intermediate resolution of $512^3$, and the rest of the grid is at the lowest resolution of $256^3$. It is important to see that, because we simulate an octant of a sphere embedded in a cartesian grid, the refined regions are always present along any line of sight. So when doing the projection from 3D to 2D, every pixel inside the shell gets data that were actually computed on its size (where it mattered).

Regarding emission, for the photons, the energy grid is logarithmically sampled with 10~points per decade over several decades in energy: from $10^{-12}$~eV to $10^{+6}$~eV for synchrotron, from $10^{-5}$~eV to $10^{+15}$~eV for inverse Compton, and from $10^{+5}$~eV to $10^{+15}$~eV for pion decay. This encompasses the energy ranges of all relevant instruments, as shown in the following section.


\section{Results}
\label{sec:results}

In this section, we present a variety of synthetic emission maps and spectra generated from our 3D simulations, in order to investigate the effects of acceleration on observational diagnostics of the SNR, and specifically the impact of both hydrodynamic and magnetic back-reactions occurring in efficient DSA. 

As in paper~II, in order to illustrate the effect of the hydrodynamic modifications, in all the plots we compare data obtained without including the back-reaction from the pressure of accelerated particles on the flow (cases labeled ``OFF") so that the shock structure is unmodified, and data obtained including the back-reaction from the pressure of accelerated particles (cases labeled ``ON") so that the shock structure is modified. It~is important to understand that our ON/OFF switch controls the hydrodynamic back-reaction only: in both cases we run the same non-linear acceleration model, so that the OFF cases are not in ``test-particle" regime in terms of magnetic field and particle spectra.

New in this paper, in order to investigate the magnetic back-reaction (see Section~\ref{sec:parameters-MFA}), we compare on each figure the two limit cases obtained with complete damping  ($\zeta=1$, the magnetic field remains at the ambient level) and with no damping  ($\zeta=0$, the magnetic field is amplified well above the ambient value). When magnetic amplification is at work, we in addition consider the effect of computing the Alfv{\'e}nic drift in the ordered ambient field ($f_A=0$, the drift is then negligible with our parameters) or in the turbulent amplified field ($f_A=1$, the drift may be important if there is much amplification).

We first show the quantities related to the plasma and particles, on which the non-thermal emission relies (Section~\ref{sec:results-diagnostics}), before showing the emission itself (Section~\ref{sec:results-emission}).

\subsection{Plasma and particles}
\label{sec:results-diagnostics}

Here we show the quantities, diagnosed from the RAMSES code, needed to compute the emission. We present all these data in the form of slices (in a plane z = 0 perpendicular to the line of sight) in order to show the inner structure of the remnant.

\subsubsection{Plasma state}
\label{sec:results-diagnostics-plasma}

The fluid density inside the shocked region is shown in Figure~\ref{fig:map-np}. The shocked region is bounded by the forward shock at the outer edge and by the reverse shock at the inner edge. The shocked ejecta are shaped by the development of the Rayleigh-Taylor instability, showing a distinctive pattern of fingers and holes near the contact discontinuity. When comparing the ON vs. OFF cases (right plots vs. left plots), it is obvious that the shocked region gets compressed when the particle pressure is taken into account, as expected (see paper~I). In the case with no net magnetic field amplification ($\zeta=1$), its width is smaller by about 1/3, and the post-shock density is higher by a factor of about~2. The thin layer of dense, recently shocked ISM is now almost in contact with the Rayleigh-Taylor fingers of the ejecta. When comparing the two limit MFA cases $\zeta=0$ vs. $\zeta=1$ (middle and bottom plots vs. top plots), no impact is visible in the OFF case, as expected for such a purely hydrodynamic diagnostic. For the ON case, as long as the Alfv{\'e}nic drift is negligible ($f_A=0$, top and middle plots), the level of back-reaction is overall the same: it is just a little higher with efficient MFA ($\zeta=0$, middle plots). This result that was not obvious, given that the back-reaction is adjusted through completely different mechanisms: pre-heating in the precursor for $\zeta=1$, vs. magnetized sub-shock for $\zeta=0$. If the Alfv{\'e}nic drift is important ($f_A=1$, bottom plots), the level of back-reaction is lower: the shock is not as modified. Quantitatively, the width of the shocked region is reduced by only about 1/4, and the post-shock density is enhanced by only about~50\%.

The magnetic field downstream of the forward shock is shown in Figure~\ref{fig:map-B_FS}. It is only shown where the ISM fraction ($f_{\rm ism}=1-f_{\rm ej}$, see Section~\ref{sec:model-hydro}) is $>2/3$, because its evolution is not computed in the ejecta (grey area). With no net MFA ($\zeta=1$, top plots), the downstream evolution of the field for a modified shock is similar to the case of an un-modified shock, but over a smaller region. In this case, $B_2$ is just the compression of the ambient $B_0$ at the shock, with ratio~$r$ dependent on back-reaction. According to Equation~(\ref{eq:B2_shock}), the ambient field $B_0=5~\mu{\rm G}$ is compressed by a factor $3.3$ for $r=4$ (the reference value) and $6.6$ for $r=8$ (the maximal value predicted by the acceleration model), leading to $B_2 \simeq 33~\mu{\rm G}$ as observed on the maps behind the shock front. Then $B_2$ decreases away from the shock, according to Equations~(\ref{eq:Br_down}) and~(\ref{eq:Bt_down}) (with the latter dominating close to the contact discontinuity, because of the drop in density). In the case with efficient MFA ($\zeta=0$, middle and bottom plots), $B_2$ is much higher: it reaches hundreds of $\mu{\rm G}$ (note the different scale). In this case, the values are slightly lower in the OFF vs. ON cases (by about 20\%), because in our MFA model weaker shocks are less efficient at generating magnetic turbulence (so computing MFA in the un-modified shock profile results in over-estimating the magnetic field). Also, in this case $\zeta=0$, not much radial variations are seen downstream of the shock, so that the variations seen in the $\zeta=1$ case must be counter-balanced by the time evolution of the shock properties: at earlier ages (that correspond to material close the the contact discontinuity) the shock was stronger, inducing more efficient MFA. 

\subsubsection{Shock history and particle losses}
\label{sec:results-diagnostics-history}

We recall that in order to compute the evolution of particle spectra and their emission, we need to reconstruct the history of the material downstream of the shock (see Section~\ref{sec:model-down}).
The synchrotron radiative loss age, as defined by Equation~(\ref{eq:tau_L}), is shown in Figure~\ref{fig:map-tL_FS}. The higher $\tau_L$, the more particles are likely to be affected by losses (depending on their mass and energy). $\tau_L$ rises downstream away from the shock, because material farther away from the shock crossed it at earlier times. The maximal value, reached close to the contact discontinuity, is slightly lower for the modified shocks (case ON, right plots, vs. case OFF, left plots), following the magnetic field. The value of $\tau_L$ is strongly dependent on $B_2$, and thus strongly depends on the MFA recipe: when $\zeta=0$ (middle and bottom plots) it reaches values about 2 orders of magnitude higher than when $\zeta=1$ (top plots). When $\zeta=0$, its value is lower by a factor of about~2 when the Alfv{\'e}nic drift is computed in the amplified field ($f_A=1$, bottom plots) compared to when it is computed in the ambient field ($f_A=0$, middle plots). 

The quantity $\tau_L$ was purposely defined to be independent of the mass $m$ of the particle, and is used for both protons and electrons. The actual loss factor $\Theta$ (not shown here), given by Equation~(\ref{eq:theta}), is inversely proportional to $m$ and thus is very different for protons and electrons. From Equations~(\ref{eq:f/f0}) and~(\ref{eq:p/p0}), we see that the higher $\Theta$, the lower the momentum at which particles will be affected. For protons $1/\Theta_p$ is very high (more than $10^{12}$ $m_p c$), so that they are not affected by radiative losses (they still suffer adiabatic losses, through the parameter~$\alpha$ in the equations). For electrons $1/\Theta_e$ is comparable with $m_p c$, so that they are strongly affected by radiative losses (on top of adiabatic losses). This implies different maximum energies for protons and electrons.

\subsubsection{Particles energy}
\label{sec:results-diagnostics-particles}

The maximum momentum of protons $p_{\mathrm{max},p}$ is shown in Figure~\ref{fig:map-pLp}. It is always the highest at the shock front, and decreases downstream because of the adiabatic losses (contrast with Figure~\ref{fig:map-tL_FS}). The values reached by $p_{\mathrm{max},p}$ are notably higher (note the different scales) when MFA is efficient: between roughly 100 and 300~TeV/c for $\zeta=0$ (middle and bottom plots), versus between only 10 and 20~TeV/c for $\zeta=1$ (top plots). This is because with a much higher magnetic field the particles are better confined in the vicinity of the shock. For the same reason, $p_{\mathrm{max},p}$ is slightly lower for modified shocks (right plots vs. left plots). In the case of efficient MFA ($\zeta=0$), $p_{\mathrm{max},p}$ is even higher when the Alfv{\'e}nic drift is computed in the amplified field ($f_A=1$, bottom plots) compared to when it is computed in the ambient field ($f_A=0$, middle plots), which is opposite to the behaviour of the magnetic field. This demonstrates the non-trivial impact of the drift in the non-linear acceleration process. 

The maximum momentum of electrons $p_{\mathrm{max},e}$ is shown in Figure~\ref{fig:map-pLe_FS}. As for protons, it is highest at the shock front and decreases downstream, and the effect is even more marked than for protons because of the combined (adiabatic + radiative) losses. In the case $\zeta=1$ (top plots), $p_{\mathrm{max},e}$ remains of the same order as $p_{\mathrm{max},p}$ (note the same scales), meaning that it is limited by the finite age and/or size of the remnant as much as by the magnetic field. In the case $\zeta=0$, $p_{\mathrm{max},e}$ is much lower than $p_{\mathrm{max},p}$, because it is limited by the radiative losses. According to Equation~(\ref{eq:t_sync-2}), after 500~yr one still has electrons of energies up to $\sim30$~TeV in a uniform $B_2=30~\mu{\rm G}$, appropriate for the un-amplified case $\zeta=1$, and up to only $\sim1$~TeV in a uniform $B_2=200~\mu{\rm G}$, appropriate for the amplified case $\zeta=0$. In reality, the downstream magnetic field is not constant, because of the spherical geometry and shock history, yet these order of magnitude values are in broad agreement with the results obtained in Figure~\ref{fig:map-pLe_FS}. In the case of efficient MFA ($\zeta=0$), $p_{\mathrm{max},e}$ is slightly higher when the Alfv{\'e}nic drift is computed in the amplified field ($f_A=1$, bottom plots) compared to when it is computed in the ambient field ($f_A=0$, middle plots), which follows from the behaviour of the magnetic field. But this is a secondary effect: the main result is that when $\zeta=0$ highly energetic electrons ($E>$~TeV) are confined very close to the shock front, which will restrict their emission.

\subsection{Non-thermal broad-band emission}
\label{sec:results-emission}

Using the data shown in the previous section, we compute the non-thermal emission in each cell, which is then projected along the line of sight (arbitrarily chosen to be the z-axis of the simulation box) to generate synthetic maps. The maps, in different energy bands, are presented in Section~\ref{sec:results-emission-maps}. Spatially-integrated spectra are also presented in Section~\ref{sec:results-emission-spectra}.

\subsubsection{Broadband maps}
\label{sec:results-emission-maps}

We have computed projected maps in 5 energy bands: \\
$\bullet$ in radio from $10^{-7}$ to $10^{-4}$~eV, that is from 30~MHz to 30~GHz, which covers the range of LOFAR to Effelsberg instruments;\\
$\bullet$ in soft X-rays from 0.3 to 10~keV, the range of Chandra, XMM-Newton, Suzaku;\\
$\bullet$ in hard X-rays from 5 to 80~keV, the range of NuSTAR (and partially that of INTEGRAL);\\
$\bullet$ in low-energy $\gamma$-rays from 0.1 to 100~GeV, the range of Fermi and AGILE;\\
$\bullet$ in high-energy $\gamma$-rays from 0.01 to 10~TeV, the range of HESS~2, MAGIC, VERITAS. \\
All the maps show the differential volume emissivity as computed by the code, in erg/cm$^3$/s/eV, as indicated on the colour scales.
Note that the angular resolution of our maps is limited by the numerical resolution of our grid, they have not been convolved with the PSF of any particular instrument. We recall that the maps presented here have 1024 pixels along each direction, at the age shown here they span 5.61~pc, hence a physical cell size of 0.0055~pc, and an angular resolution of about $1''$ at a distance $d=1$~kpc. 

The synchrotron emission from electrons extends from the radio to the X-rays. It is shown in the radio band in Figure~\ref{fig:map-NTsy-radio} and in the soft X-ray band in Figure~\ref{fig:map-NTsy-Xsoft}. 
The morphology of the remnant in radio does not vary much depending on the various back-reaction cases. The maximum of emission is always well behind the shock, close to the contact discontinuity, where low energy (GeV) electrons have accumulated. This allows us to see the imprint of the Rayleigh-Taylor instability, as a negative image of the fingers. The emissivity is higher for $\zeta=0$ vs. $\zeta=1$ (note the different scales), especially when $f_A=0$, because of the much higher value of $B_2$: electrons radiate more strongly in the amplified field, while at the same time they are not affected much by radiative losses at these energies. The morphology of the remnant in X-rays is noticeably different. The maximum of emission is now close to the shock, where the high energy (multi TeV) electrons are confined (see Figure~\ref{fig:map-pLe_FS}), and decreases rapidly downstream (note that the scales now span two orders of magnitude, compared to one for the radio). We recall that the emissivity cut-off happens to be in the X-ray band (see also Figure~\ref{fig:spc-back-zeta}), and therefore critically depends on the fate of the electrons near $p_{\mathrm{max}}$. The effect is much enhanced with efficient MFA ($\zeta=0$) because of the stronger losses in the amplified $B_2$: very thin rims are visible immediately behind the shock, regardless of how the Alfv{\'e}nic drift is computed. As a result of the rapid drop of emissivity downstream of the shock, no more physical structures are visible inside the shell. Note that the overall emissivity levels are not as much affected by $\zeta$, as for the radio (the same scale was used for all MFA cases). Moreover, there is a stronger contrast between the edge and the center of the shell in the ON case, so that the rims look even thinner. The morphology of the remnant in the hard X-ray band (not shown here) is similar to the soft X-ray map, but with much lower normalizations as we are sampling the steeply falling cut-off of the synchrotron emissivity.

The inverse Compton emission from electrons extends over the $\gamma$-ray bands. It is shown around GeV energies in Figure~\ref{fig:map-NTic-Gsoft} and around TeV energies in Figure~\ref{fig:map-NTic-Ghard}. First we see that the emissivity is lower for the ON vs. OFF cases, as previously. This emission does not explicitly depend on the magnetic field as the synchrotron emission, but it is still affected by the different MFA cases, because it depends on the number of emitting electrons left after losses. As a result, the emissivity is lower for $\zeta=0$ vs. $\zeta=1$, by about one order of magnitude at GeV energies and almost two orders of magnitude at TeV energies (note the different scales). MFA also changes the emission pattern, at both GeV and TeV energies, depending on where the emitting electrons are: widespread in the shocked ambient medium when $\zeta=1$ (so that the Rayleigh-Taylor fingers are visible), and close to the forward shock when $\zeta=0$ (so that inner structures disappear). The effect is more pronounced at higher energies, because high-energy photons are emitted by high-energy electrons that are the most affected by losses, and thus restricted to the vicinity of the forward shock where they are accelerated. We note that this effect does not depend much on the Alfv{\'e}nic drift.

The pion decay emission from protons extends over the $\gamma$-ray bands. It is shown around GeV energies in Figure~\ref{fig:map-NTpi-Gsoft} and around TeV energies in Figure~\ref{fig:map-NTpi-Ghard}. Less variations are seen than for the leptonic emissions (note that the scale is the same for all maps in a given energy band). The emission extends in most of the downstream region, it never peaks at the shock, it always reaches the contact discontinuity and therefore highlights the instabilities shaping it. There is not much difference between the OFF and ON cases (except of course for the compression of the emitting region). There are some differences in emissivity levels between the different MFA cases, that depend on the energy band. At GeV energies, the emissivity is slightly reduced for $\zeta=0$, for both drift cases, and more so when $f_A=1$. This reflects a decreasing number of emitting protons, as spectra get different curvatures with different back-reaction recipes. At TeV energies, the emissivity is slightly reduced for $\zeta=0$ and $f_A=1$ because in that case spectra are steeper, but slightly enhanced for $\zeta=0$ and $f_A=0$, mostly because of the higher $p_{\mathrm{max},p}$ (see Figure~\ref{fig:map-pLp}). 

\subsubsection{Integrated spectra}
\label{sec:results-emission-spectra}

Finally, we show the broad-band photon distribution in Figure~\ref{fig:spc-back-zeta}, in order to compare the different emission processes for the different acceleration scenarios. 
The energy bands used to produce the previous maps are labeled. For clarity, we only show here the spatially integrated spectrum, although we have computed it in every pixel of the map. We recall that, because these emission spectra integrate the contributions of different parts of the remnant, that contain particles that were accelerated at different times in different conditions, there is not a single emitting population behind these spectra. 
For convenience, here we have multiplied the results by the emitting volume and by the emitted energy, to get the (isotropic) emissivity $E^2 {\rm d}F/{\rm d}E$ in erg/s, as in commonly done. To get the flux from the SNR at the Earth, we need to multiply this emissivity by~8 (as only one octant was simulated) and divide by the illuminated surface $4 \pi d^2$ at distance~$d$. Numerically, we get a translation factor of $7\times10^{-44}$/cm$^2$ at 1~kpc. For comparison, fluxes observed for Tycho's SNR (at a distance of $2-5$~kpc) at various wavelengths are summarized in \cite{Edmon2011a}: in radio at 1.4~GHz: $1.6\times10^{-13}$ erg/cm$^2$/s, in X-rays at 10~keV: $3.2\times10^{-11}$ erg/cm$^2$/s, in $\gamma$-rays above 1~TeV: $3\times10^{-13}$ erg/cm$^2$/s. Missing is the detection by Fermi \citep{Giordano2012a}, for $\gamma$-rays at 1~GeV: $1.6\times10^{-12}$ erg/cm$^2$/s. So the fluxes we obtain are of the correct order of magnitude, although we have not attempted a fit of any particular object at this point.

We already observed that, for all three emission processes, the emissivity is on average higher towards the edge rather than in the center of the shell, regardless of the back-reaction effects. This is expected since energetic particles are in the vicinity of the forward shock, plus there is a limb-brightening effect in projection. 
The emission is always lower for the ON vs. OFF cases (solid vs. dashed lines), firstly because the emitting region is smaller and there are less emitting particles. Note however that the emission in the OFF case is unrealistic since it involves a lot of accelerated particles without hydrodynamic back-reaction. 

More relevant to observations is the comparison of different magnetic back-reaction recipes in the ON case. 
There are important variations in emissivity level and cut-off energy depending on the amplification of the magnetic field ($\zeta=0$, thick lines, vs. $\zeta=1$, thin lines). With a high $B_2$ (obtained when $\zeta=0$), the emissivity level gets higher for synchrotron in radio (by a factor of up to almost~10), because the magnetic field is higher and there are as many low energy electrons. It gets lower for inverse Compton in $\gamma$-rays (by a factor of up to more than 100) because there are much less high-energy electrons, and it stays almost the same for pion decay in $\gamma$-rays (within a factor of about~2) because this hadronic process is not directly dependent on the magnetic field. With a high $B_2$, the cut-off energy stays the same for synchrotron because, in all cases considered, the magnetic field is high enough so that the maximum energy of electrons is loss-limited.\footnote{An electron of energy~$E$ in a magnetic field~$B$ emits synchrotron photons having energy $h \nu \propto B E^2$, and in the loss-limited case and assuming Bohm diffusion the maximum energy of electrons goes as $E_\mathrm{max} \propto B^{-\frac{1}{2}} u_S$, so that the maximum energy of photons goes as $h \nu_\mathrm{max} \propto u_S^2$ independent of~$B$ \citep{Aharonian1999a}.\label{note:loss-limited}} It gets lower for inverse Compton (by almost a factor of~10) because of the lower $p_{\mathrm{max},e}$ caused by strong radiative losses, and it gets higher for pion decay (by more than a factor of~10) because of the higher $p_{\mathrm{max},p}$ caused by better confinement at the shock. Note that all these general behaviours are not affected by the exact value of the Alfv{\'e}nic drift ($f_A=0$ for top plots, $f_A=1$ for bottom plots). The latter does alter the cut-off energies, but in the same way for both kinds of particles (towards slightly higher energies for a higher drift). Also note that the pion-decay plateau of emission is slightly flatter for $f_A=1$, which reflects a steeper spectrum of the emitting protons. 


\section{Discussion}
\label{sec:discussion}

In this section, we recap our main results that are mostly pertaining to cosmic-ray acceleration at the blast wave of young type Ia SNRs like Tycho, linking them with existing or future observations.

\subsection{The key role the magnetic field}
\label{sec:discussion-MF}

The most important effect highlighted in our simulations is the critical role of the magnetic field, that can be much affected by the accelerated protons, and that in turns much affects the accelerated electrons.

\subsubsection{Amplification in the upstream medium}
\label{sec:discussion-MF-up}

In a CR-modified shock, some important processes are occurring ahead of the shock front, in the shock precursor. As streaming particles can themselves generate the magnetic turbulence that scatters them off, this makes a second back-reaction loop in DSA, in addition to the back-reaction exerted through their pressure on the fluid. The fate of the leptonic component of the accelerated particles critically depends on the value of~$B_2$ downstream of the shock. When MFA is efficient (that is when the magnetic waves are growing and are not damped, the case $\zeta=0$), $B_2$ can easily be one order of magnitude higher than what would be expected from a mere compression of the ambient field at the shock front, regardless of the exact value of the shock compression ratio (Figure~\ref{fig:map-B_FS}). The values we obtain, in the range of 200 to 300 $\mu$G, are consistent with what has been inferred from observations of young SNRs: for instance, \citealt{Parizot2006a} estimated the downstream field of Tycho's SNR to be between $200$ and $400~\mu$G. Our work is also in agreement with modeling of this particular SNR: \cite{Volk2008a} adopted the values of $240$ and $360~\mu$G, \cite{Morlino2012a} calculated values of up to $300~\mu$G, and \cite{Slane2014a} found a current value of $\simeq 180~\mu$G. Because of radiative losses, these strong magnetic fields drastically limit the lifetime of multi-TeV electrons that are thus confined very close to the shock (Figure~\ref{fig:map-pLe_FS}). As as result, the synchrotron emission in X-rays produces a very thin rim immediately downstream of the shock (Figure~\ref{fig:map-NTsy-Xsoft}). This is in agreement with what is observed in several young SNRs, in particular Tycho \citep{Hwang2002a,Bamba2005b,Cassam-Chenai2007a}. 

MFA has mostly been discussed in the context of the synchrotron X-ray emission, but the inverse Compton $\gamma-$ray emission also relies on the presence of high-energy electrons. When $B_2$ is amplified, this emission is also concentrated towards the shock at energies close to the cut-off, which is around the TeV (Figure~\ref{fig:map-NTic-Ghard}). So, in principle, the same kind of diagnostic for MFA would be possible in $\gamma$-rays as in X-rays, if the post-shock region could be resolved. Also note that, further downstream, the non-thermal emission may bear the imprint of the Rayleigh-Taylor instabilities, depending on how far the particles can travel, which for electrons depends on the magnetic field and therefore on MFA. For the synchrotron emission, we observe that radio images always show the turbulent interior of the remnant (Figure~\ref{fig:map-NTsy-radio}), whereas X-ray images never do (Figure~\ref{fig:map-NTsy-Xsoft}), in agreement with observations. More interestingly, for the inverse Compton emission, we observe that, in the TeV range, the visibility of inner structures depends on MFA: they disappear when $B_2$ is efficiently amplified, leaving only the outer rim already mentioned (Figure~\ref{fig:map-NTic-Ghard}). Unfortunately, current $\gamma$-ray instruments are far from having the same angular resolution as current X-ray instruments, and cannot resolve a SNR shell with as a much details as we do in our simulations. However, the planned Cherenkov Telescope Array (CTA) observatory will aim at improving the spatial resolution in the TeV band \citep{Actis2011a}. 

We recall that all these results depend on the value of two free parameters in our model: the level of wave damping $\zeta$ (Equations~\ref{eq:Pw,p-res} and~\ref{eq:Pth,p-alfven}), and the factor $f_A$ determining the Alfv{\'e}nic drift (Equation~\ref{eq:f_A}), which both range between~0 and~1. Out of these two parameters, $\zeta$~is clearly the most important: $f_A$~has no noticeable effect when $\zeta$ is close to~1, and it brings a correction when $\zeta$ is close to~0. With these two parameters, we have tried to offer a realistic, yet reasonably simple, modeling of the fate of the magnetic field around the shock front, suitable for the inclusion in large-scale simulations of SNRs. However, the process of MFA is still poorly understood, and ongoing studies keep unravelling new possible mechanisms with their share of complications (see \citealt{Schure2012a} for a review of recent developments). Being difficult to address directly, yet having a direct impact on the observational results, MFA definitively appears to be a critical topic in the studies of DSA. We believe that we have bracketed a fair range of scenarios relevant to young SNRs, although somewhat different recipes for the evolution of the magnetic field in the precursor could be proposed, and could possibly alter the value of $B_2$ by a factor of~two or so. We made some additional tests, using an alternative recipe for MFA proposed by \cite{Caprioli2012a} (see Section~\ref{sec:model-MFA-amplification}). We observed that the general features of the maps and spectra are of the same kind as reported here. This model produces even higher magnetic fields, and so the general behaviours discussed here are enhanced. However, this model leads to much less back-reaction (for the same injection parameters), and so less modified shocks and steeper spectra (in line with the modeling of Tycho's SNR by \citealt{Morlino2012a}). What this shows, is that the two back-reaction loops of DSA (on the shock structure and on the magnetic field, via respectively the pressure and the streaming of energetic particles in the upstream), are not connected in any simple way. As a result, both effects should be assessed individually in the modeling of a SNR.

\subsubsection{Evolution in the downstream medium}
\label{sec:discussion-MF-down}

We have so far focused on the physical processes occurring in the upstream region, since this is where the amplification is believed to occur. However, the subsequent fate of the magnetic field in the downstream region is not entirely settled. \cite{Pohl2005a} argued that a high magnetic field could be efficiently damped just downstream of the shock, which would invalidate the standard interpretation of non-thermal filaments as being loss-limited. We refer the reader to \cite{Marcowith2010a} for a critical analysis of this possibility, and to \cite{Cassam-Chenai2007a} and \cite{Rettig2012a} for a comparison of the two scenarios. 

Further downstream, the magnetic field is expected to peak close to the contact discontinuity (e.g.~\citealt{Cassam-Chenai2005a}). It can be furthermore stretched and amplified by the development of Rayleigh-Taylor instabilities \citep{Jun1996a} and may become dynamically important. Our current model does not allow an investigation of these effects, that require MHD simulations. On a related matter, we note that, in our modeling, we do not let the accelerated particles interact with the contact discontinuity (from which they are very close initially), or go past the contact discontinuity (where they could in principle illuminate the dense ejecta). The first point would probably require a better treatment of the magnetic field close to the contact discontinuity as well, and the second point should probably be studied jointly with particle acceleration at the reverse shock (on this topic, we refer the reader to the recent work by \cite{Telezhinsky2012a}, and to the discussion in paper~II). In this paper, we focused our attention to the shocked ISM, where the processes we consider are inevitable and indeed well observed.

\subsection{Protons vs. electrons}
\label{sec:discussion-protons-electrons}

As stated in the introduction, the particles that are the easiest to track are the electrons, but the ones that are the most important for CR theory are the protons, since they make the bulk of Galactic CRs and can reach higher energies.

\subsubsection{Disentangling the emissions}
\label{sec:discussion-protons-electrons-emissions}

The emission at $\gamma$-ray energies is of paramount importance for particle acceleration studies, because energetic protons only contribute in this band (through pion decay), but unfortunately, electrons also radiate in the same energy band (through inverse Compton effect). If MFA is not efficient ($\zeta=1$), then it would be very hard to distinguish the emission mechanism based on the morphology alone (compare top row of Figure~\ref{fig:map-NTic-Gsoft} to~Figure~\ref{fig:map-NTpi-Gsoft}, and of Figure~\ref{fig:map-NTic-Ghard} to~Figure~\ref{fig:map-NTpi-Ghard}). 
If MFA is efficient ($\zeta=0$), then the inverse Compton emission could be distinguished as the component starting right at the shock front and peaking just behind it, with the pion decay emission peaking further behind, and extending far enough that it highlights the turbulent interior of the remnant -- provided the angular resolution is good enough. At the moment, given the instrumental limitations, distinguishing the two scenarios is only possible via spectral features. These are the most visible at the lowest and highest energies. In between, the shape of the spectrum depends on the shape of the parent distribution, which is much affected by non-linear effects, plus by losses for electrons, so that any modeling can hardly be conclusive. At the lower end, the emission from protons cannot extend much below $\simeq 100$~MeV (a threshold corresponding to the mass of the $\pi_0$ particle), when the emission from electrons just decreases gradually. At the higher end, the emission from protons can extend well above $10$~TeV, when the emission from electrons cannot. We note that, by reducing the cut-off energy of the inverse Compton emission, and increasing the cut-off energy of the pion-decay emission, efficient MFA (case $\zeta=0$) greatly helps to separate the leptonic and hadronic contributions (see again Figure~\ref{fig:spc-back-zeta}). As a result, a clear observational detection of SNRs in the 10--100~TeV range would be strong evidence for the presence of very energetic protons, as well as good evidence for efficient MFA. It is therefore of paramount importance to extend the energy coverage of existing instruments. The planned facility CTA will be able to probe the TeV cut-off in SNRs, and thus identify the emission.

Although the situation is still confused at $\gamma$-ray energies for now, it shall be noted that observations in the radio to X-ray domains can greatly help to shed light on the mechanisms at play. The properties of the synchrotron emission can be used to constrain the underlying electron distribution, and therefore the possible contribution through inverse Compton (for a given ambient photon field). In particular, observations of thin X-ray synchrotron rims suggest high magnetic fields, that disfavour inverse Compton emission as a viable explanation of the $\gamma$-ray emission. This is in particular the case for Tycho's SNR \citep{Giordano2012a}. Moreover, the electrons are not believed to be able to generate these high magnetic fields that make them radiate: instabilities are driven by protons and other heavy nuclei. Therefore, the radiation from electrons offers indirect diagnostics on the presence of energetic protons, and on the efficiency of particle acceleration at the blast wave.

\subsubsection{The problem of the maximum energy}
\label{sec:discussion-protons-electrons-Emax}

Measuring efficient DSA in SNRs is an important step towards establishing them as cosmic-ray accelerators in the Galaxy, however that does not tell the maximum energy achieved by the particles in these objects. For electrons, it can be directly inferred from studying the hard non-thermal X-ray emission \citep{Reynolds1999a}, but it is limited by losses and can barely reach 100~TeV. For protons the situation is less clear. For Tycho's SNR, VERITAS observations do not indicate a cut-off in the $\gamma$-ray spectrum \citep{Acciari2011a}, so the maximum energy of protons can only be estimated based on the age and size of the system -- assuming a certain diffusion law. For this remnant, \cite{Parizot2006a} obtained likely values between 100 and 600 TeV, reaching 2~PeV only in an "extreme case" with all parameters pushed to their limits. We note that the amplification of the magnetic field has an opposite effect on the maximum momentum of protons, compared to electrons: a higher~$B$ (case $\zeta=0$), both upstream and downstream of the shock, helps to confine particles close to the velocity discontinuity, where they can gain energy. In our simple model the acceleration timescale is directly proportional to $1/B$ on each side of the shock, where $B$ is the total field. Things can actually be more complicated: confining particles up to the highest energies requires having magnetic waves at all the relevant resonant wavelengths, and so amplifying the field does not always guarantee reaching higher energies \citep{Ellison2008a}. In any case, even when $p_{\mathrm{max},p}$ is boosted by MFA, it is still lower than the CR knee energy in our simulations, by a factor of almost ten, at less than 400~TeV/c. In their modeling of Tycho's SNR, \cite{Morlino2012a} inferred a similar limit of 500~TeV/c, while \cite{Slane2014a} found even lower values in the range 40 to 100~TeV/c. Although somewhat disappointing, we note that these results are in agreement with recent investigations of the timescales involved in the process, by the very first proponents of the MFA mechanism themselves \citep{Bell2013a}. 

\subsection{Environmental effects}
\label{sec:discussion-environment}

In this paper we considered a uniform medium, appropriate for a type~Ia. We used typical values of the density and magnetic field, but these will of course vary from remnant to remnant, or even in the vicinity of a given remnant. Here we discuss how such variations affect our results, and how our model could be extended to handle more complicated cases.

\subsubsection{The effect of external density and magnetic field}
\label{sec:discussion-environment-uniform}

We note that, with our chosen parameters, the emissivity levels in $\gamma$-rays happen to be comparable for the leptonic and hadronic emissions (see Figure~\ref{fig:spc-back-zeta}). However, all sorts of outcomes are possible, depending on the local conditions. For pion-decay, the emissivity scales directly with the ambient density, which can vary by several orders of magnitude in different regions of the Galaxy (or even different regions around a given SNR), and with the density of accelerated protons, which depends on the ambient density as well as on the poorly constrained injection ratio at the shock. For inverse Compton, as well as for synchrotron, the emissivity scales directly with the density of accelerated electrons, which was parametrized with the ratio $K_{ep}$ of electrons over protons, which again is poorly constrained in the sources. In the end, the best way to distinguish the emissions is therefore through their spectral shape, as discussed in Section~\ref{sec:discussion-protons-electrons-emissions}.

The impact of the ambient magnetic field is more complicated, as it impacts both acceleration (of protons) and losses (of electrons), on top of controlling the synchrotron emissivity of electrons. In general, a higher magnetic field presumably helps confining particles at the shock and reaching higher energies for protons, while it may limit the maximum energy of electrons through radiative losses. In the case with efficient MFA ($\zeta=0$), the resulting field is completely dominated by the amplified field, so that the actual value of $B_0$ does not matter much. Any uncertainty regarding the value of~$B_0$ is likely small compared to the huge difference between $B_0$ and~$B_2$. In the case with no net MFA ($\zeta=1$), variations of $B_0$ can be seen on the spectra. In particular, as soon as $B_0$ is reduced by a factor of about two -- or if the compression at the shock is lower than was assumed with Equation~(\ref{eq:B2_shock}), in the case of a more ordered field -- then the maximum energy of electrons may no longer be loss-limited, so that note~\ref{note:loss-limited} does not hold and the synchrotron cut-off energy becomes dependent on~$B$. However, this case is probably not the most relevant one to observations, as discussed in Section~\ref{sec:discussion-MF-up}.

To study in more details the impact of ambient parameters, such as $n_0$ and $B_0$, and of other free model parameters, such as~$\xi$ and~$\zeta$, \cite{Kosenko2014a} conducted an extensive parametric study using our code presented in paper~I. This resulted in the production of a library of models of SNRs undergoing (efficient) DSA, that can be used either to assess the possible outcomes of acceleration when the ambient parameters are sufficiently known, or to constrain these ambient parameters when the acceleration efficiency could be measured. In the future we will extend this library to include the most salient observational features discussed in paper~II and this paper.

\subsubsection{The contribution of escaping particles}
\label{sec:discussion-cloud}

In this paper we were only interested in the particles accelerated at the shock and subsequently trapped inside the shell. However, protons reaching $p_\mathrm{max}$ eventually escape the system upstream of the shock. The corresponding energy loss is included in the acceleration model, and their energy distribution can be computed \citep{Caprioli2010b}, but their emission is not included in the simulations presented in this paper. This would be especially interesting when dense targets are present in the immediate vicinity of the remnant, a situation already observed and modeled (e.g. \citealt{Fujita2009c,Li2010a,Ohira2011b}). This situation is more likely for remnants of core-collapse supernovae, that are the final stage of massive stars, that were born inside molecular clouds. Particles accelerated at the blast wave may irradiate neighbouring clouds, providing another kind of diagnostic on their acceleration and diffusion mechanisms (see \citealt{Casanova2010b} and references therein). We postpone this study to a dedicated paper. 


\section{Conclusion}
\label{sec:conclusion}

We have developed a model that simulates the morphological and spectral evolution of a young type Ia SNR undergoing efficient particle acceleration. Starting from Chevalier's profiles, the time-evolution of the shocked structure is computed using the 3D hydrodynamical code RAMSES, in a comoving grid of variable resolution. The acceleration of particles at the forward shock is computed with Blasi's non-linear semi-analytical model that provides the level of back-reaction on the fluid. Time-dependent shock modifications are implemented in the hydrodynamic code through the use of a variable adiabatic index, which is lowered at the shock front according to the contribution of the particle pressure to the total pressure. We are thus effectively working with a pseudo-fluid, which represents both thermal and non-thermal populations. At the end of the hydrodynamic simulation, each cell located downstream of the forward shock is processed to compute the evolution of the magnetic field, the spectra of remaining particles after losses, and their non-thermal emission: pion decay for protons, synchrotron and inverse Compton for electrons. Finally the emission is projected along the line of sight, integrating variations due to the geometry and the history of the shocked ISM. This allows us to produce realistic synthetic maps of a typical remnant, in any energy band from radio to X-rays and $\gamma$-rays.

A~major improvement of our model, compared to our study of the thermal emission in paper~II, concerns the fate of the magnetic field. We have implemented recipes for the amplification and transport of the magnetic field in the vicinity of the shock, that allows to probe the impact of CR-driven instabilities on their own spectra and on their emission spectra. We have considered two limit cases for the evolution of the magnetic field in the shock precursor: either the amplified field is entirely damped in the plasma, so that the shock propagates through a pre-heated medium, or the amplified field is advected to the (sub-)shock, which is then a magnetized shock. Interestingly, the net effect in terms of shock modifications is roughly the same for both models -- unless the speed of the waves that scatters particles is computed in the amplified field, in which case the second model leads to less modified shocks than the first. Regardless of this complication induced by the Alfv{\'e}nic drift, when MFA is efficient, very high values of the magnetic field are obtained downstream of the shock ($>100~\mu$G). This markedly alters the spectra of electrons, especially around their maximum energy: because of the strong radiative losses, the most energetic ($>$~TeV) electrons are confined close to the shock front. As a result, their emissions, at the cut-off of the synchrotron in X-rays, and at the cut-off of the inverse Compton in $\gamma$-rays, produce thin rims immediately behind the shock front, with no other structure visible inside the shell -- in contrast with leptonic emission in radio, and with hadronic emissions in $\gamma$-rays, which show the imprint of the Rayleigh-Taylor instabilities. The cut-off energy itself is significantly reduced for the inverse Compton emission (from almost 10~TeV to about 1~TeV), while it stays the same for the synchrotron emission (around 10~keV) as long as the electron's energy is loss-limited. On the other hand, an enhanced magnetic field can boost the maximum momentum of protons -- although it still falls short from the energy of the knee in the local CR spectrum in all our simulations, with a maximum of about $400$~TeV at the shock front. As a result, the cut-off energy of the pion decay emission is significantly increased (from about 10~TeV to about 100~TeV). In conclusion, MFA has a strong impact on the X-ray emission from electrons, and can help distinguish the $\gamma$-ray emissions from electrons and protons, by altering both the radial distribution and the cut-off in this energy band. 

An important goal of SNR observations is to assess the presence of energetic protons at the shock, that are believed to be the constituents of Galactic cosmic-rays. Observing these protons directly proved to be a difficult task, because they do not radiate as efficiently as electrons. However, they can show up in two other ways: first they impact the dynamics of the shock wave (see paper~I), and therefore the thermal X-ray emission from the plasma (see paper~II); and second they impact the evolution of the magnetic field, and therefore the non-thermal emission from electrons in radio, X-rays and $\gamma$-rays (this paper). This work completes our modeling of the emission from SNRs, providing a framework for the study of particle acceleration from broadband observations of young shells. Next, we will apply this model to make a detailed modeling of some specific object such as Tycho's SNR or other type~Ia SNRs. In parallel, we shall extend our model to more complex environmental situations, such as a shock expanding inside a wind or interacting with a molecular cloud, of relevance for the case of a core-collapse SNRs. More generally, our work will help for the interpretation of X-ray and $\gamma$-ray observations with current instruments (such as the recently launched Astro-H), as well as for the preparation of future missions (such as the proposed Athena+). 


\section*{Acknowledgements}
This work has been partially funded by the ACCELRSN project ANR-07-JCJC-0008 in France, and has been supported in Canada by the Natural Sciences and Engineering Research Council (NSERC) through the Canada Research Chairs program and by the Canadian Institute for Theoretical Astrophysics (CITA) through the National Fellows Program.
The code used was largely developed on the Nuit computing cluster at the University of Manitoba, funded by the Canada Foundation for Innovation (CFI) and the Manitoba Research Innovation Funds (MRIF). The high-resolution simulations presented here were performed on the national computing facilities provided by WestGrid (Western Canada Research Grid).
We thank Paul Edmon for providing us with a version of the code AURA used for computing the non-thermal emission. 
We thank the anonymous referee for their useful comments that helped us to substantially improve the quality of our manuscript.


\bibliographystyle{apj}

\newpage

\newcommand{\figsizemap}{13cm}

\begin{figure}[t]
\centering
\includegraphics[width=\figsizemap]{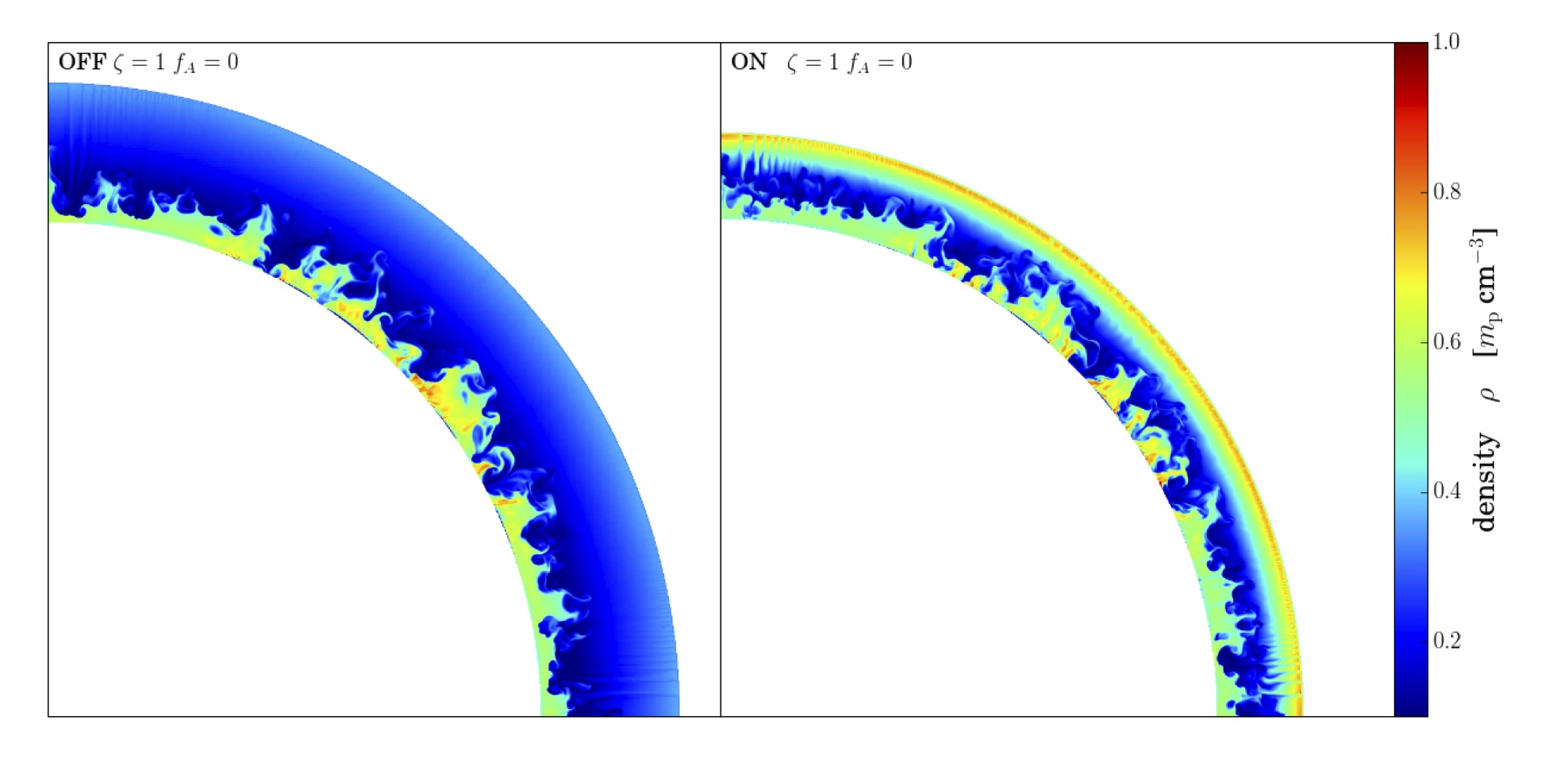}
\includegraphics[width=\figsizemap]{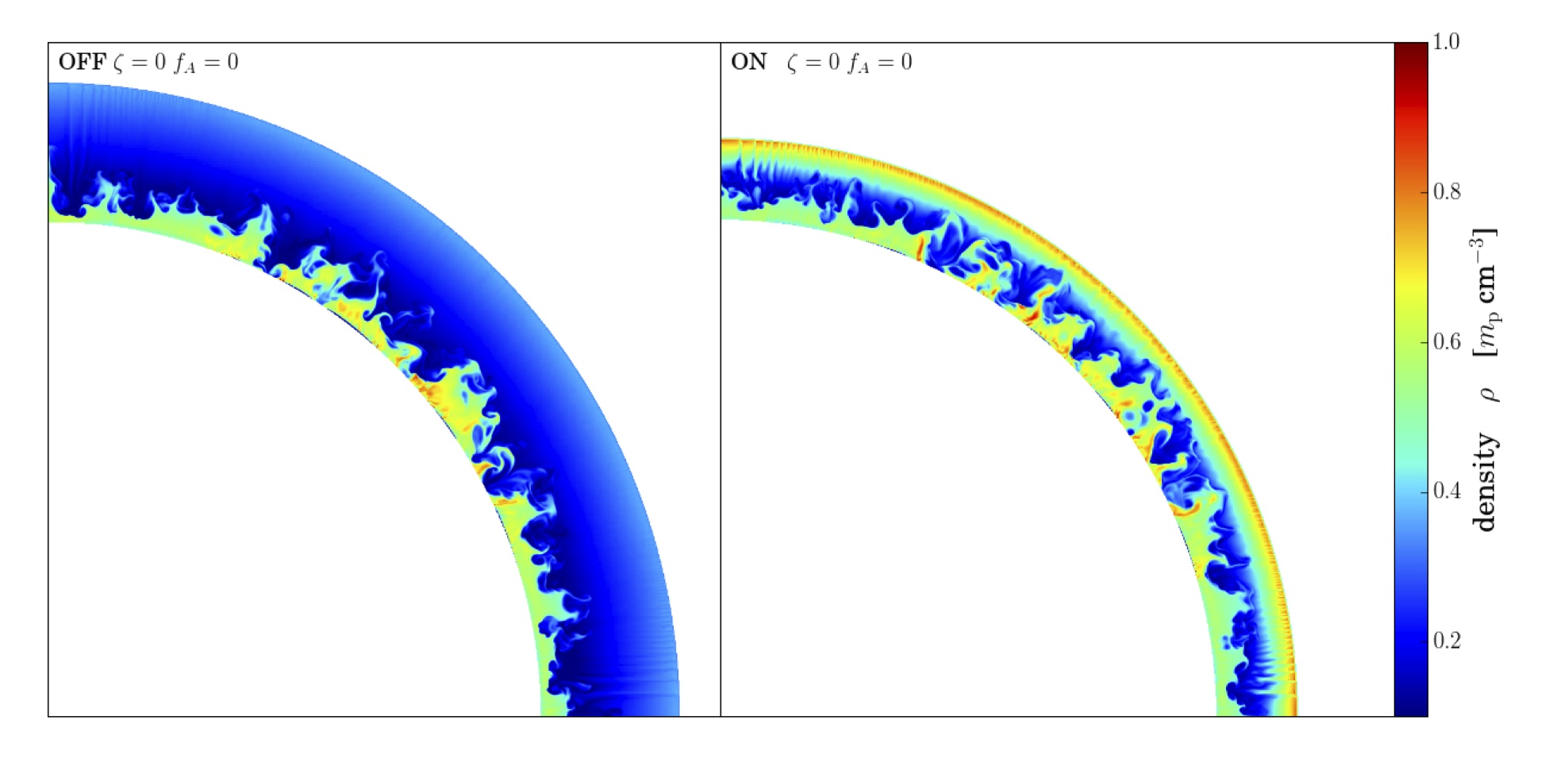}
\includegraphics[width=\figsizemap]{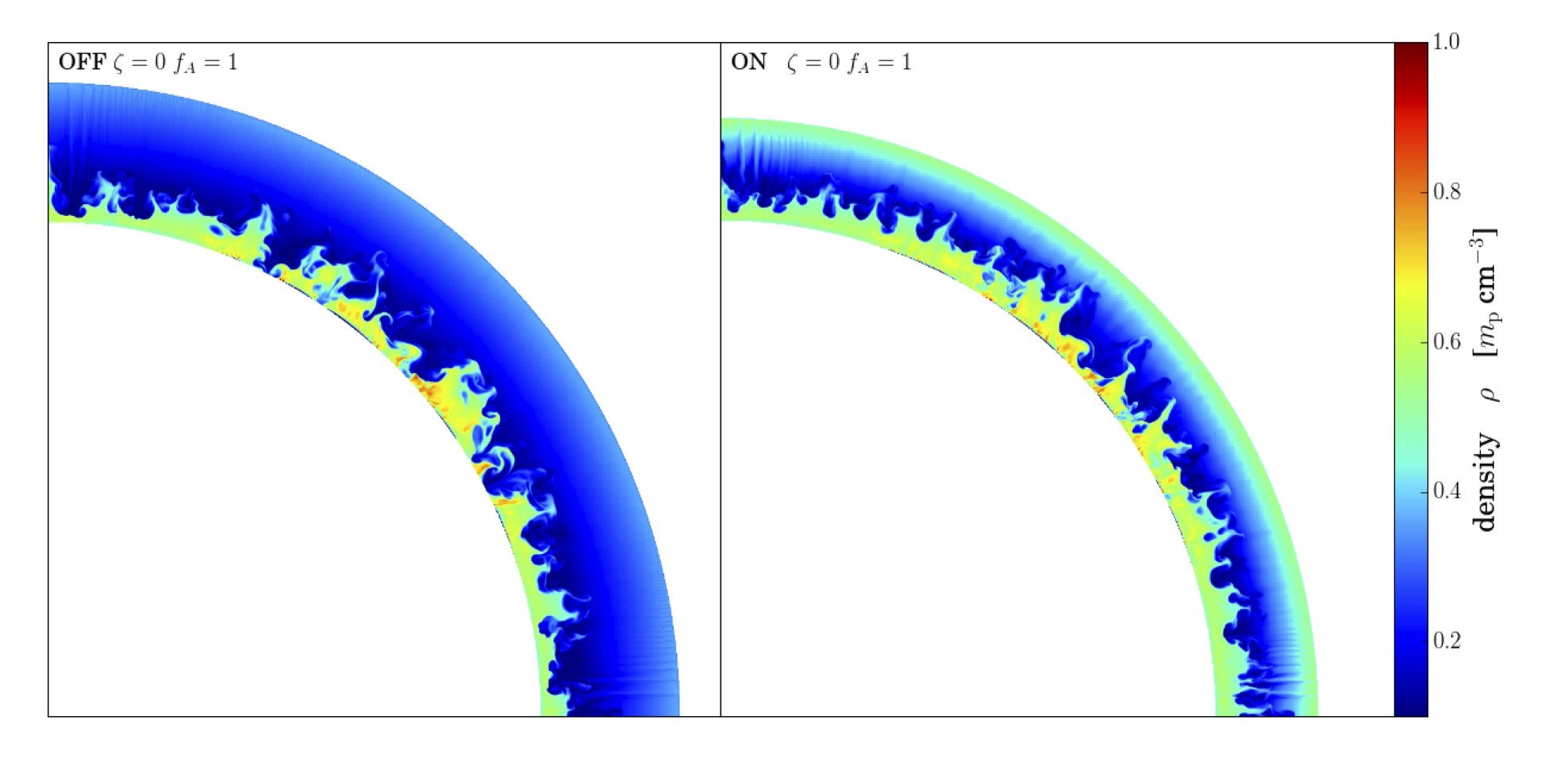}
\caption{Slices of the fluid density in the shocked region, 500 years after the supernova. We assume 1.4~solar~masses were ejected with a kinetic energy of $10^{51}$ erg in a uniform ISM of density $0.1\,{\rm cm^{-3}}$ (see Section \ref{sec:parameters-SNR} for details). Each plot covers 5.61~pc with a resolution of 1024 pixels in each direction. The right plots (ON cases) include the hydrodynamic back-reaction of accelerated particles. The three rows show the effect of the magnetic back-reaction in different cases: with no net amplification ($\zeta=1$ and thus $f_A=0$, top), with efficient amplification but still negligible Alfv{\'e}nic drift ($\zeta=0$ and $f_A=0$, middle), and with efficient amplification and strong Alfv{\'e}nic drift ($\zeta=0$ and $f_A=1$, bottom).
\label{fig:map-np}}
\end{figure}

\begin{figure}[t]
\centering
\includegraphics[width=\figsizemap]{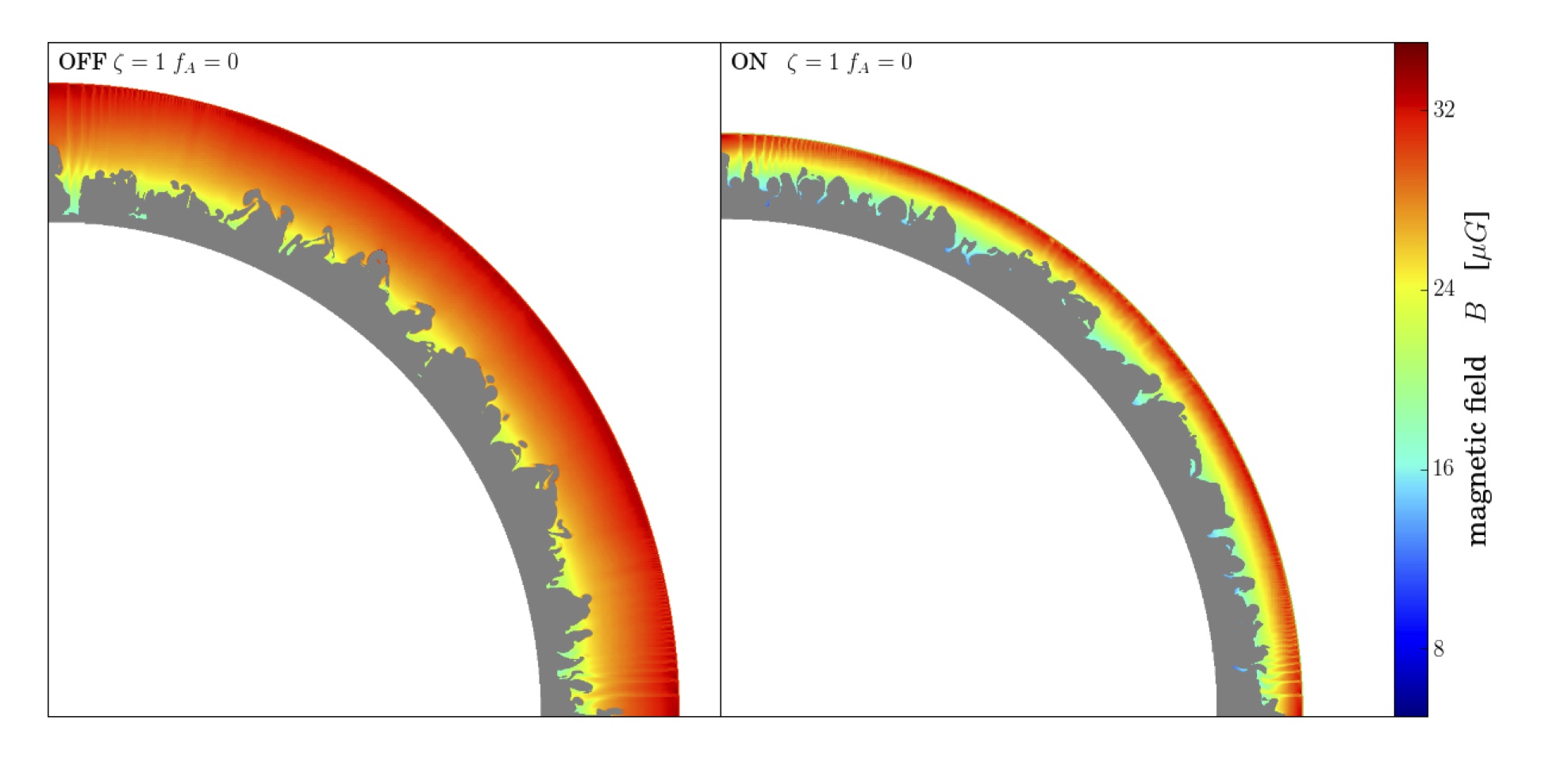}
\includegraphics[width=\figsizemap]{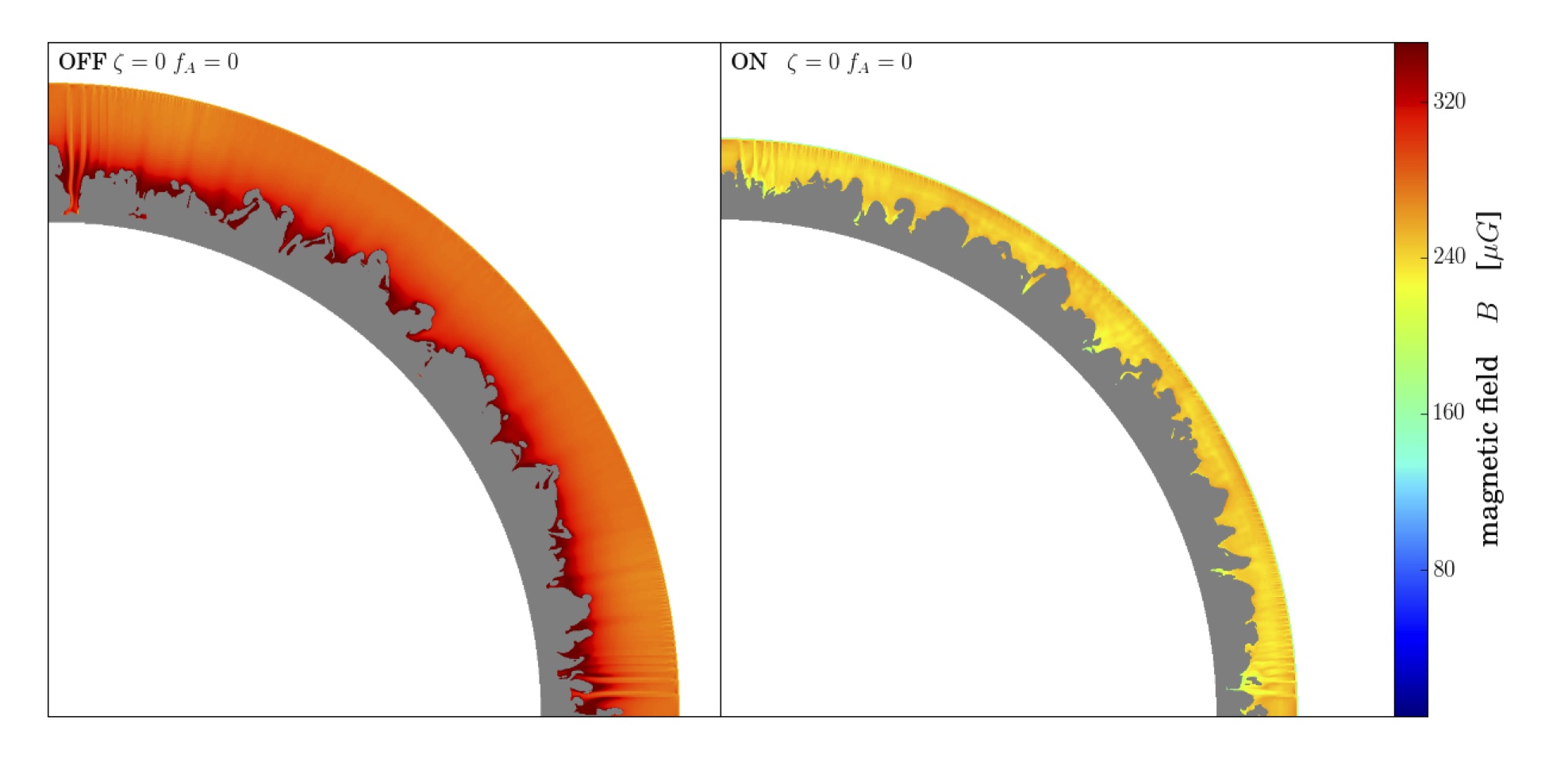}
\includegraphics[width=\figsizemap]{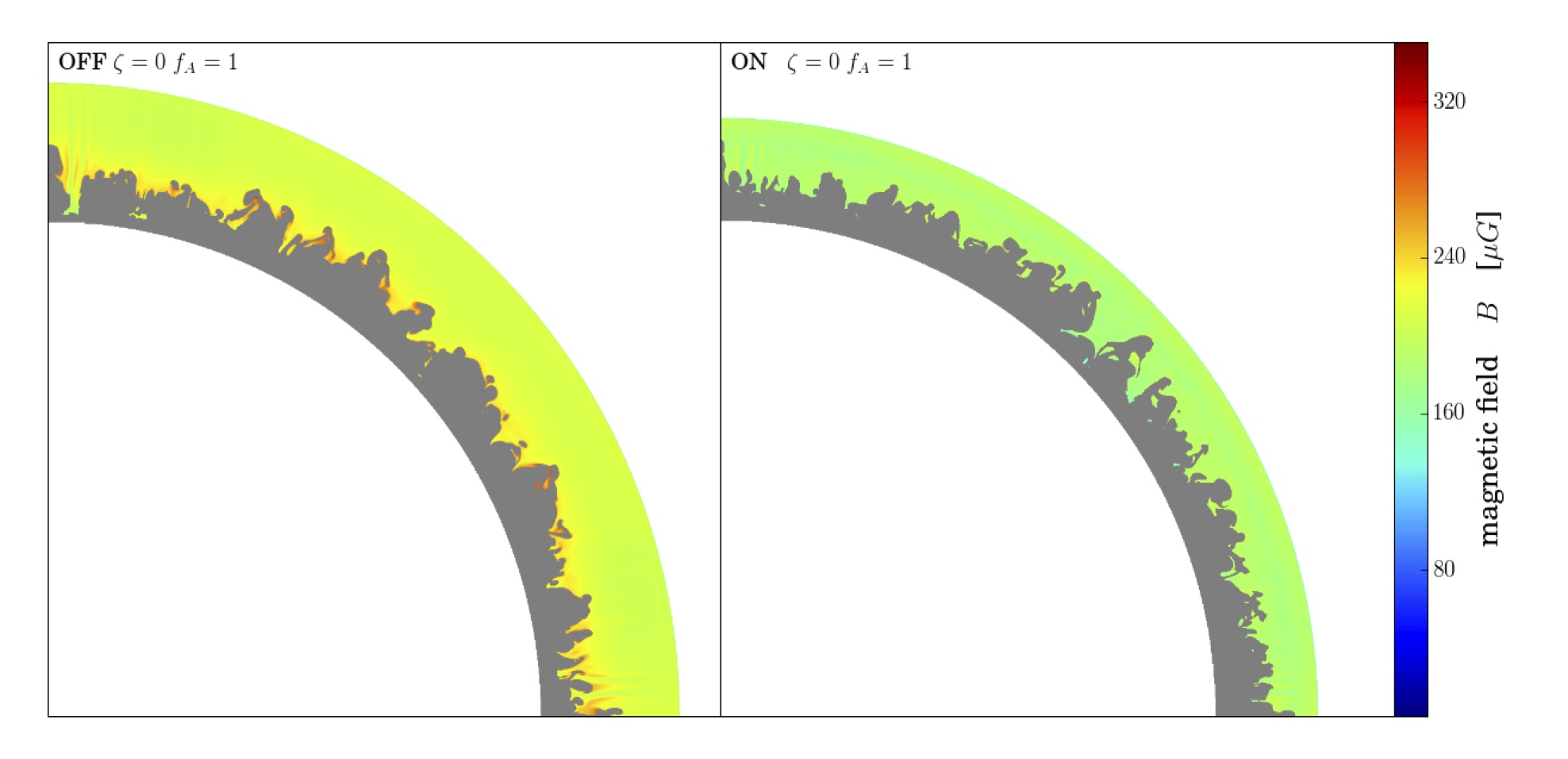}
\caption{Slices of the magnetic field in the shocked ISM. Model parameters are the same as for Figure~1. Note that a different scale is used for the case with no net amplification of the magnetic field ($\zeta=1$, top panel). No computation is performed in the shocked ejecta (grey area).
\label{fig:map-B_FS}}
\end{figure}

\begin{figure}[t]
\centering
\includegraphics[width=\figsizemap]{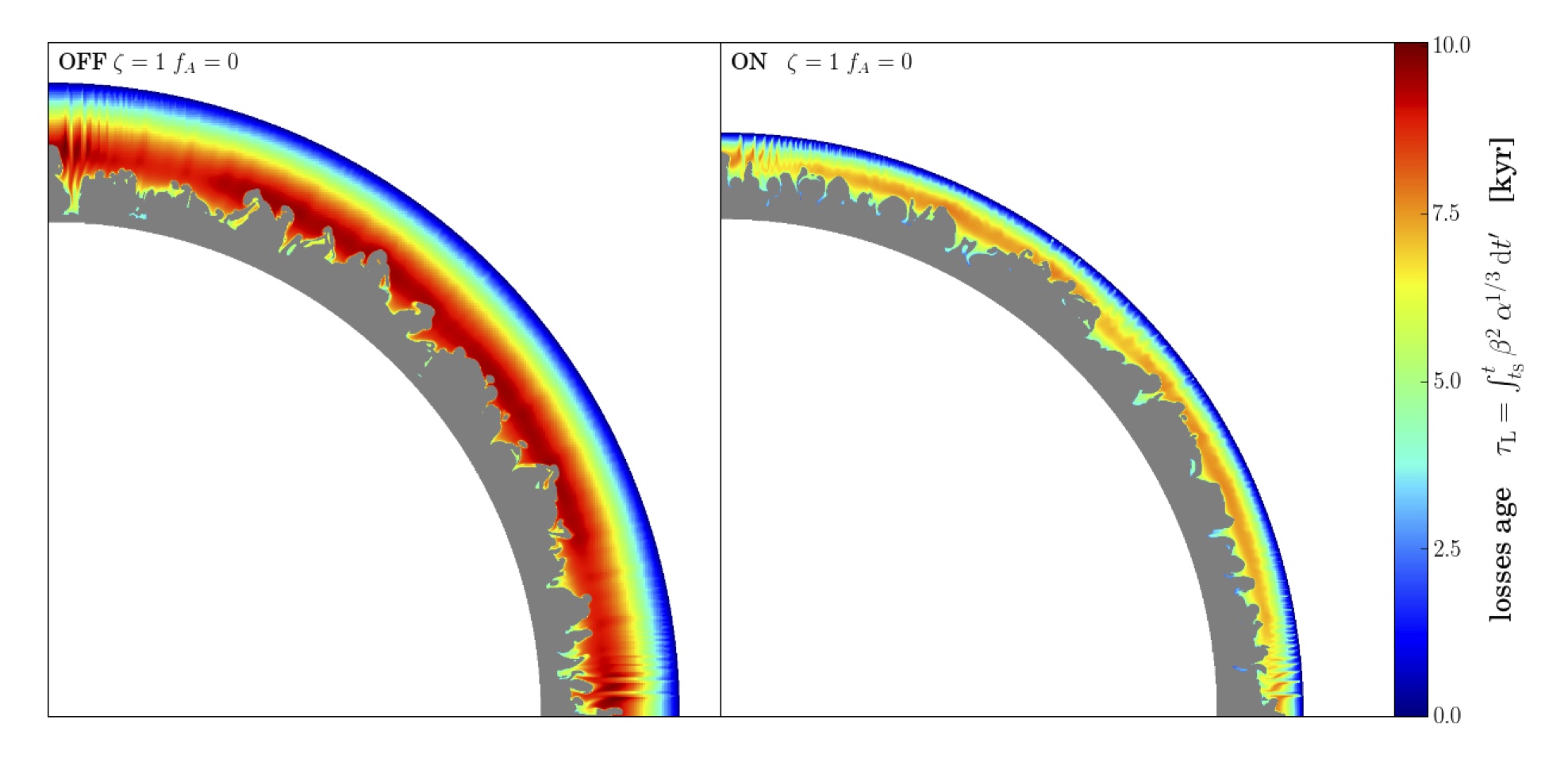}
\includegraphics[width=\figsizemap]{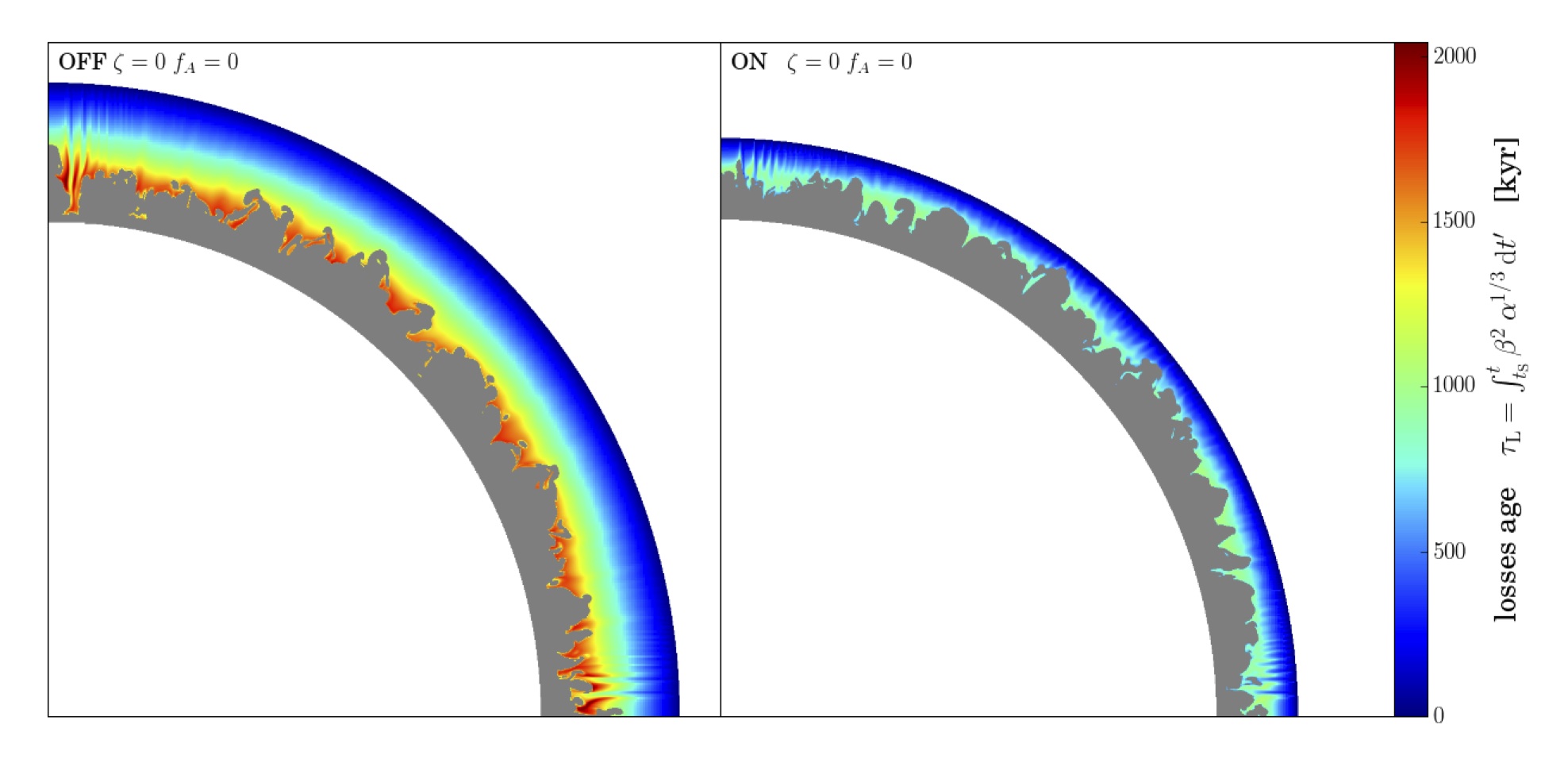}
\includegraphics[width=\figsizemap]{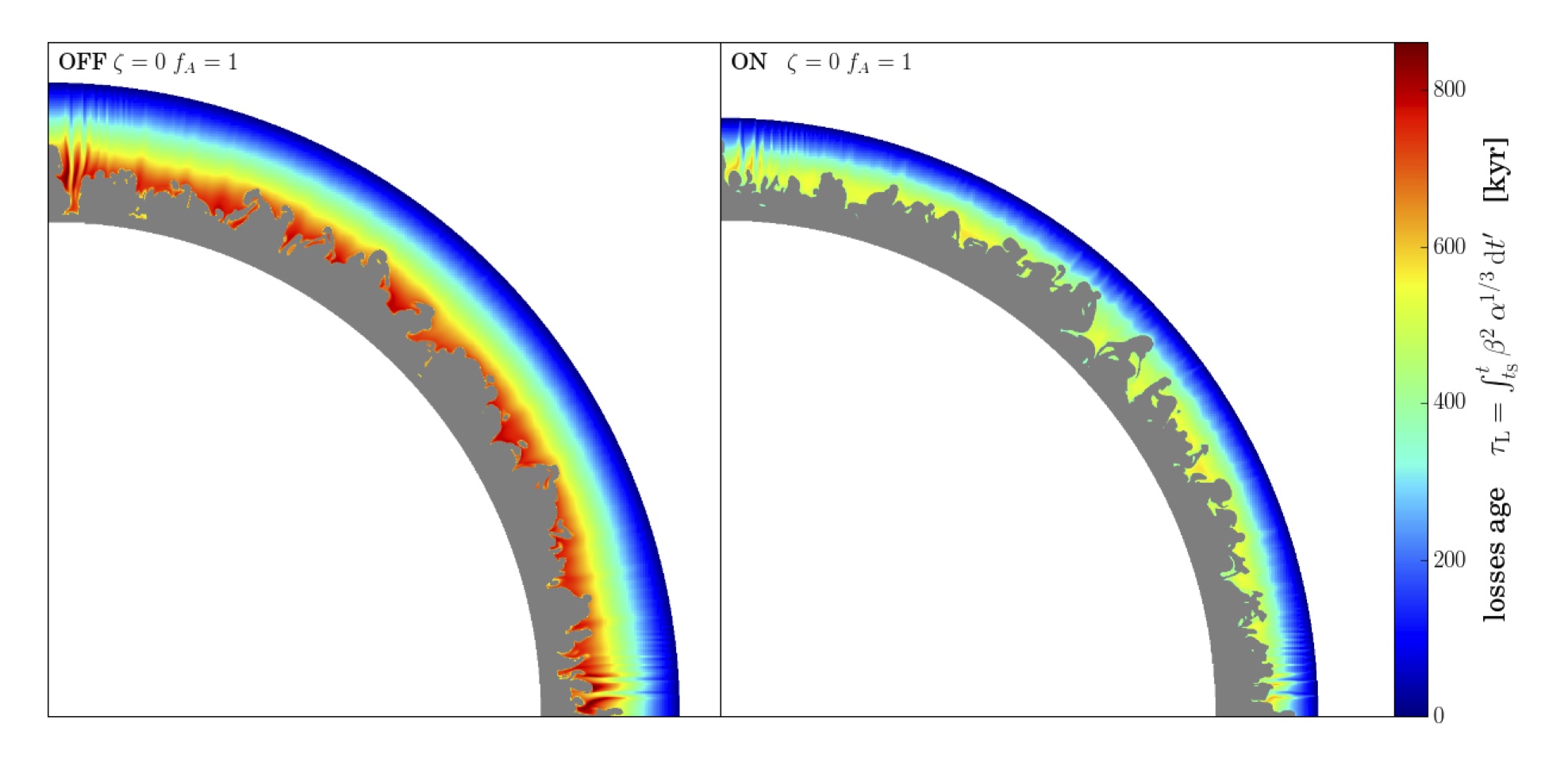}
\caption{Slices of the synchrotron radiative loss age in the shocked ISM, as defined by Equation~(\ref{eq:tau_L}). Model parameters are the same as for Figure~1. Note that a different scale is used for each row. 
\label{fig:map-tL_FS}}
\end{figure}

\begin{figure}[t]
\centering
\includegraphics[width=\figsizemap]{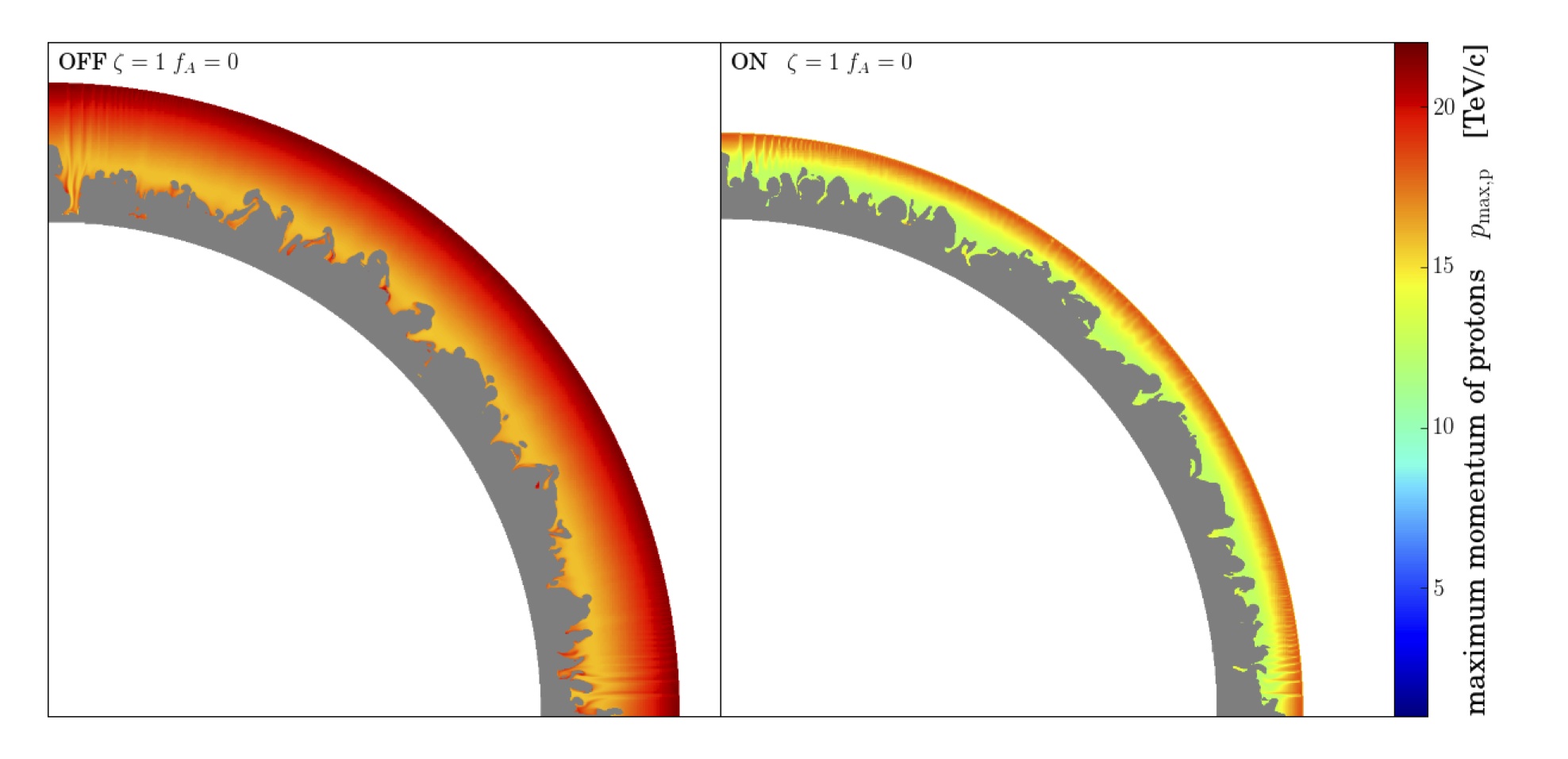}
\includegraphics[width=\figsizemap]{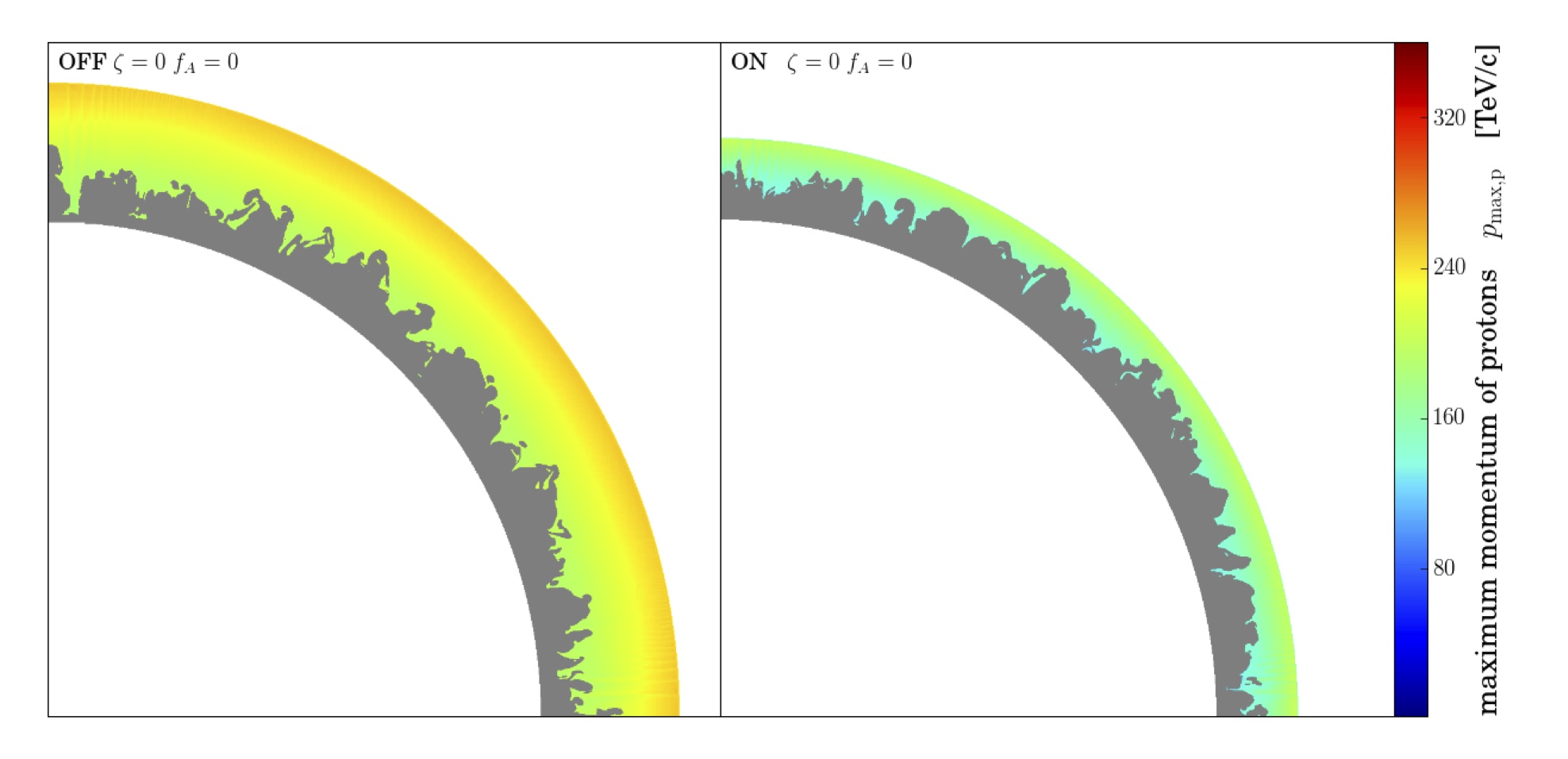}
\includegraphics[width=\figsizemap]{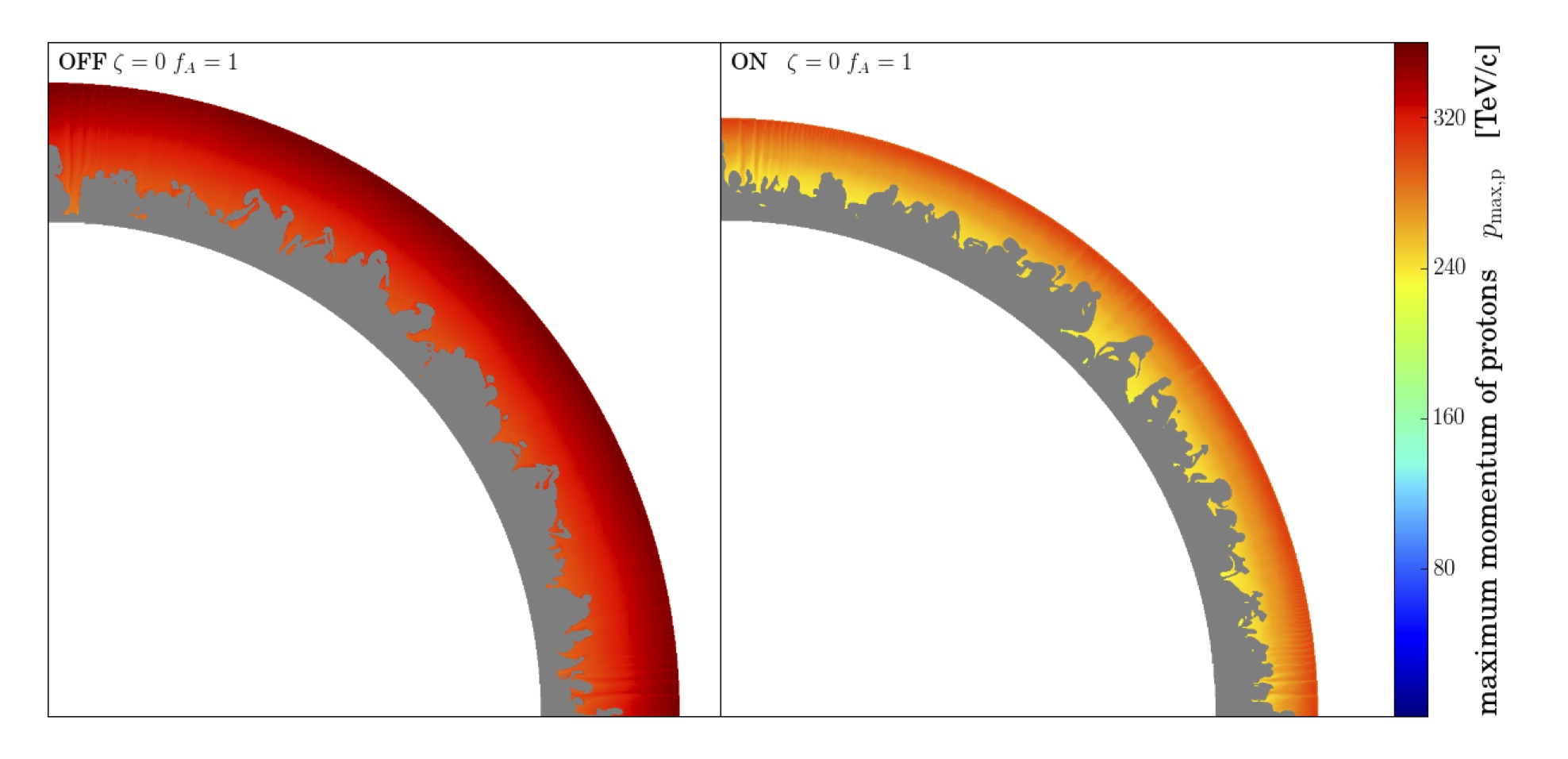}
\caption{Slices of the maximum energy of protons in the shocked ISM, after losses. Model parameters are the same as for Figure~1. Note that a different scale is used for the case with no net amplification of the magnetic field ($\zeta=1$, top panel). As only acceleration at the forward shock is considered, no computation is performed in the shocked ejecta (grey area).
\label{fig:map-pLp}}
\end{figure}

\begin{figure}[t]
\centering
\includegraphics[width=\figsizemap]{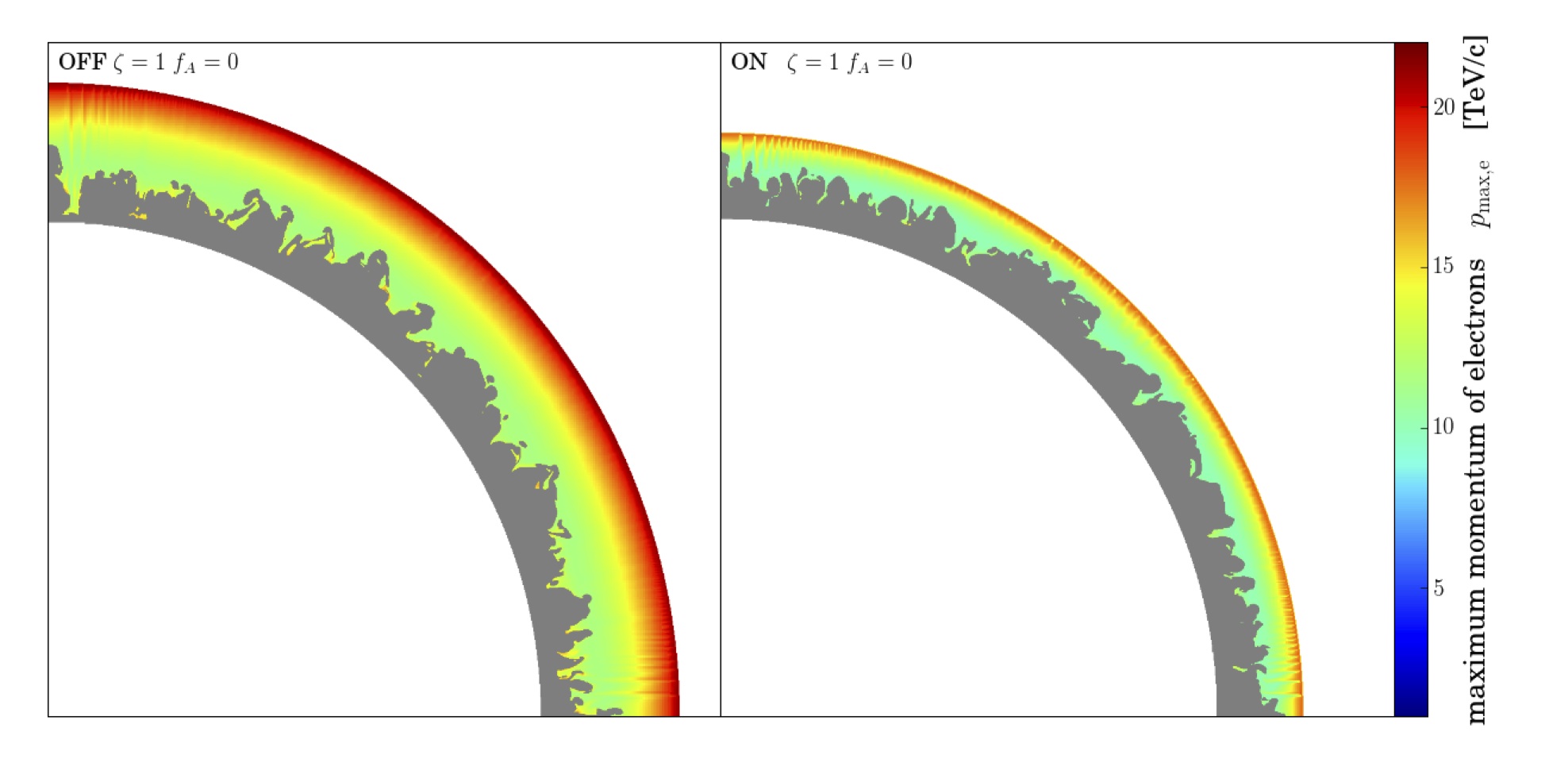}
\includegraphics[width=\figsizemap]{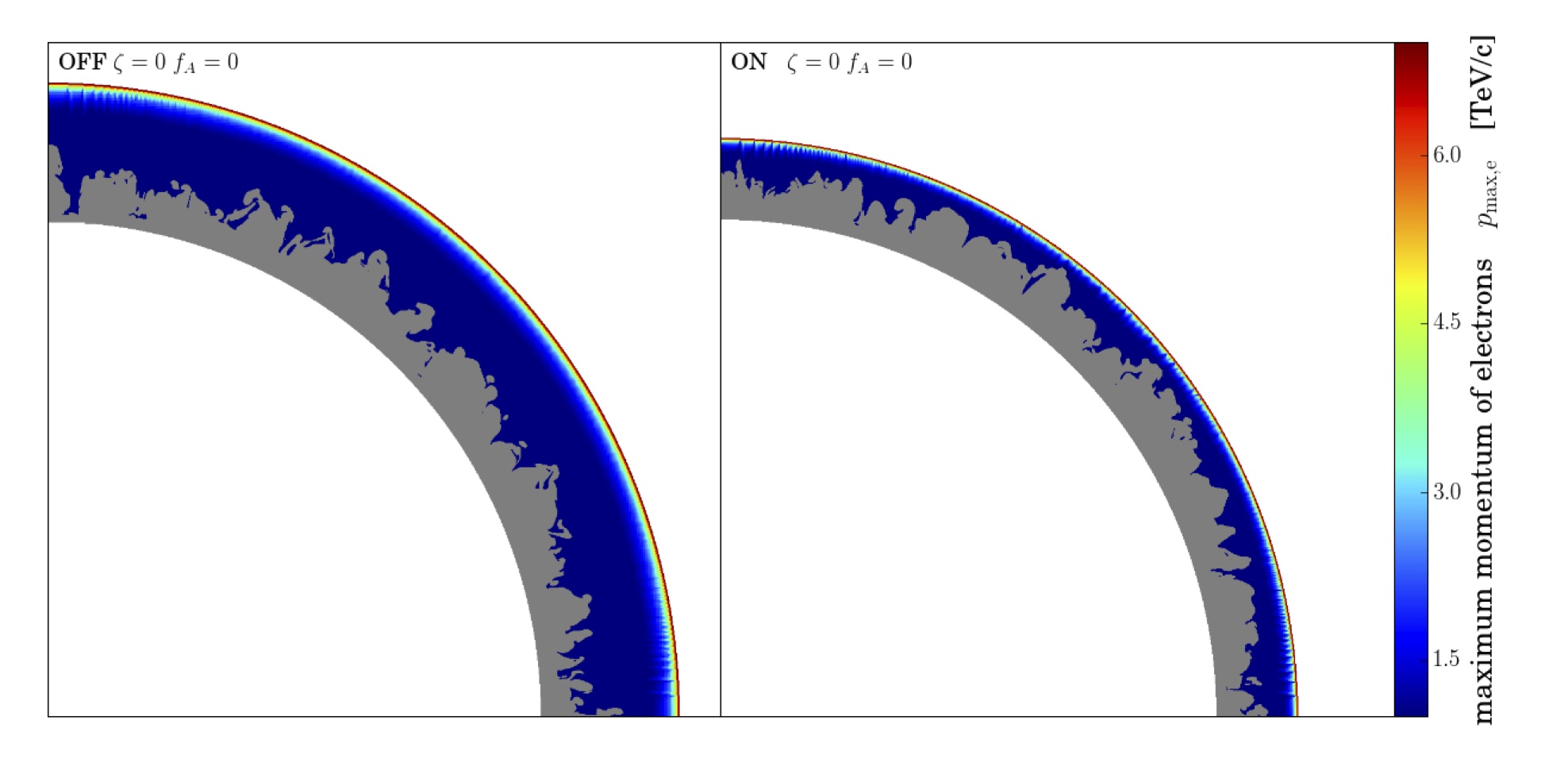}
\includegraphics[width=\figsizemap]{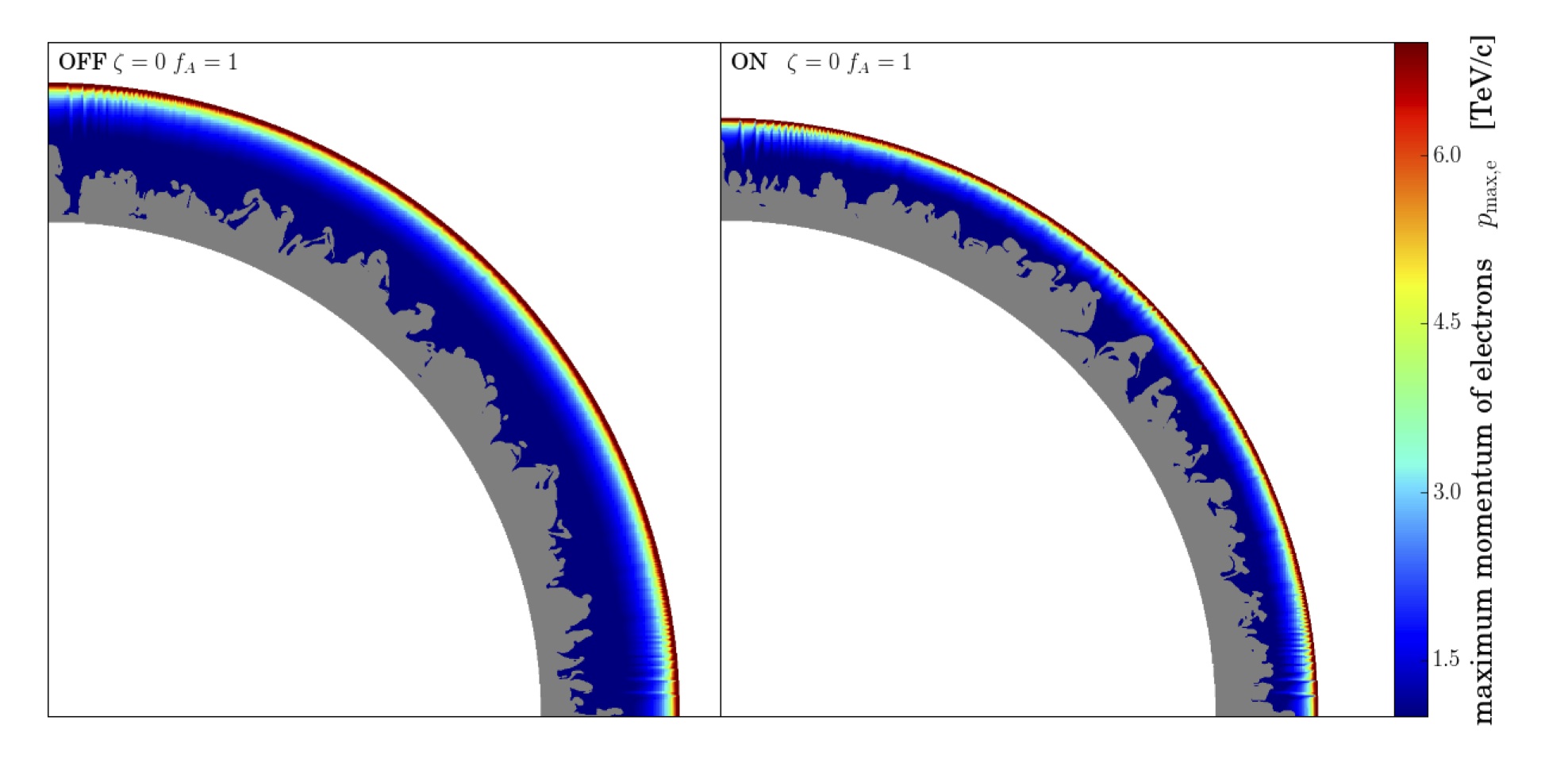}
\caption{Slices of the maximum energy of electrons in the shocked ISM, after losses. Model parameters are the same as for Figure~1. Note that a different scale is used for the case with no net amplification of the magnetic field ($\zeta=1$, top panel). As only acceleration at the forward shock is considered, no computation is performed in the shocked ejecta (grey area).
\label{fig:map-pLe_FS}}
\end{figure}

\begin{figure}[t]
\centering
\includegraphics[width=\figsizemap]{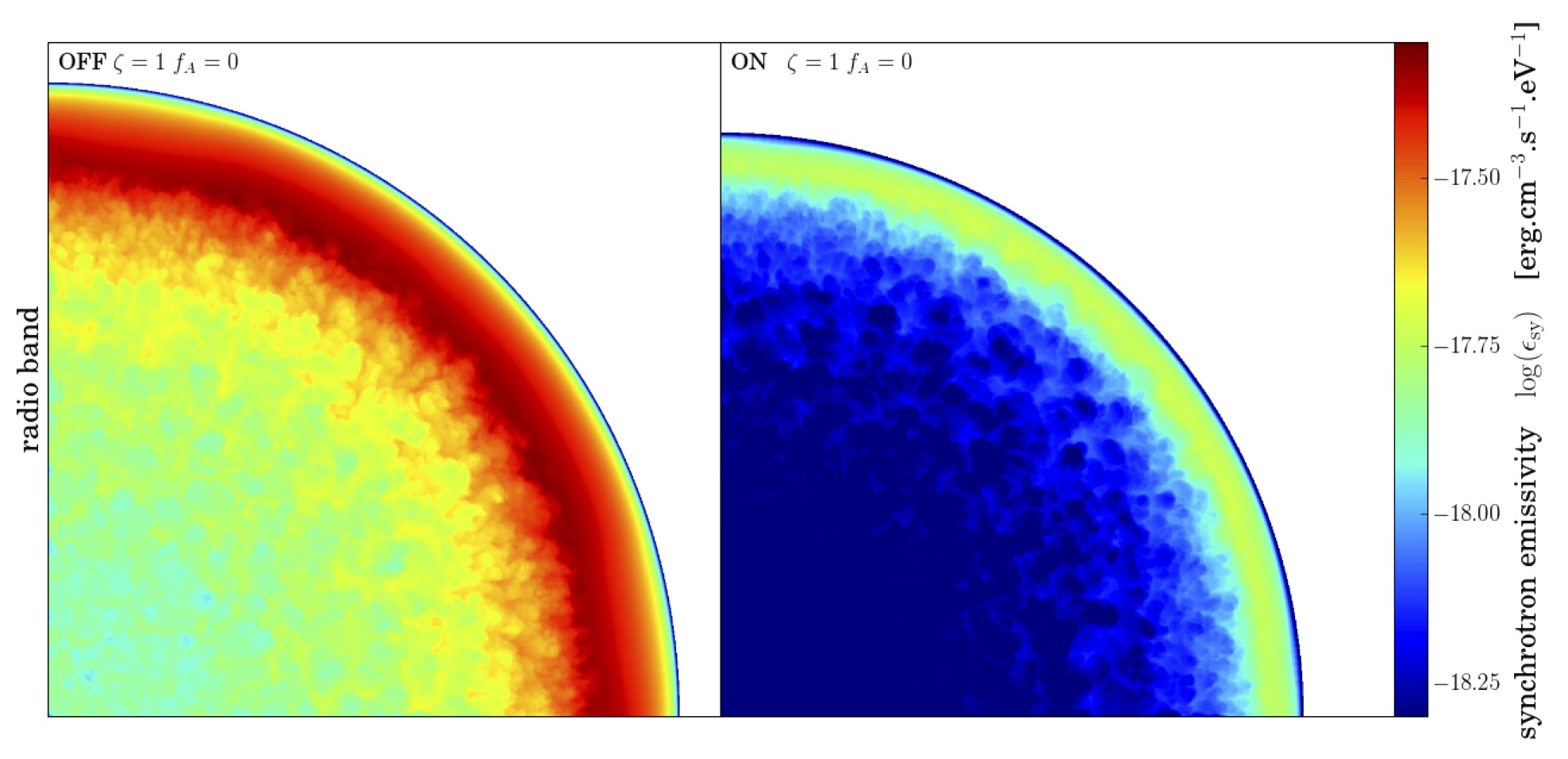}
\includegraphics[width=\figsizemap]{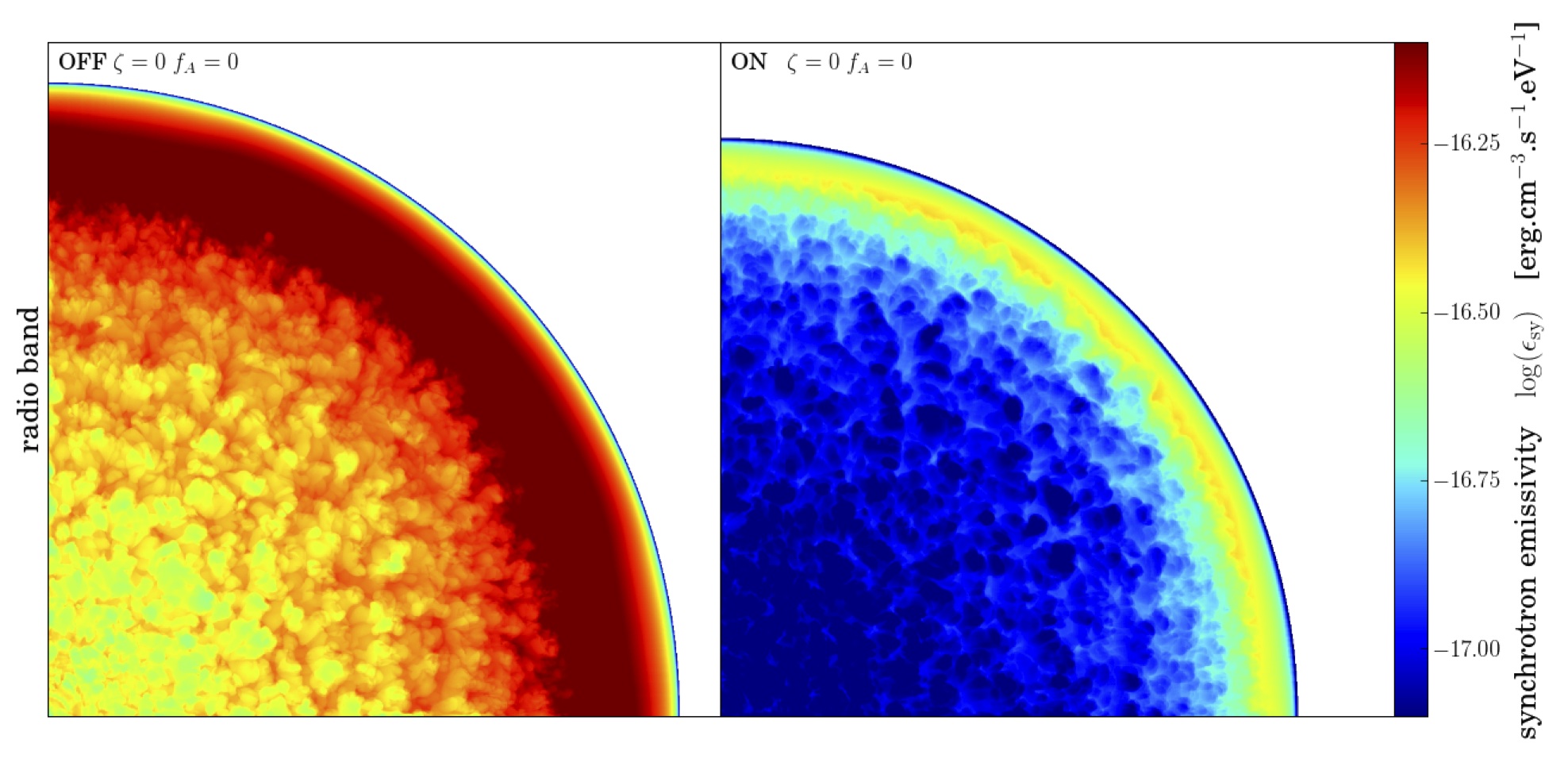}
\includegraphics[width=\figsizemap]{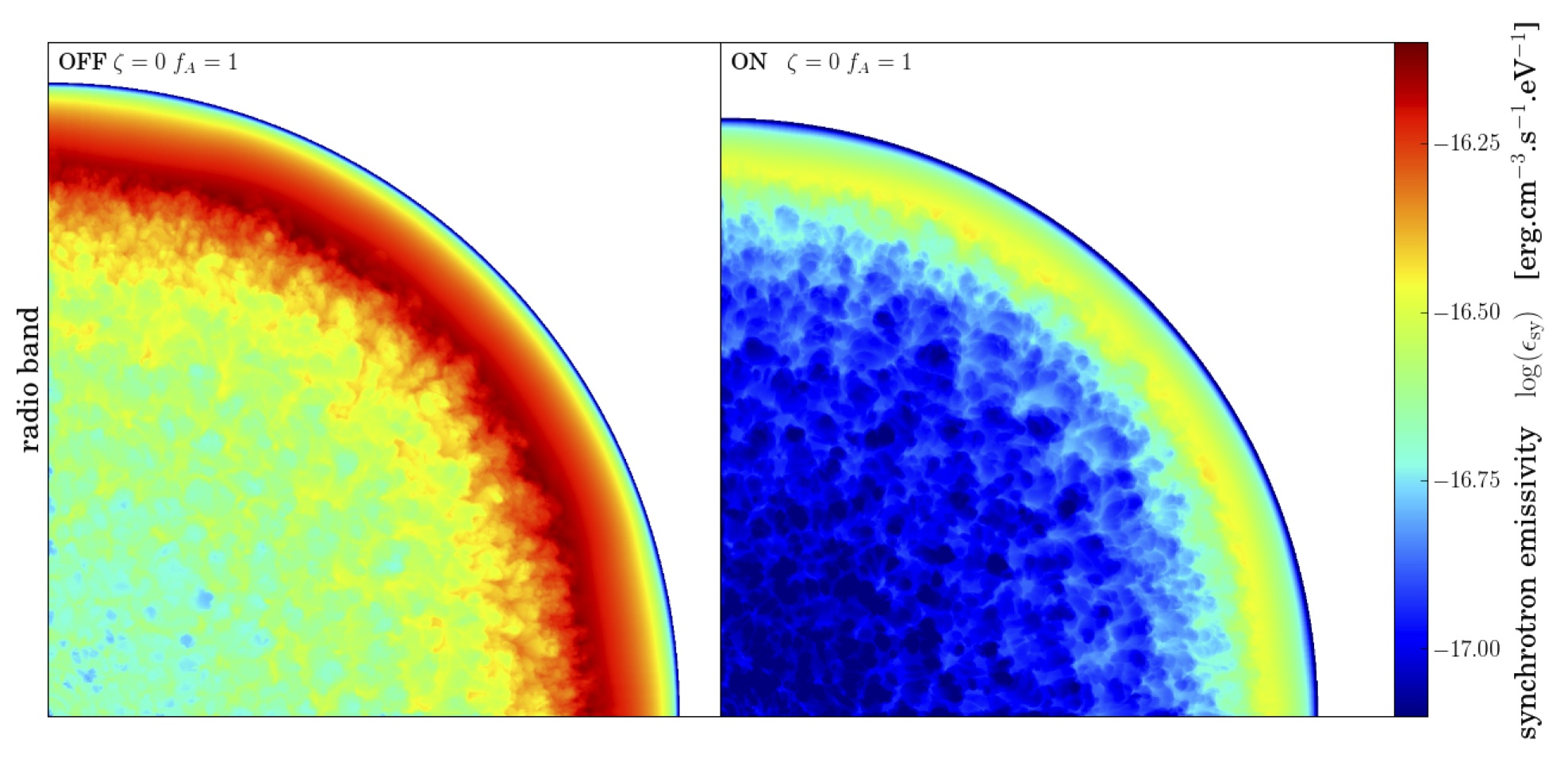}
\caption{Projected maps of the synchrotron emission in the $10^{-7}$ -- $10^{-4}$~eV radio band. Note that a different scale is used for the case with no net amplification of the magnetic field ($\zeta=1$, top panel).
\label{fig:map-NTsy-radio}}
\end{figure}

\begin{figure}[t]
\centering
\includegraphics[width=\figsizemap]{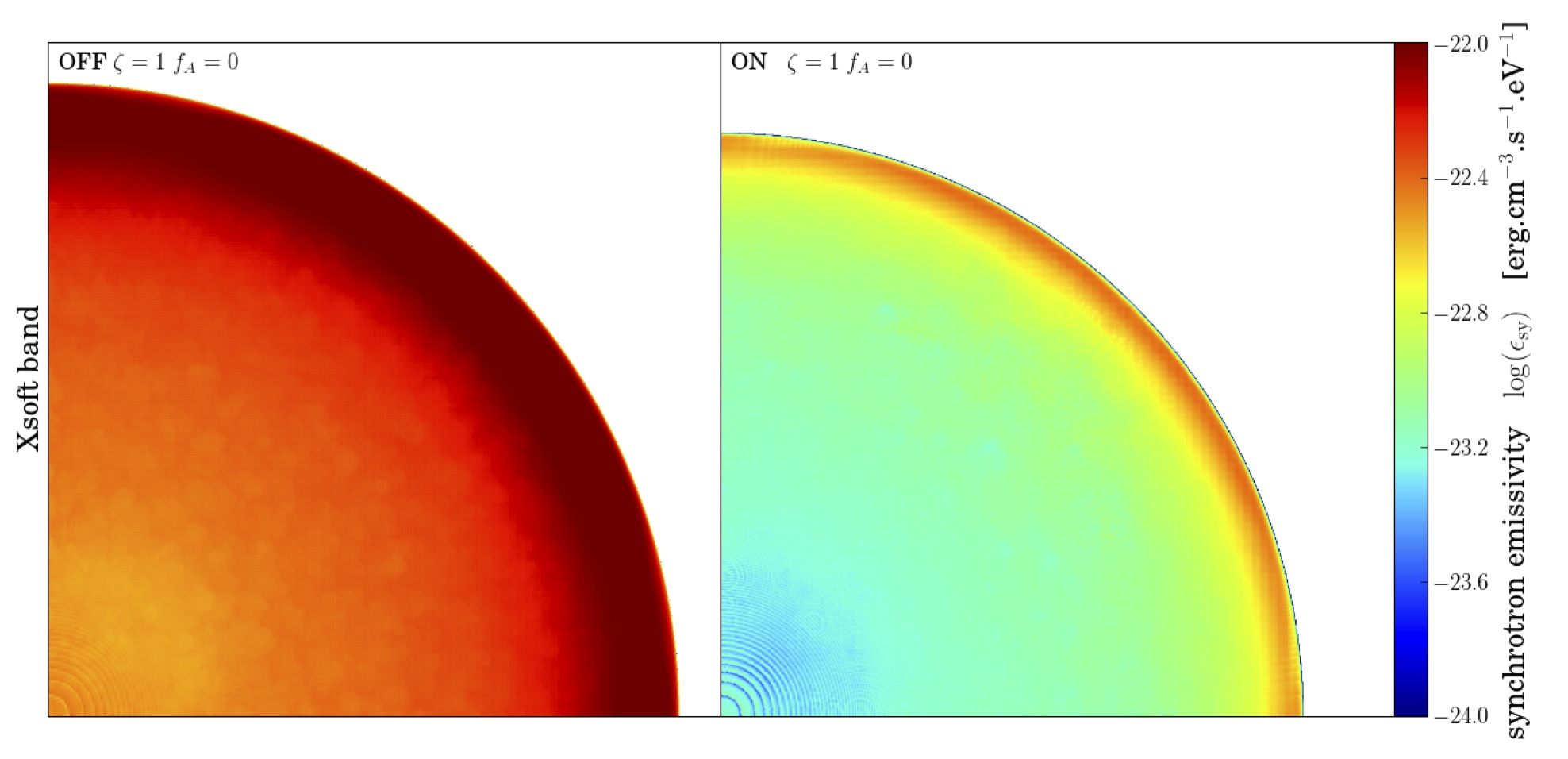}
\includegraphics[width=\figsizemap]{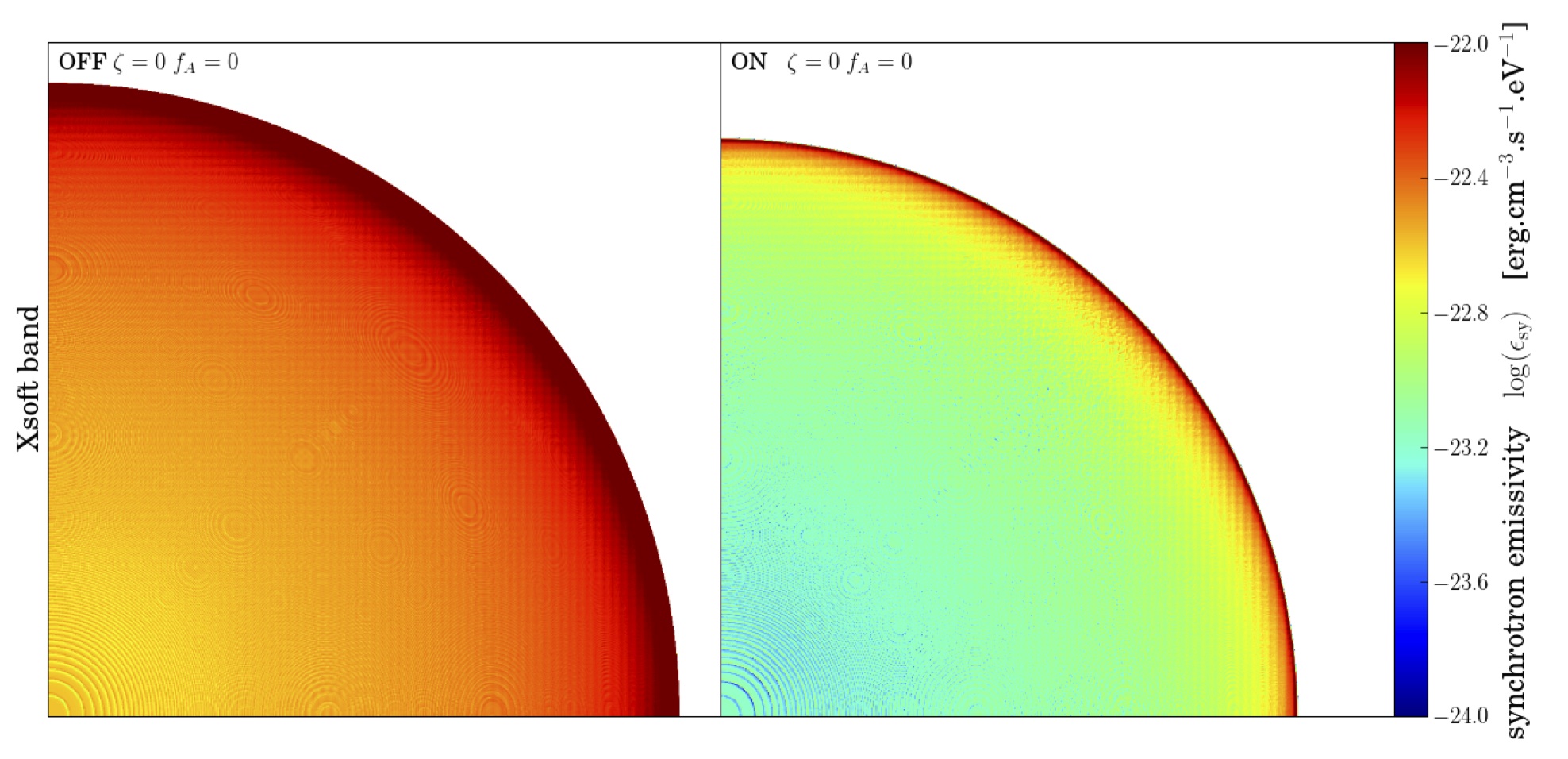}
\includegraphics[width=\figsizemap]{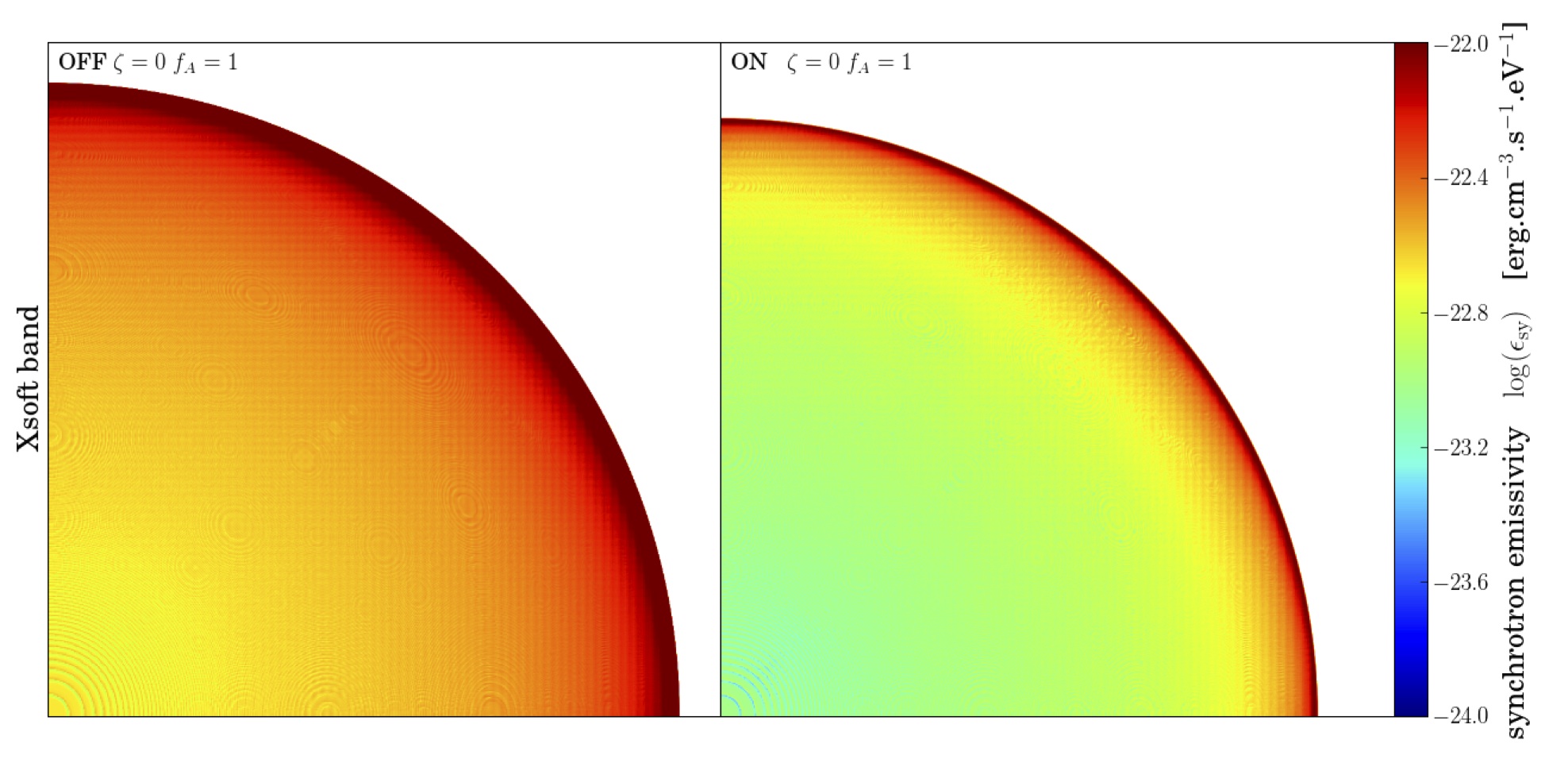}
\caption{Projected maps of the synchrotron emission in the 0.3 -- 10~keV X-ray band.
\label{fig:map-NTsy-Xsoft}}
\end{figure}

\begin{figure}[t]
\centering
\includegraphics[width=\figsizemap]{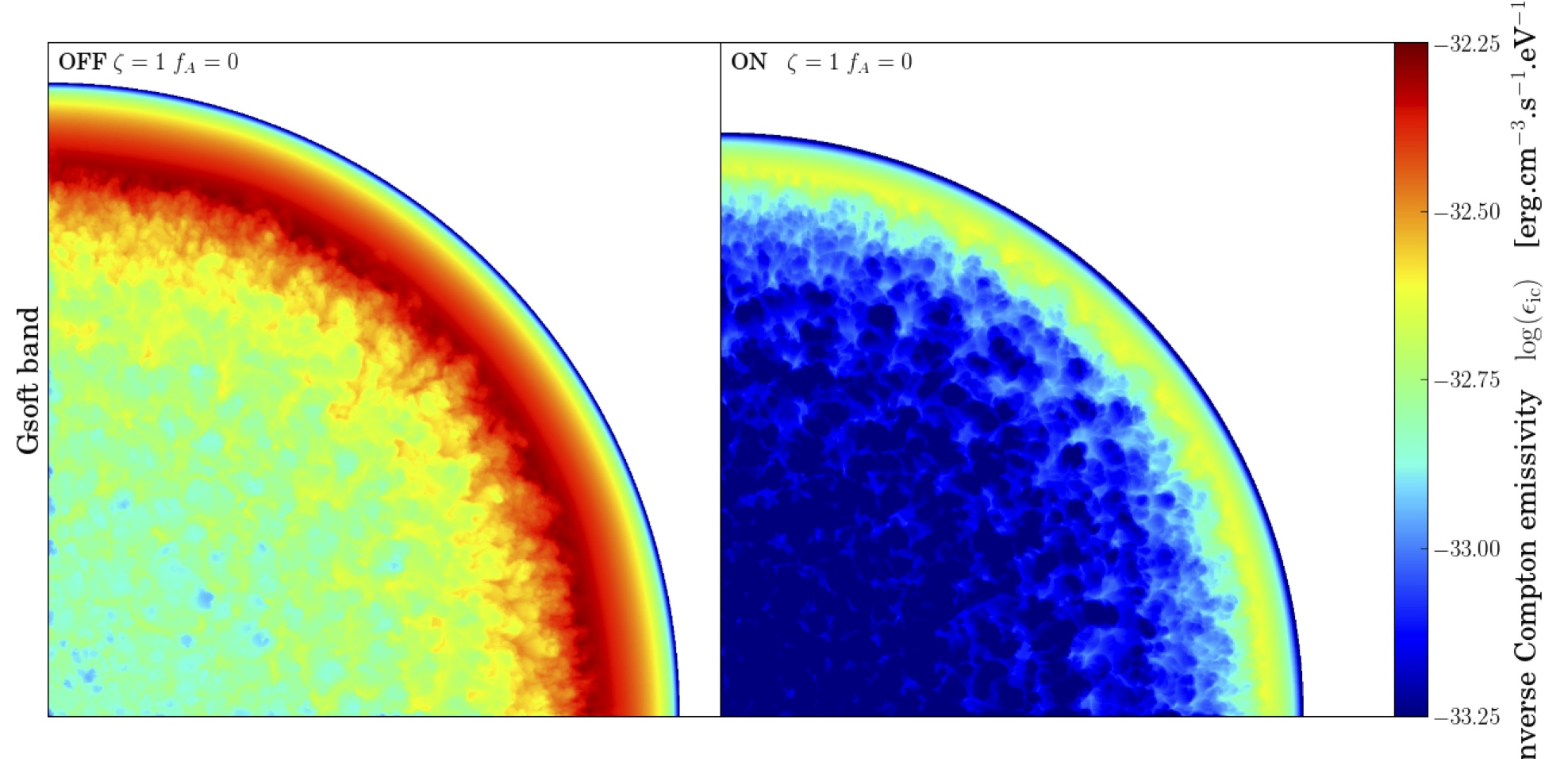}
\includegraphics[width=\figsizemap]{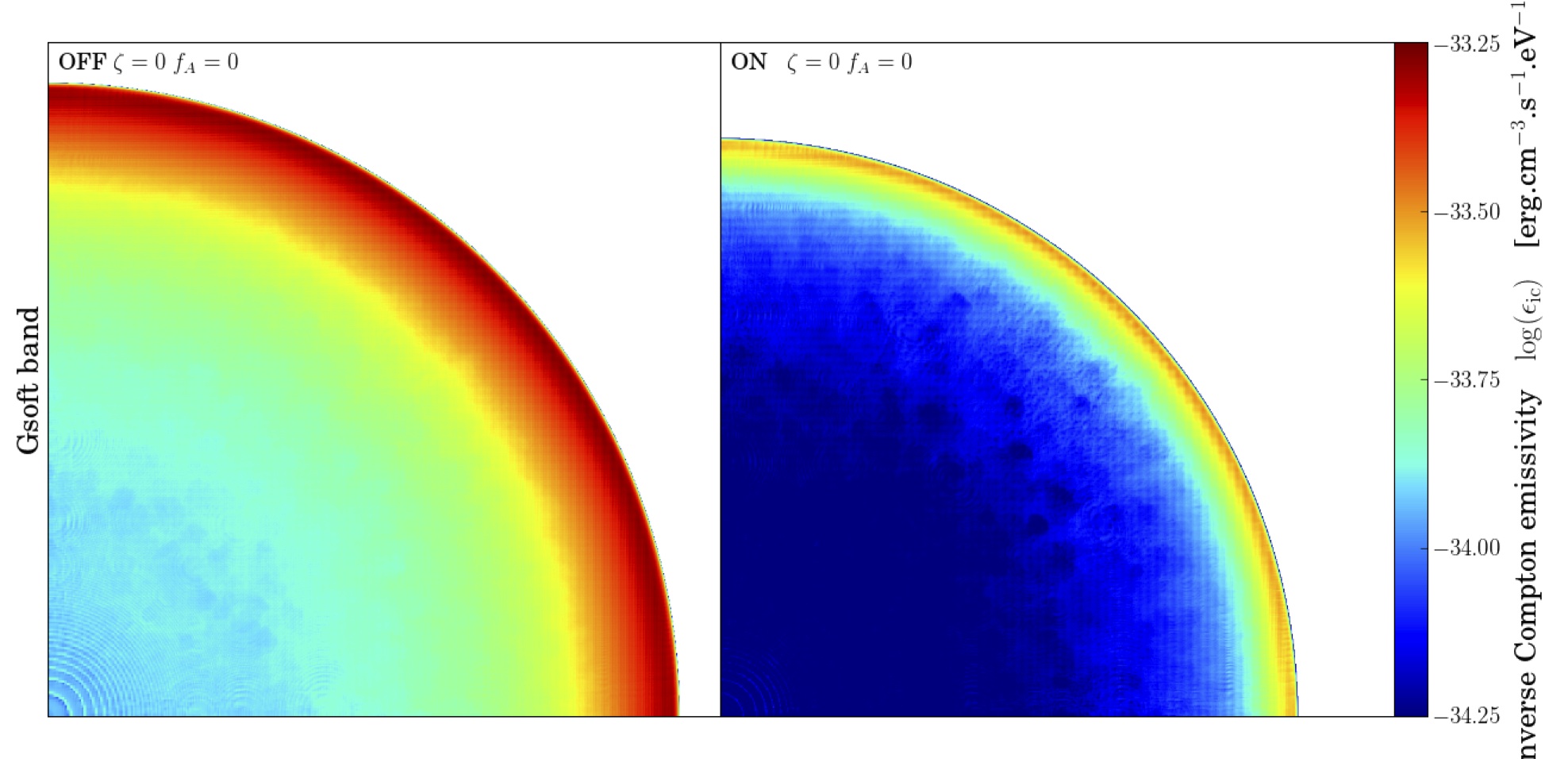}
\includegraphics[width=\figsizemap]{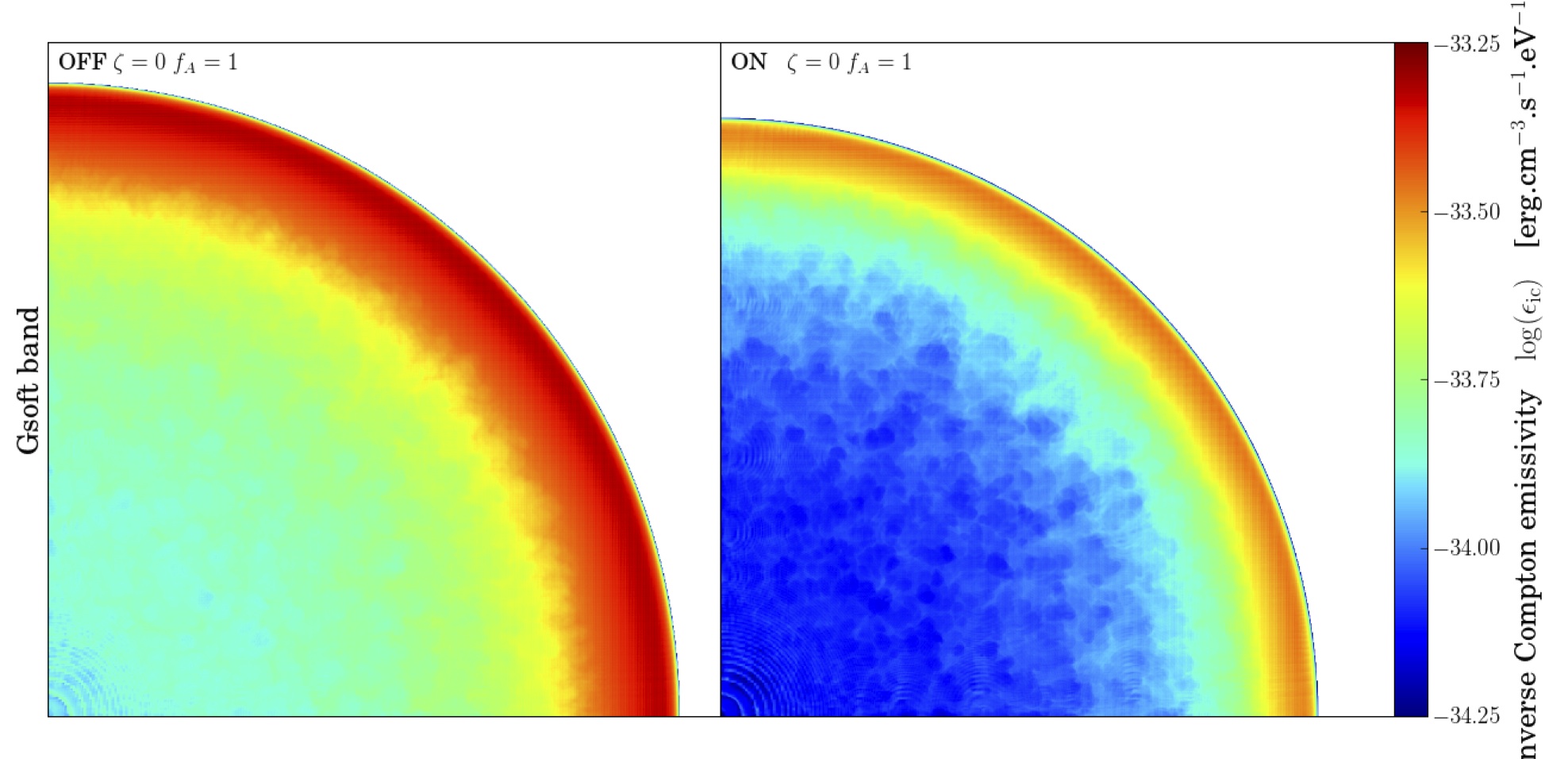}
\caption{Projected maps of the inverse Compton emission in the 0.1 -- 100~GeV $\gamma$-ray band. Note that a different scale is used for the case with no net amplification of the magnetic field ($\zeta=1$, top panel).
\label{fig:map-NTic-Gsoft}}
\end{figure}

\begin{figure}[t]
\centering
\includegraphics[width=\figsizemap]{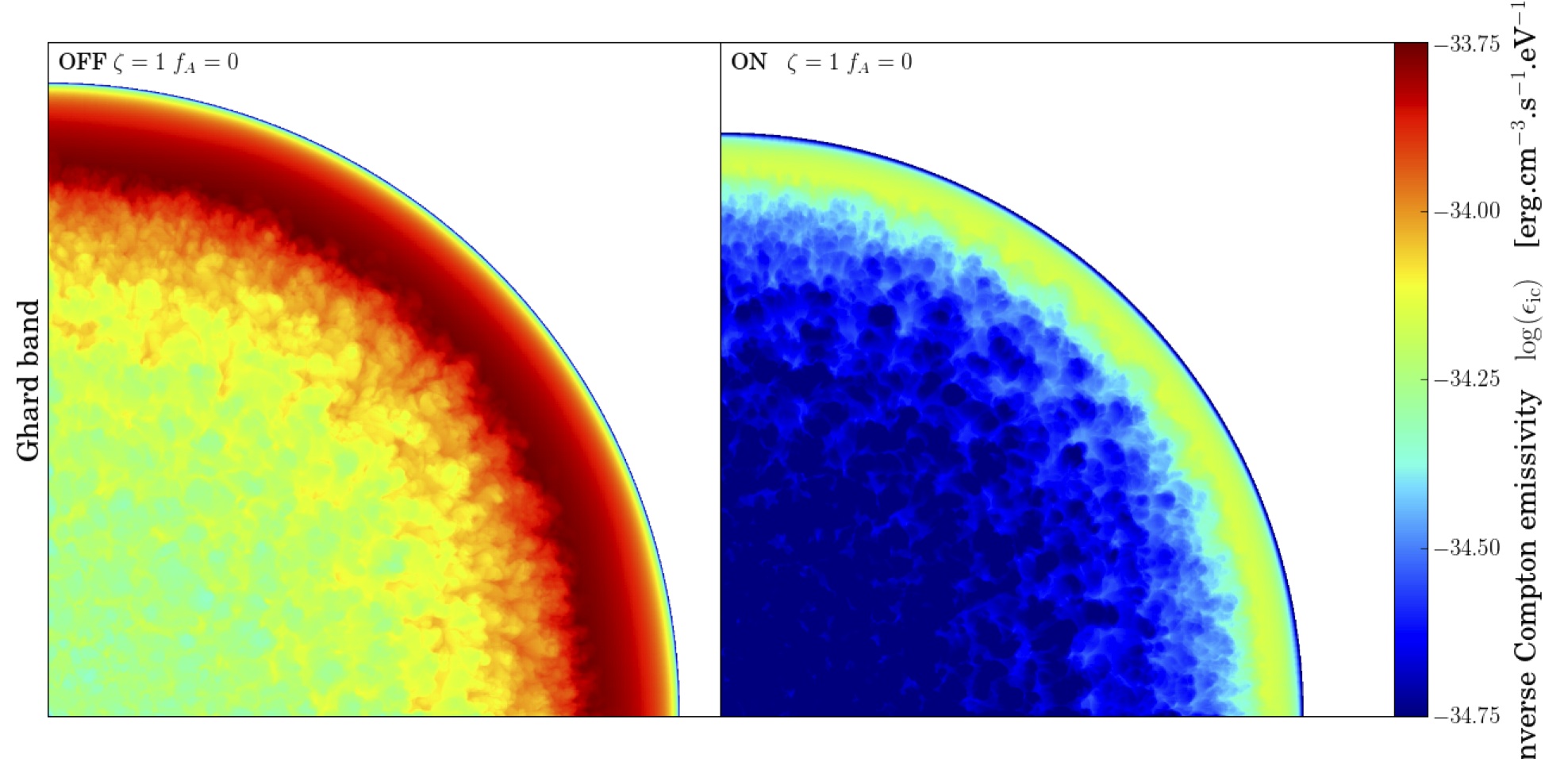}
\includegraphics[width=\figsizemap]{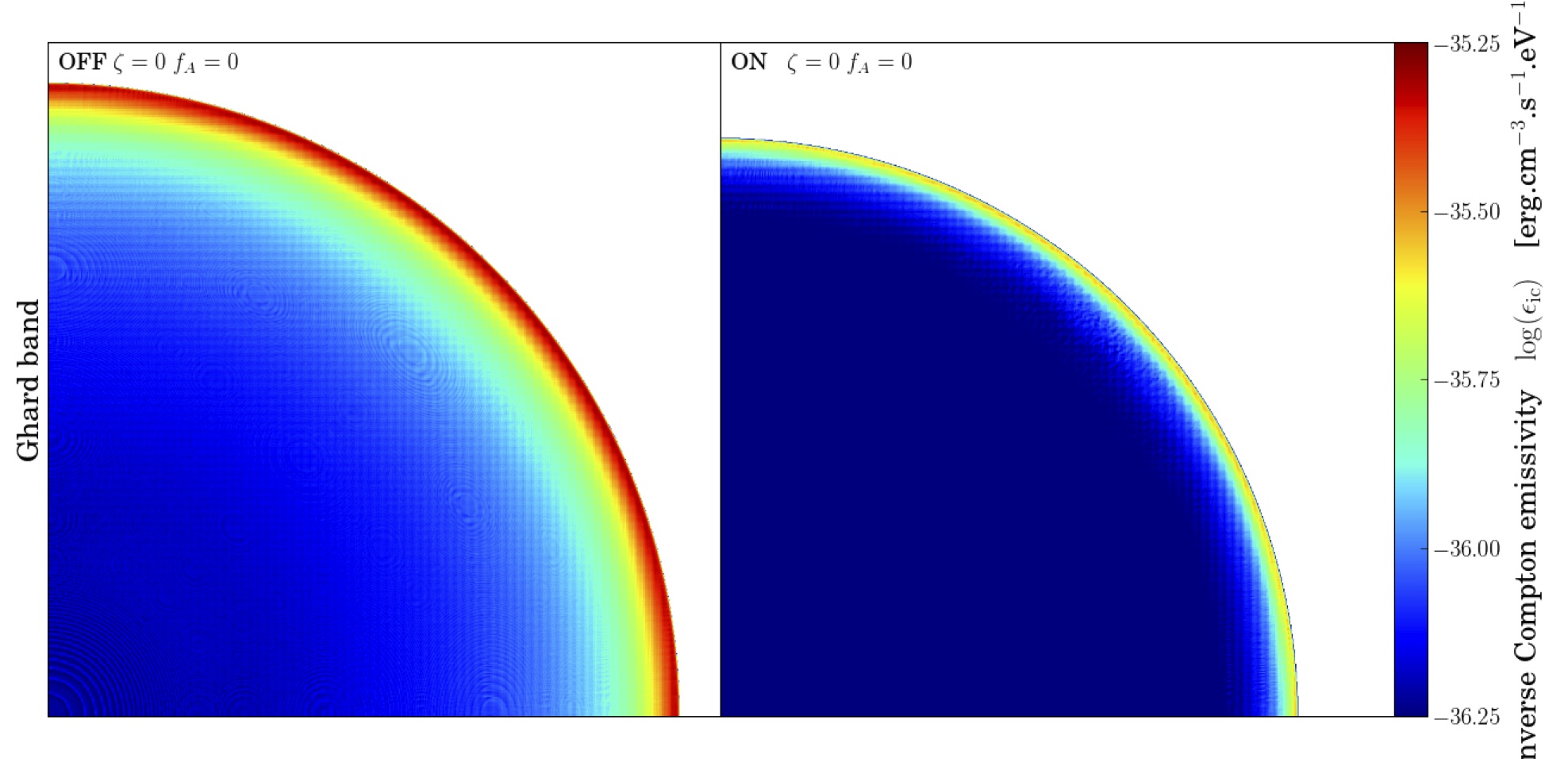}
\includegraphics[width=\figsizemap]{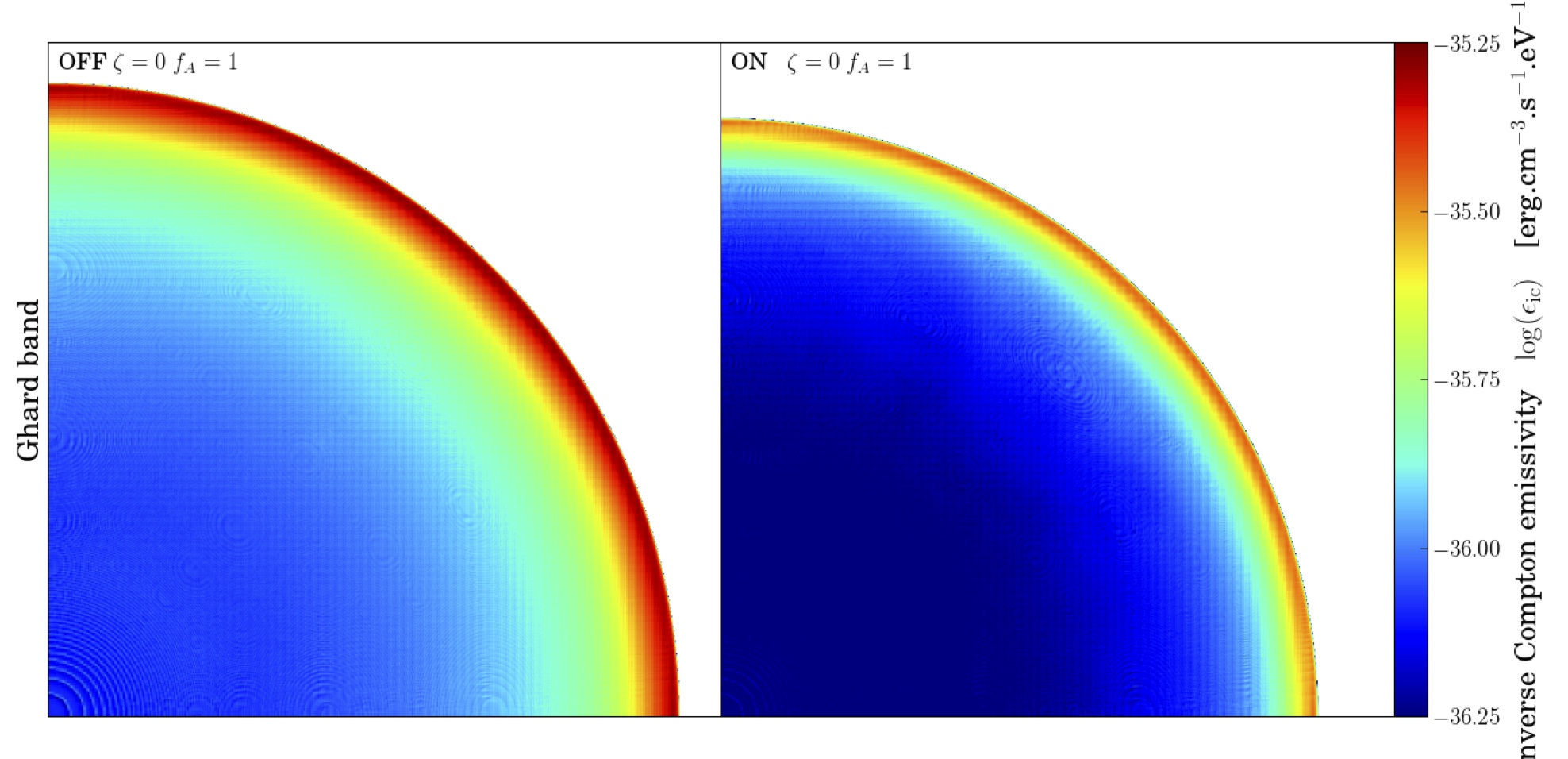}
\caption{Projected maps of the inverse Compton emission in the 0.01 -- 10~TeV $\gamma$-ray band. Note that a different scale is used for the case with no net amplification of the magnetic field ($\zeta=1$, top panel).
\label{fig:map-NTic-Ghard}}
\end{figure}

\begin{figure}[t]
\centering
\includegraphics[width=\figsizemap]{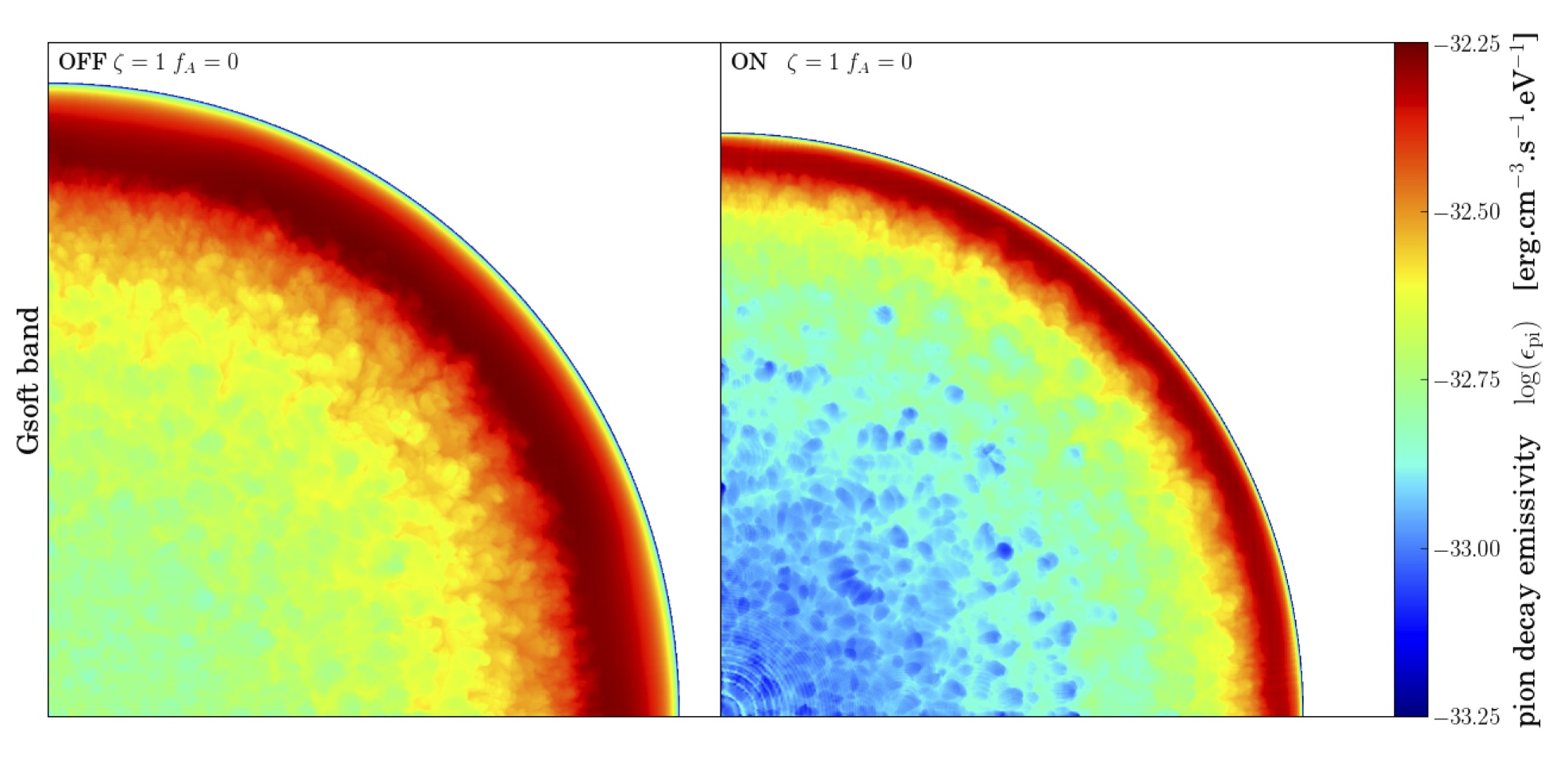}
\includegraphics[width=\figsizemap]{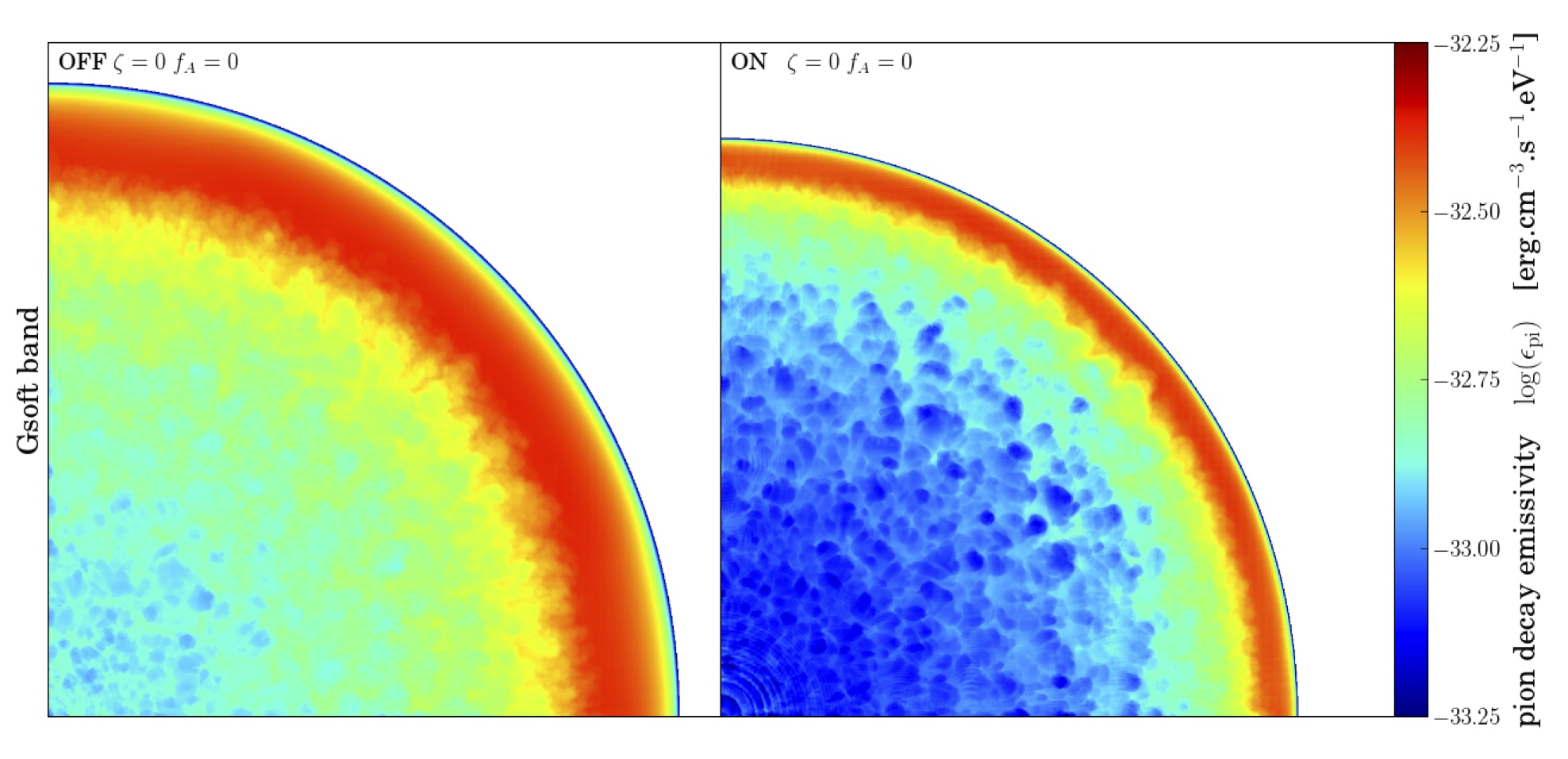}
\includegraphics[width=\figsizemap]{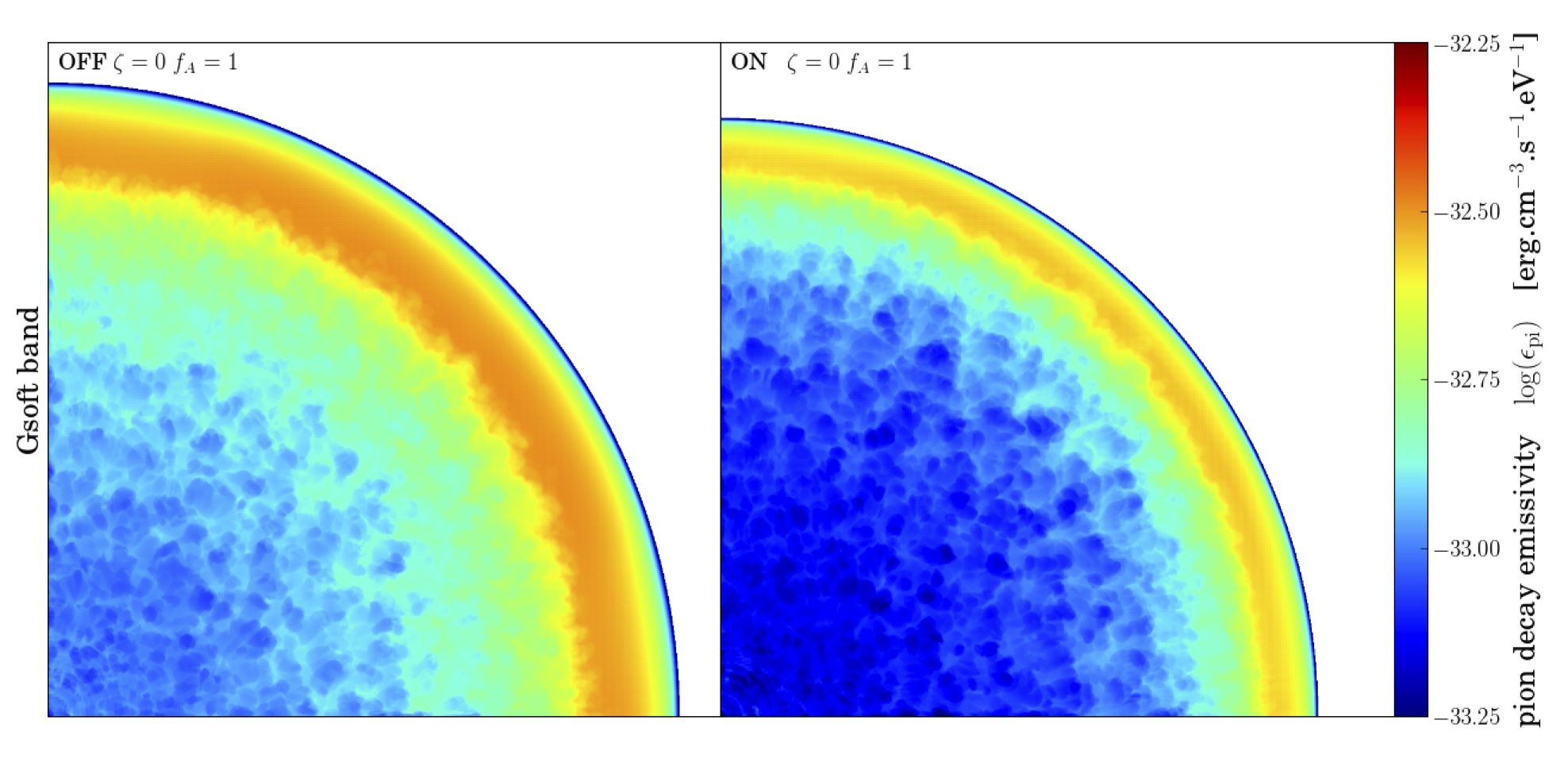}
\caption{Projected maps of the pion decay emission in the 0.1 -- 100~GeV $\gamma$-ray band.
\label{fig:map-NTpi-Gsoft}}
\end{figure}

\begin{figure}[t]
\centering
\includegraphics[width=\figsizemap]{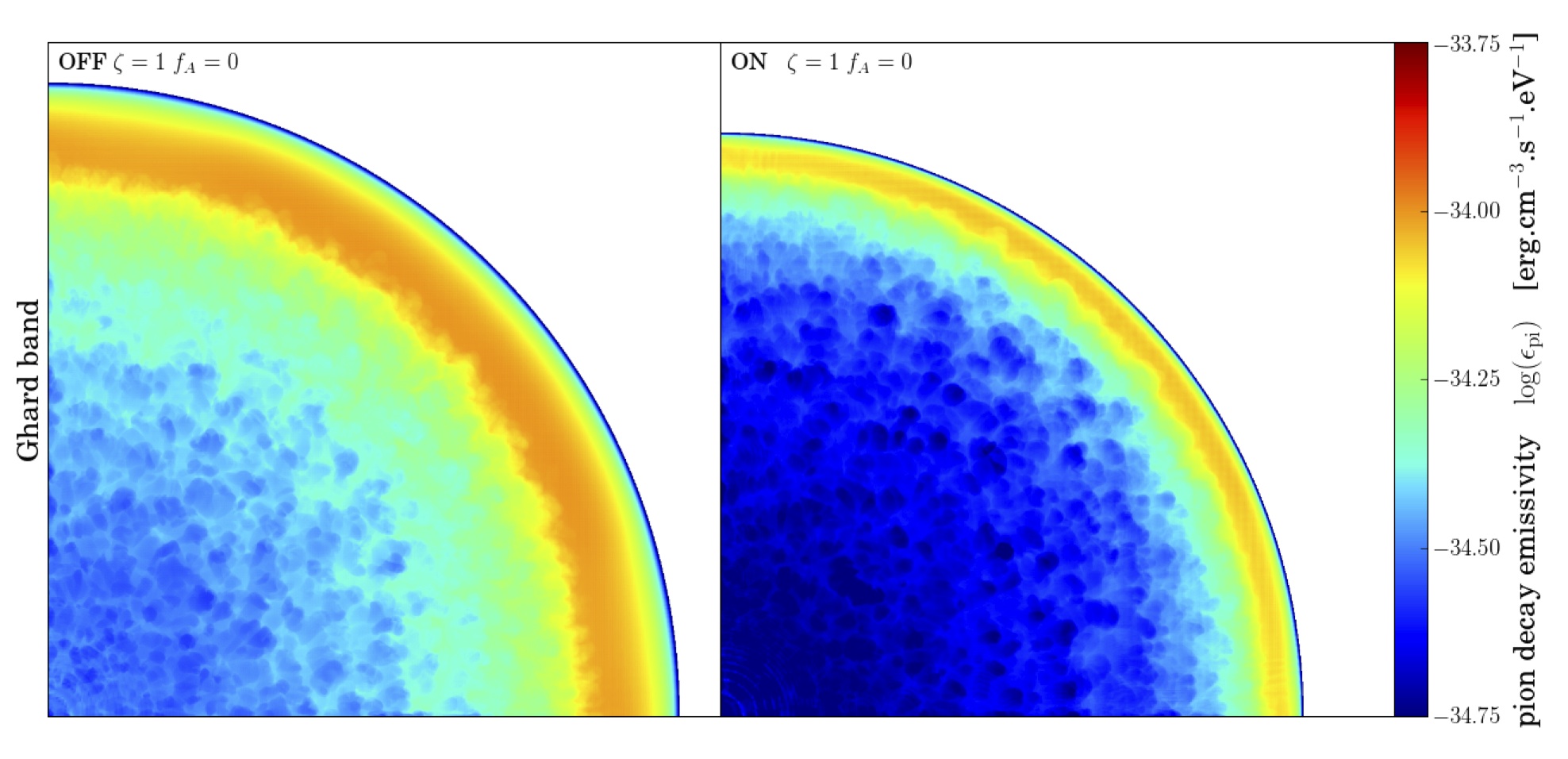}
\includegraphics[width=\figsizemap]{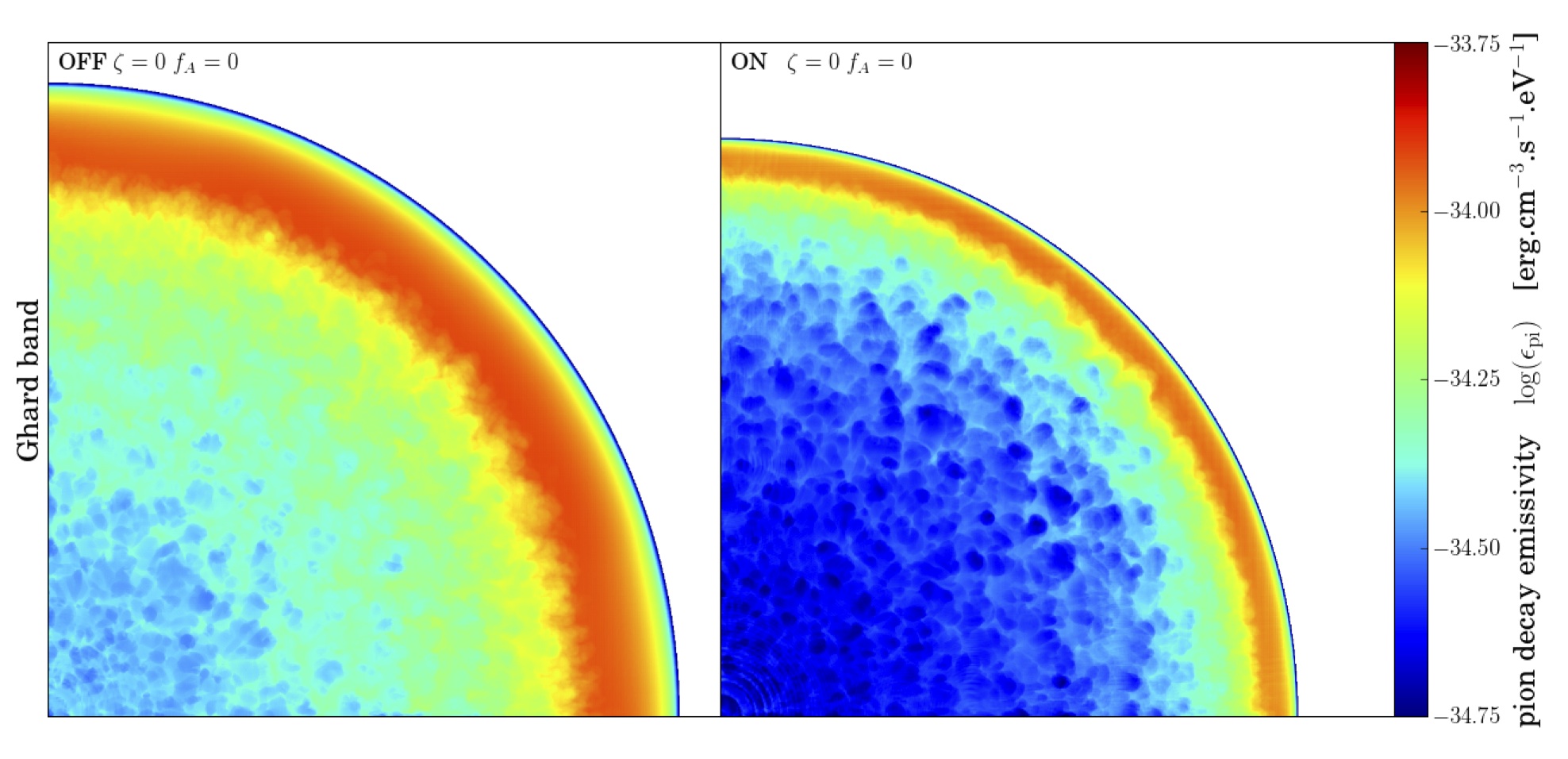}
\includegraphics[width=\figsizemap]{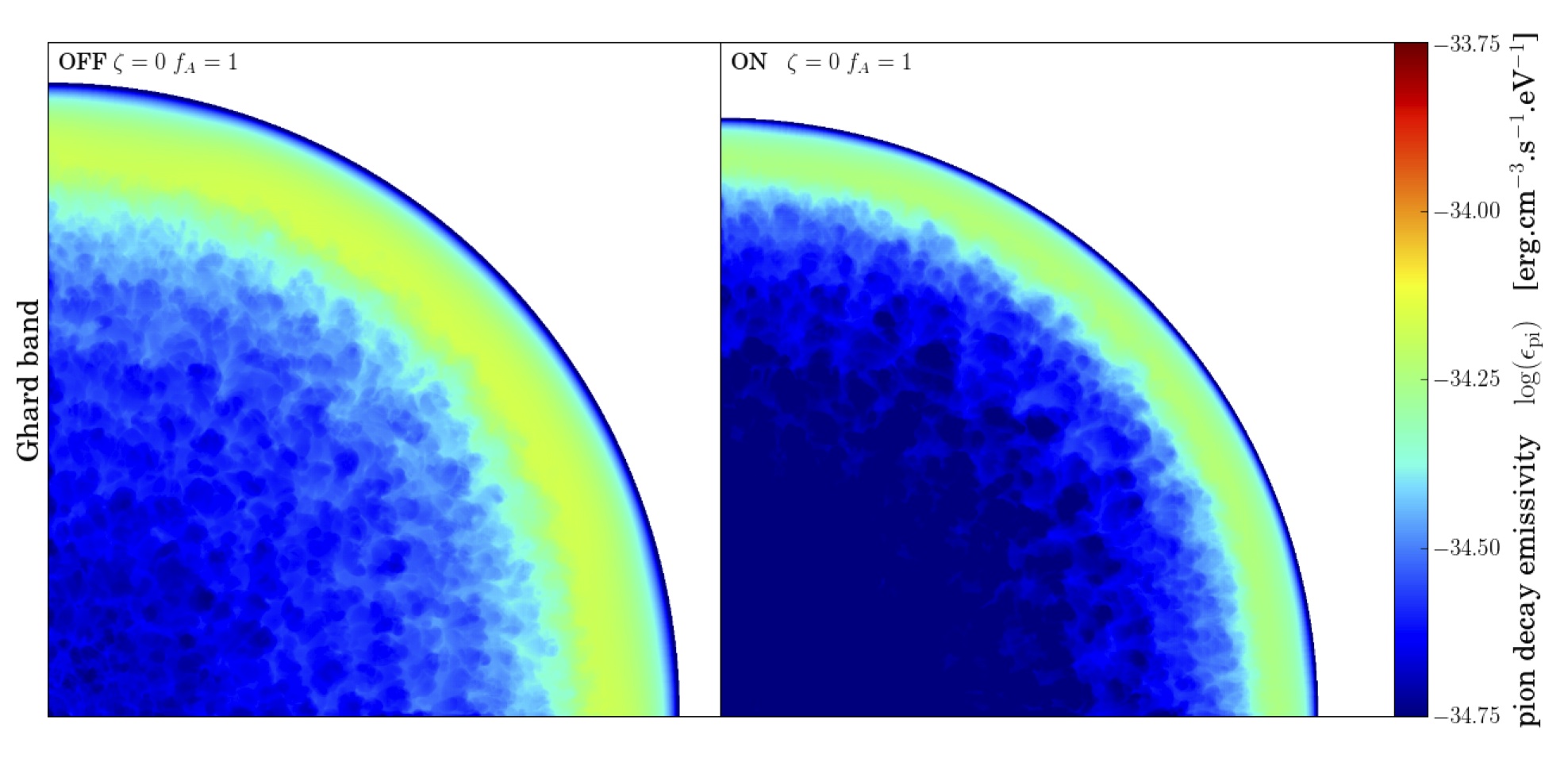}
\caption{Projected maps of the pion decay emission in the 0.01 -- 10~TeV $\gamma$-ray band.
\label{fig:map-NTpi-Ghard}}
\end{figure}

\newcommand{\figsizespc}{16cm}

\begin{figure}[t]
\centering
\includegraphics[width=\figsizespc]{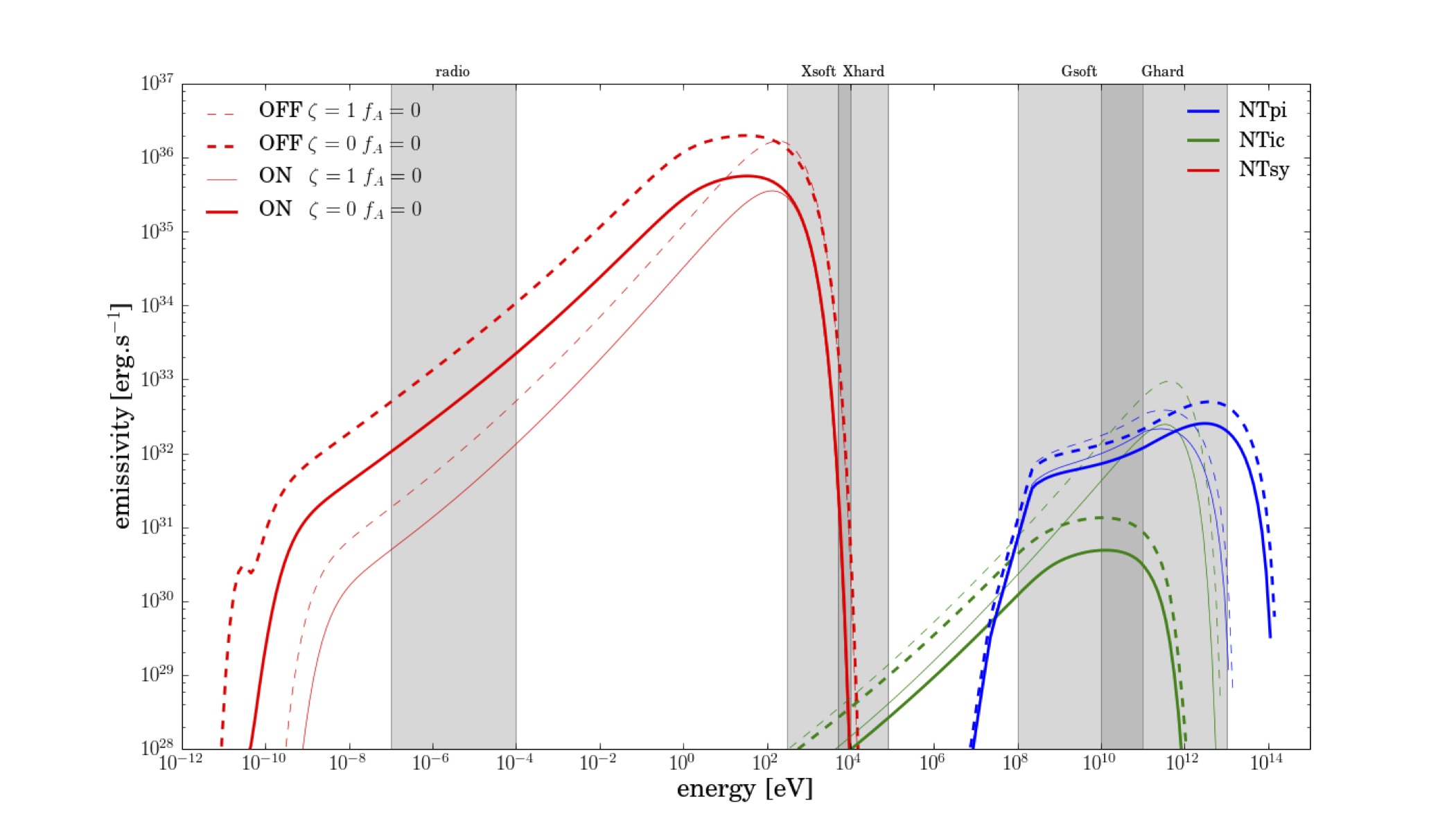}
\includegraphics[width=\figsizespc]{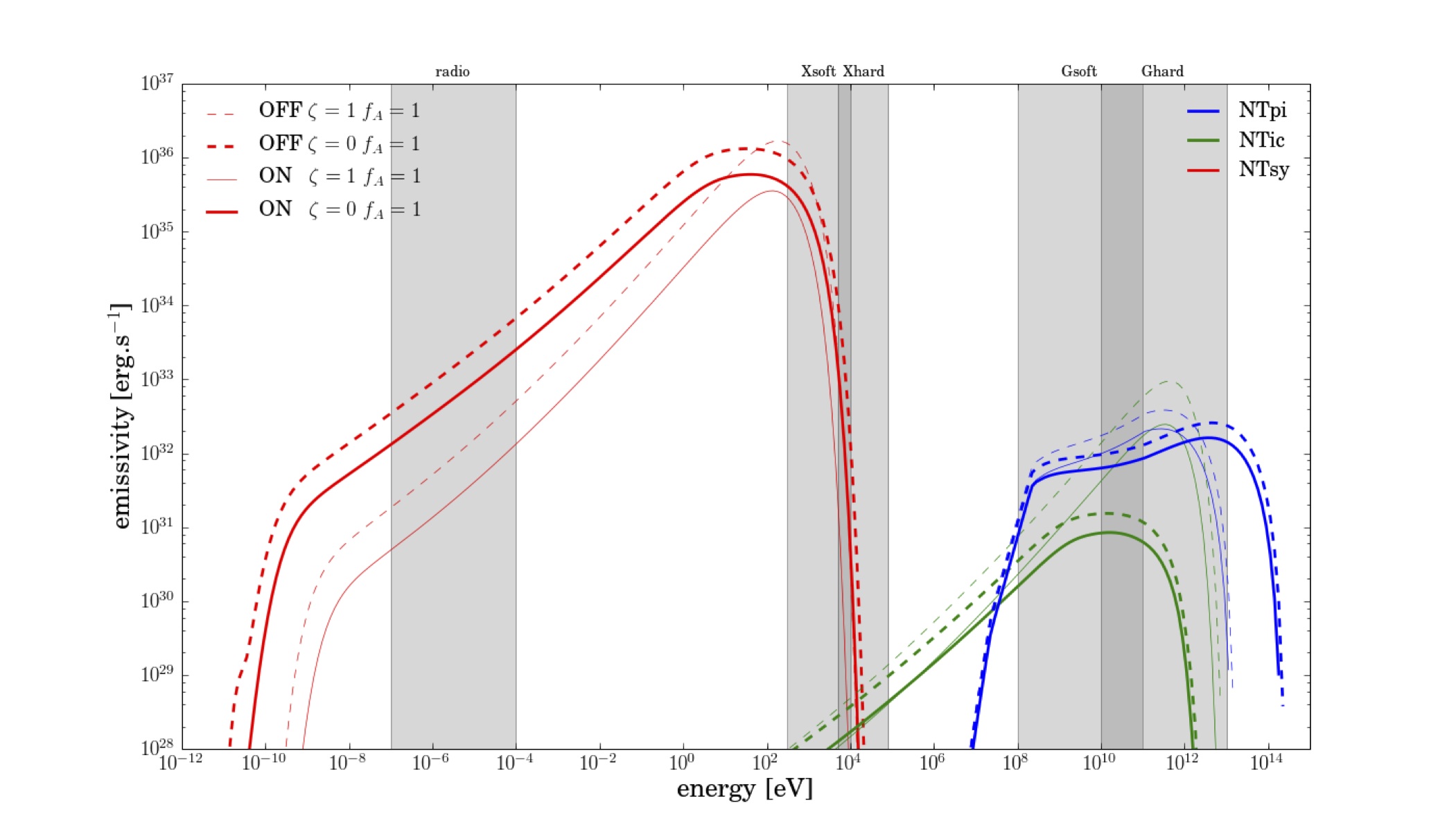}
\caption{Broadband spectra, integrated over the whole remnant, of the synchrotron emission (red), the inverse Compton emission (green), and the pion decay emission (blue). The top panel is with negligible Alfv{\'e}nic drift ($f_A=0$), while the bottom panel is with strong Alfv{\'e}nic drift ($f_A=1$). Each panel compares the cases with or without hydrodynamic back-reaction (ON vs. OFF), and with or without magnetic back-reaction ($\zeta=0$ vs. $\zeta=1$).
\label{fig:spc-back-zeta}}
\end{figure}

\end{document}